\documentclass[aps,pra,superscriptaddress,longbibliography,12pt]{revtex4-2}

\usepackage{amssymb,amsmath,bbm,ragged2e,graphicx,physics,enumerate,color,multirow,booktabs,bm,titlesec,threeparttable,diagbox}
\usepackage[caption=false]{subfig}
\usepackage[export]{adjustbox}
\usepackage[table]{xcolor}

\newcommand{\bathh}{\mathrm{h}}
\newcommand{\bathc}{\mathrm{c}}

\newcommand{\rhogg}{\rho_\mathrm{gg}}
\newcommand{\rhoge}{\rho_\mathrm{ge}}
\newcommand{\rhoeg}{\rho_\mathrm{eg}}
\newcommand{\rhoee}{\rho_\mathrm{ee}}

\newcommand{\tilderhogg}{\tilde{\rho}_\mathrm{gg}}
\newcommand{\tilderhoge}{\tilde{\rho}_\mathrm{ge}}
\newcommand{\tilderhoeg}{\tilde{\rho}_\mathrm{eg}}
\newcommand{\tilderhoee}{\tilde{\rho}_\mathrm{ee}}

\newcommand{\ketg}{\ket{\mathrm{g}}}
\newcommand{\kete}{\ket{\mathrm{e}}}
\newcommand{\brag}{\bra{\mathrm{g}}}
\newcommand{\brae}{\bra{\mathrm{e}}}
\newcommand{\Pg}{p_{\mathrm{g}}}
\newcommand{\Pe}{p_{\mathrm{e}}}

\begin{document}
\title{Roles of the Internal Coupling: From Equilibrium to Non-Equilibrium Dynamics}
\author{Jingyi Gao\\
Department of Physics, the University of Tokyo\\5-1-5 Kashiwanoha, Kashiwa, Chiba, 277-8574, Japan.\\
Naomichi Hatano\\
Institute of Industrial Science, the University of Tokyo\\5-1-5 Kashiwanoha, Kashiwa, Chiba, 277-8574, Japan}
\date{\today}

\begin{abstract}
We identify the precise thermodynamic roles of internal coupling in quantum thermal machines by analyzing the quantum Otto cycle across thermalized Gibbs-state limit cycles (GSLC) and finite-time non-equilibrating limit cycles (NELC). By systematically comparing the internally coupled system with standard and dressed-spectrum models, we resolve previous misconceptions regarding the physical origins of performance variations. First, we establish that the internally coupled system outperforms the standard system solely due to the Hamiltonian spectrum update, which intrinsically broadens the operating regimes. Second, we reveal that time-dependent efficiency and coefficient of performance (COP) are not caused by general quantum coherence, but specifically by eigenbasis mismatch arising from the non-commutativity of Hamiltonians across different strokes. 
The mismatch acts as quantum internal friction, reducing efficiency and COP and generating dissipative thermal modes. This results in the power-efficiency trade-off law, which we should find in realistic engines. However, we would not find the trade-off law when coherence exists but the eigenbasis mismatch does not exist, because there would be no friction, and hence the efficiency and COP would remain completely time-independent. 
Third, we isolate the distinct impact of coherence on power output. Validated by the global approach of the GKSL master equation, we demonstrate that coherence inherently reduces power by suppressing kinetic thermalization rates, regardless of eigenbasis mismatch. 
\end{abstract}

\maketitle

\section{Introduction}
\label{intro}

Quantum thermal machines have attracted extensive attention in quantum thermodynamics as platforms for studying energy conversion at the microscopic scale~\cite{EPJB_2021, Mukherjee_2021, PhysRevLett.119.090603, PhysRevE.101.052129}. Among various models, the quantum Otto cycle provides a particularly transparent framework for analyzing work extraction, heat flows, and thermodynamic performance in driven quantum systems~\cite{Piccitto_2022, PhysRevE.109.014122, e19040136}. In standard formulations of the quantum Otto cycle, the working medium is typically assumed to possess well-separated energy levels without internal coupling, so that the cycle dynamics are determined solely by externally controlled strokes and interactions with heat baths~\cite{annurev:/content/journals/10.1146/annurev-physchem-040513-103724, PhysRevLett.112.030602, PhysRevE.68.016101}. Under these assumptions, fundamental performance bounds such as the Otto efficiency and the coefficient of performance (COP) follow directly from the level spacings and bath temperatures~\cite{Beretta_2012, Abah_2016, PhysRevB.101.054513, PhysRevResearch.5.023066}.

Realistic quantum systems often involve non-negligible internal couplings between energy levels, arising from coherent interactions, external fields, or intrinsic device architectures. Such internal couplings can strongly influence quantum thermodynamic behavior and may serve as an additional resource for modifying the performance of thermal machines~\cite{PhysRevE.96.032110, PhysRevE.106.064133, PhysRevA.98.042102}. Several studies have investigated the performance of coupled thermal machines~\cite{PhysRevE.83.031135, PhysRevE.109.064129, PhysRevResearch.6.023172, PhysRevE.92.032124}, but they primarily focus on couplings among multiple subsystems, in which the thermodynamic behavior depends not only on the coupling strength but also on system size and the number of subsystems. Hitherto, the intrinsic role of internal coupling itself in determining the operational regime, performance bounds, and steady-state behavior of quantum Otto cycles remains insufficiently understood.

In this work, we investigate the roles that internal coupling plays in the operation and performance of quantum Otto thermal machines using a minimal single-qubit model. By focusing on a single-qubit system with internal coupling, we isolate the intrinsic thermodynamic role of the coupling without the additional effects introduced by system size or multiple subsystems. By comparing three types of quantum Otto cycles from thermalized to non-equilibrium conditions, we find that the internal coupling influences the operating regimes, efficiency, coefficient of performance (COP), and power of the quantum Otto cycle in three ways. 

First, the internal coupling broadens the operating regimes and improves the thermal performance by updating the system Hamiltonian. Second, the internal coupling requires the eigenbasis transformation in the isochoric process and leads to off-diagonal coherence, which directly results in lower power in the finite-time cycle. Third, when the eigenbasis mismatches in each stroke, the efficiency and the COP are reduced, revealing the power-efficiency trade-off relation.

\begin{figure*}[h!]
\centering
	\subfloat[Standard system]{\makebox[0.3\textwidth][c]{{\includegraphics[scale=0.7]{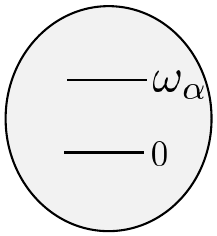}\label{StandardSystem}}}}
	\subfloat[Internally coupled system]{\makebox[0.3\textwidth][c]{{\includegraphics[scale=0.7]{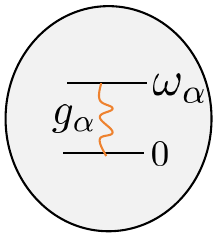}\label{CoupledSystem}}}}
	\subfloat[Dressed-spectrum system]{\makebox[0.3\textwidth][c]{{\includegraphics[scale=0.7]{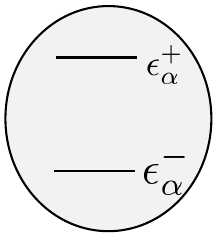}\label{DressedSystem}}}}
	\caption{Comparison of the internal systems in the isochoric process $\alpha=\bathh,\bathc$ among (a)~the standard Otto cycle, (b)~the internally coupled Otto cycle, and (c)~the dressed-spectrum Otto cycle. Both (a)~the standard system and (c)~ the dressed-spectrum system follow the protocol of the standard Otto cycle without internal coupling, as shown in Fig.~\ref{standardcycle}, while the Otto cycle based on the internally coupled system runs differently, as shown in Fig.~\ref{cyclecoupling}.}
	\label{Systems}
\end{figure*}

The present paper is organized as follows. 
In Sec.~\ref{Sec1}, we present a generic analysis of three types of quantum Otto cycles. 
In Sec.~\ref{Sec1-1}, we review the standard Otto cycle without internal coupling, as shown in Figs.~\ref{Systems}~\subref{StandardSystem} and \ref{standardcycle}, representing the conventional case where coupling is neglected. Here, the operating-regime boundaries, efficiency, and coefficient of performance (COP) take generic forms determined solely by the bare energy levels and heat-bath temperatures, remaining independent of the system state and interaction time. 
In Sec.~\ref{Sec1-2}, we analyze the dressed-spectrum Otto cycle based on effective energy levels obtained from the diagonalized Hamiltonian, as shown in Fig.~\ref{Systems}~\subref{DressedSystem} and~\ref{dressedcycle}. This model incorporates internal coupling strictly as a Hamiltonian update while neglecting off-diagonal coherence and eigenbasis transformations across strokes. Consequently, like the standard cycle, its efficiency, COP, and operating regimes retain state- and time-independent generic forms. 
Finally, in Sec.~\ref{Sec1-3}, we introduce the realistic internally coupled Otto cycle, as shown in Fig.~\ref{Systems}~\subref{CoupledSystem} and~\ref{cyclecoupling}]. By adjusting both energy levels and coupling strengths during work production, the non-commutativity of Hamiltonians across strokes renders the work stroke non-adiabatic, departing from the simplified behavior of the former two models.

\begin{figure*}[h!]
\centering
	\subfloat[Gibbs-state limit cycle (GSLC)]{\includegraphics[width=0.4\textwidth]{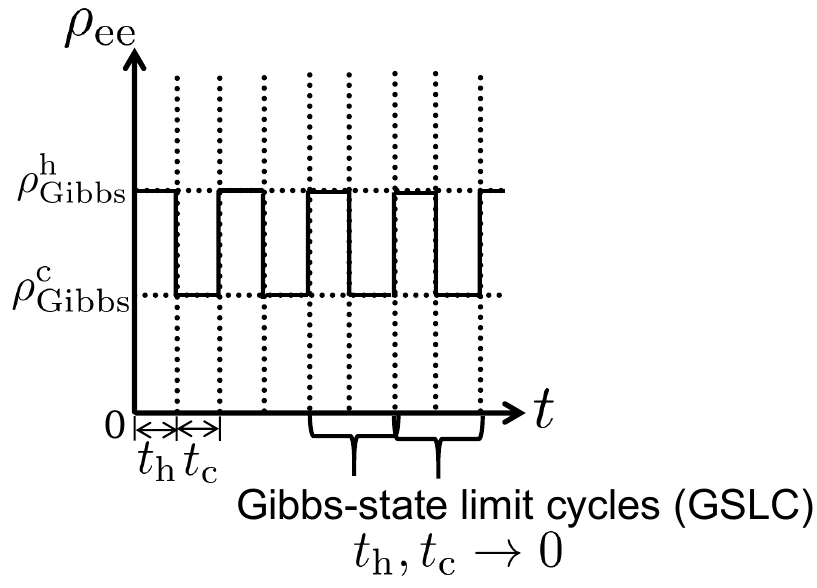}\label{GSLCDiagram}}
	\subfloat[equilibrating limit cycle (ELC)]{\includegraphics[width=0.6\textwidth]{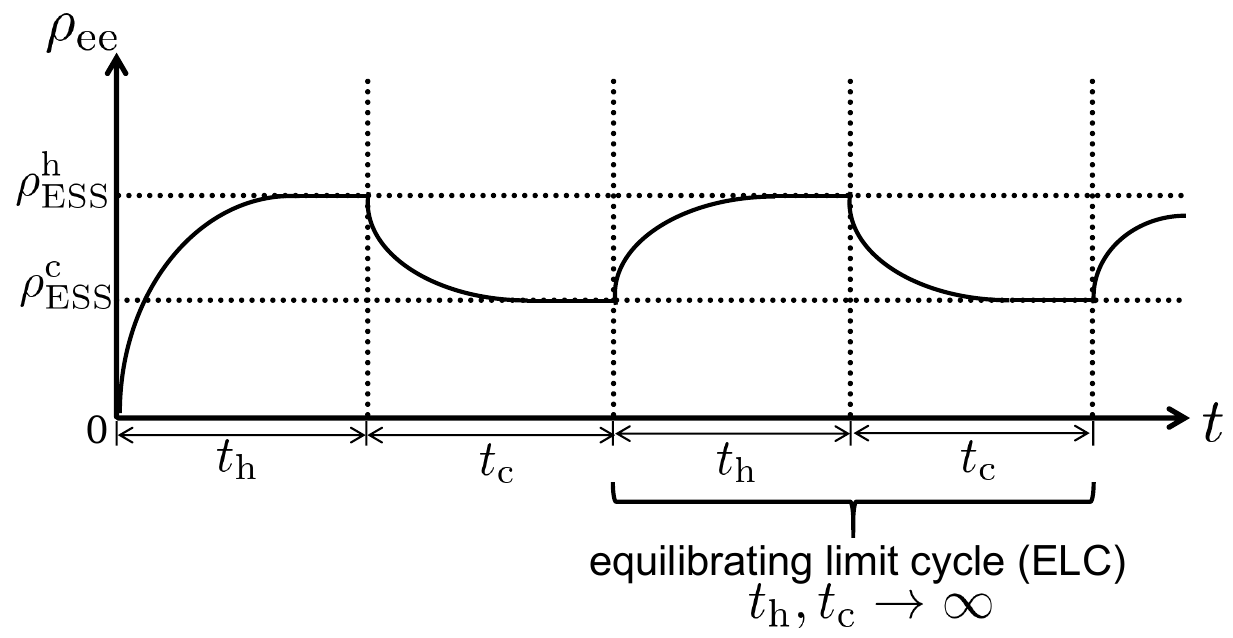}\label{ELCDiagram}}
	
	\subfloat[non-equilibrating limit cycle (NELC)]{\includegraphics[width=0.6\textwidth]{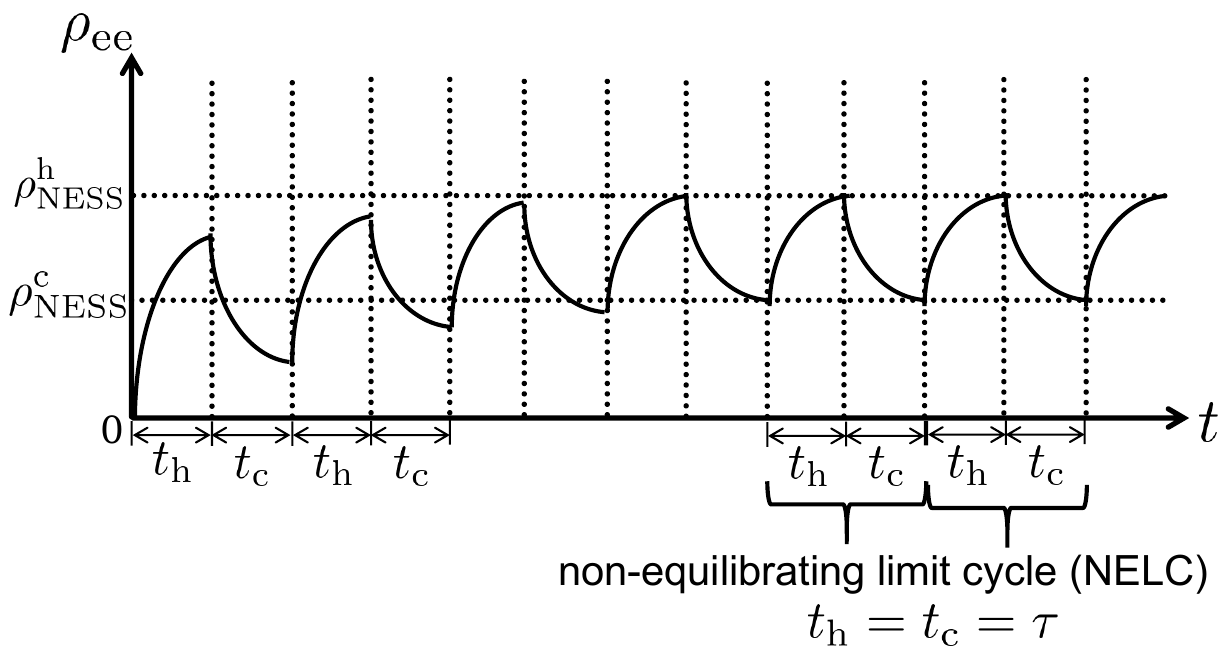}\label{NELCDiagram}}
	\caption{Schematic illustration of the three types of limit cycles for the internally coupled system: 
(a) Gibbs-state limit cycle (GSLC), where the system is assumed to equilibrate instantaneously with the heat baths; 
(b) equilibrium limit cycle (ELC), obtained for sufficiently long interaction times; 
(c) non-equilibrating limit cycle (NELC), arising when the interaction time is too short for the system to reach equilibrium during the isochoric strokes. Both the GSLC and the ELC are thermalized systems, and should show the same phenomena. The NELC is a steady limit cycle that alternates between two states, neither of which is a steady state.}
	\label{Diagrams}
\end{figure*}

In Sec.~\ref{Sec2}, we reveal the dual roles of internal coupling under thermalized conditions. On one hand, it broadens operating regimes and enhances performance compared to the standard system; on the other hand, it introduces dissipative thermal modes that reduce efficiency and COP relative to the dressed-spectrum one. 
In Sec.~\ref{Sec2-1}, assuming rapid equilibration during isochoric strokes, we construct the Gibbs-state limit cycle (GSLC) for the internally coupled system, as shown in Fig.~\ref{Diagrams}~\subref{GSLCDiagram} and contrast it with the state-independent GSLC results of the standard and dressed-spectrum systems. 
In Sec.~\ref{Sec2-2}, we demonstrate that internal coupling expands operational regimes by updating the Hamiltonian spectrum. However, due to Hamiltonian non-commutativity in work strokes, the internally coupled GSLC exhibits distinct regime boundaries from the dressed-spectrum one. The internally coupled system introduced an emergent heater mode and an expanded accelerator region. By introducing a parameter $\Lambda$ related to the internal friction $P_{\rm fric}$, we clarify that this passive alteration is caused by Hamiltonian non-commutativity rather than state coherence itself. When the Hamiltonians commute across strokes, the internally coupled GSLC reproduces the performance of the dressed-spectrum system despite non-zero internal coupling. 
In Sec.~\ref{Sec2-3}, we first derive conditions under which internal coupling improves efficiency and COP over the standard GSLC. Furthermore, we confirm that non-commutativity acts as quantum internal friction, placing the dressed-spectrum performance as an upper bound for the internally coupled GSLC. This bound is reached when the work-stroke Hamiltonians commute. Importantly, tuning the coupling to eliminate non-commutativity collapses the internally coupled GSLC back to standard system performance. This indicates that internal friction is an indispensable thermodynamic cost incurred when using internal coupling to enhance performance via a modified Hamiltonian.

In Sec.~\ref{Sec3}, moving beyond ideal rapid thermalization, we analyze the non-equilibrating limit cycle (NELC) under finite-time interactions, as shown in Fig.~\ref{Diagrams}~\subref{NELCDiagram}, and compare it with the dressed-spectrum NELC, as detailed in Appendix~\ref{A3}. 
In the NELC regime, the dressed-spectrum system consistently exhibits superior power and efficiency than the internally coupled one; however, it is unrealistic. Because the efficiency and COP of the standard and dressed-spectrum models depend solely on energy levels and bath temperatures, they remain time-independent and evade the classical power-efficiency trade-off law~\cite{PhysRevLett.127.190604, Koyuk_2019, PhysRevLett.120.190602, PhysRevLett.117.190601, PhysRevE.106.024137, Dechant_2019}. This indicates that finite-time operation alone does not dictate the power-efficiency trade-off in quantum Otto cycles. Instead, eigenbasis mismatch caused by the non-commutative Hamiltonian acts as quantum internal friction that dissipates energy as waste heat, serving as the true microscopic driver of the trade-off. As this friction vanishes, the efficiency approaches the dressed-spectrum bound, eliminating the efficiency penalty. Nevertheless, even without eigenbasis mismatch, the power output of the internally coupled NELC remains strictly bounded below the dressed-spectrum limit. Validated by the global GKSL master equation, internal coupling leads to off-diagonal coherence and suppresses kinetic thermalization rates, thereby reducing power output. These fundamental insights highlight the indispensable roles of realistic internal coupling in finite-time quantum thermal devices.

\section{Three Types of Otto Cycles }
\label{Sec1}

To analyze the influence of internal coupling, we compare three types of Otto cycles. 
Our focal case is the internally coupled system, as illustrated in Fig.~\ref{Systems}~\subref{CoupledSystem}, which includes the inevitable internal coupling and is significant in realistic applications. 
The standard Otto cycle switches between two system Hamiltonians, which are defined by the bare energy level $\omega_\alpha$ without the internal coupling, as illustrated in Fig.~\ref{Systems}~\subref{StandardSystem}. It mimics the case in which we neglect the internal coupling. 
The dressed-spectrum Otto cycle is similar to the standard Otto cycle, while defined by the effective energy levels $\epsilon_\alpha^\pm$ (defined in Eq.~\eqref{dressedenergy} below), as illustrated in Fig.~\ref{Systems}~\subref{DressedSystem}. It treats the internal coupling as the Hamiltonian-spectrum update and switches between two dressed-spectrum Hamiltonians obtained by diagonalizing the internally coupled Hamiltonian. 

In this section, we analyze these three cycles in general, beginning with the standard Otto cycle without internal coupling, then the dressed-spectrum Otto cycle, and finally the internally coupled Otto cycle. 
In Sec.~\ref{Sec1-1}, we review the operation regimes and the thermal performance of the standard Otto cycle with the bare energy level $\omega_\alpha$ without the internal coupling, as illustrated in Fig.~\ref{Systems}~\subref{StandardSystem}. The operating-regime boundaries,  efficiency, and the coefficient of performance (COP) in the standard Otto cycle depend on the bare energy level $\omega_\alpha$ and the bath temperatures $\beta_\alpha$, . 

In Sec.~\ref{Sec1-2}, as illustrated in Fig.~\ref{Systems}~\subref{DressedSystem}, we explain the dressed-spectrum Otto cycle, which considers the internal coupling as the update of the Hamiltonian spectrum and operates as the standard Otto cycle with the effective energy levels $\epsilon_\alpha^\pm$ as defined in Eq.~\eqref{dressedenergy} below. 
Similarly to the standard Otto cycle, the efficiency and the COP in the dressed-spectrum Otto cycle only depend on the effective energy gap $\tilde{\omega}_\alpha$ as defined in Eq.~\eqref{effectiveenergygap} below and the bath temperatures $\beta_\alpha$. Due to the update of Hamiltonian spectrum, the operation regimes of the dressed-spectrum Otto cycle differ from the standard Otto cycle.

In Sec.~\ref{Sec1-3}, we finally derive the efficiency and the COP of the Otto cycle with internal coupling, as illustrated in Fig.~\ref{Systems}~\subref{CoupledSystem}. Unlike the standard and dressed-spectrum Otto cycles, it includes coherence terms arising from the non-commutativity between the isochoric processes $\alpha=\bathh,\bathc$. The non-commutativity causes the internally coupled Otto cycle to exhibit thermal modes and performance different from those of the standard and dressed-spectrum Otto cycles. 
We claim that these three cycles are distinct and compare the efficiency and COP of the internally coupled system with those of the standard and dressed-spectrum systems. 

\subsection{Standard Otto Cycle}
\label{Sec1-1}
\begin{figure*}[h!]
\centering
	\includegraphics[width=0.5\textwidth]{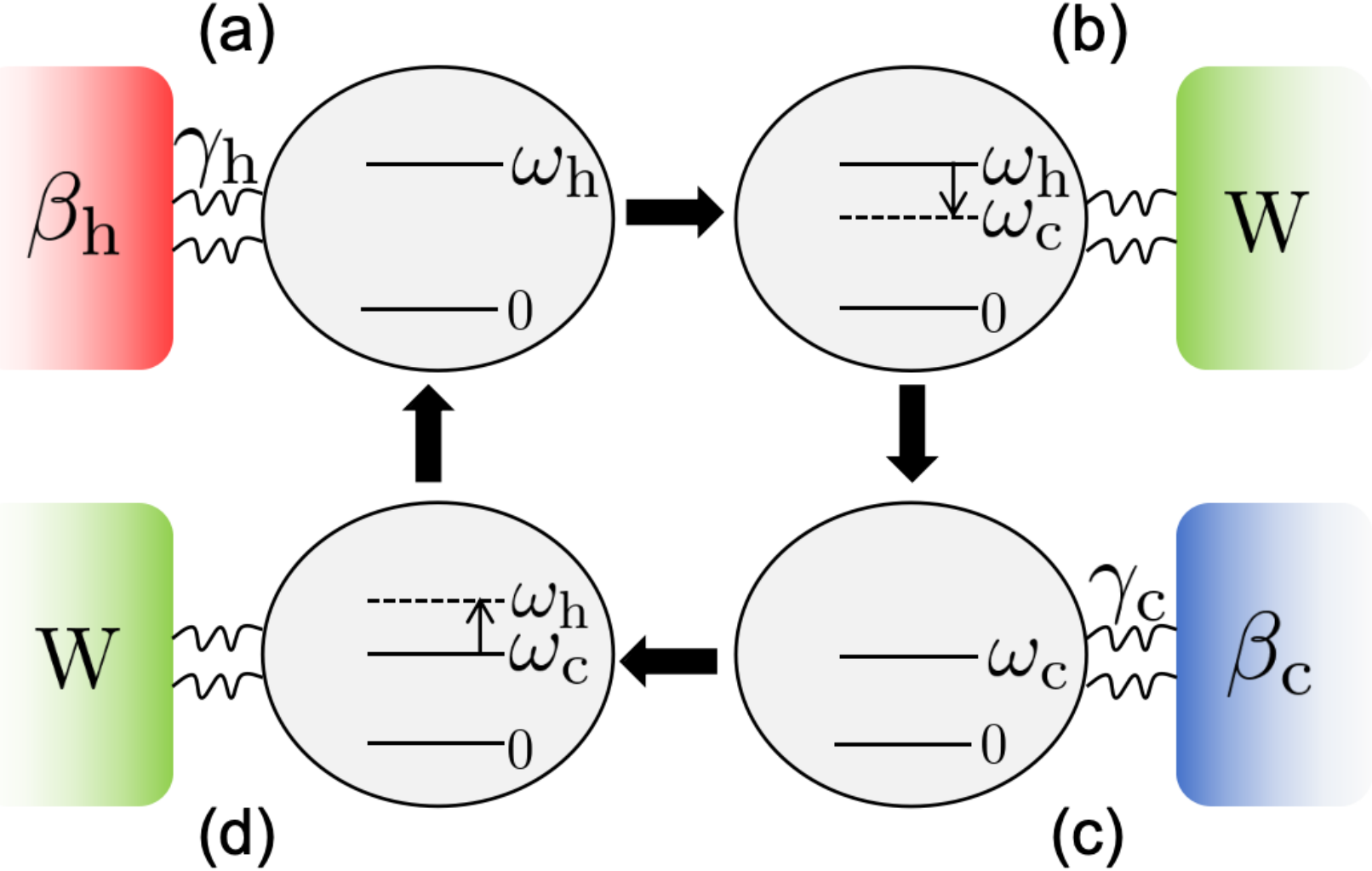}
	\caption{Schematic view of the standard Otto cycle without any internal couplings. The internal system is a two-level system with excited energy level alternates between $\omega_{\bathh}$ and $\omega_{\bathc}$. In the isochoric processes (a) and (c), it interacts with heat baths with the inverse temperature $\beta_{\bathh}$ and $\beta_{\bathc}$, respectively. In the adiabatic processes (b) and (d), it interacts with the work storage denoted by $\rm W$. We assume that the interaction time in the adiabatic processes is negligible.}
	\label{standardcycle}
\end{figure*}

The standard quantum Otto cycle consists of four strokes: a pair of isochoric processes and a pair of adiabatic processes, as illustrated in Fig.~\ref{standardcycle}. 
During each isochoric stroke, labeled (a) and (c) in the figure, we keep the Hamiltonian of the internal two-level system constant, and the system state evolves due to interaction with the external baths with the inverse temperatures $\beta_{\bathh}$ and $\beta_{\bathc}$, leading to heat flows. Conversely, during each adiabatic stroke, labeled (b) and (d) in the figure, the system state remains unchanged, while we vary the system Hamiltonian, leading to work production or extraction. 

For each of the isochoric processes (a) and (c) in Fig.~\ref{standardcycle}, the total Hamiltonian is composed of three parts: the Hamiltonian of the internal two-level system, the Hamiltonian of the heat baths with the inverse temperature $\beta_\alpha$, and the interaction Hamiltonian:
\begin{align}
H = H_S^\alpha+H_B^\alpha+H_I^\alpha,
\end{align}
where $\alpha = \bathh, \bathc$ denotes the isochoric processes (a) and (c) in Fig.~\ref{standardcycle}, respectively. The system Hamiltonian is given by
\begin{align}
\label{HSnocp}
H_S^\alpha = \omega_\alpha \sigma_S^+ \sigma_S^-= \begin{pmatrix}
0 &0 \\
0 &\omega_\alpha
\end{pmatrix},
\end{align}
where $\sigma_S^\pm$ are the spin raising and lowering operators. The heat baths are assumed to be bosonic, and thus the Hamiltonian $H_B^\alpha$ of each bath $\alpha=\bathh,\bathc$ is
\begin{align}
\label{HB}
H_B^\alpha = \sum_k \hat{b}_\alpha^\dagger \hat{b}_\alpha,
\end{align}
where $\{\hat{b}_\alpha^\dagger, \hat{b}_\alpha \}$ are the boson creation and annihilation operators. 
The interaction Hamiltonian between the system and the bath $\alpha$ takes the standard form:
\begin{align}
\label{HI}
H_I^\alpha = \sum_{k,\alpha} V_{k,\alpha} \sigma_S^x (\hat{b}_\alpha +\hat{b}_\alpha^\dagger),
\end{align}
where $\sigma_S^x = \sigma_S^++\sigma_S^-$.

For the state of the system, let us introduce the notation 
\begin{align}
\label{rho}
\rho_\alpha = \begin{bmatrix}\rhogg^\alpha &\rhoge^\alpha\\
\rhoeg^\alpha &\rhoee^\alpha\end{bmatrix}
 \text{ for } \alpha = \bathh,\bathc.
 \end{align}
The system state is $\rho_\bathh$ after the isochoric process in Fig.~\ref{standardcycle}~(a), while $\rho_\bathc$ after the process in Fig.~\ref{standardcycle}~(c). The heat absorbed by the system from each bath $\alpha = \bathh, \bathc$ during the corresponding isochoric process is given by
\begin{align}
\label{Qhstandard}
{Q_\bathh^{\rm st}} &= \Tr[H_S^\bathh (\rho_\bathh - \rho_\bathc)] 
		= \omega_{\bathh} (\rhoee^\bathh-\rhoee^\bathc),\\
\label{Qcstandard}	
{Q_\bathc^{\rm st}} &= \Tr[H_S^\bathc (\rho_\bathc - \rho_\bathh)] 
		 =\omega_{\bathc} (\rhoee^\bathc-\rhoee^\bathh),
\end{align}
where the superscript $\rm st$ denotes quantities related to the standard Otto cycle. 
Note that we define them such that the heat flows from the bath to the system when it is positive. Note also that the off-diagonal coherence terms of the system state $\rho^\alpha$ do not contribute at all. 

For the adiabatic processes in Fig.~\ref{standardcycle}~(b) and (d), the work extracted from the environment into the internal system is given by
\begin{align}
\label{W1standard}
{W_1^{\rm st}} &= \Tr[(H_S^\bathc-H_S^\bathh)\rho_\bathh] 
	= (\omega_{\bathc}-\omega_{\bathh})\rhoee^\bathh,\\
\label{W2standard}
{W_2^{\rm st}} &=  \Tr[(H_S^\bathh-H_S^\bathc)\rho_\bathc] 
	= (\omega_{\bathh}-\omega_{\bathc})\rhoee^\bathc.
\end{align}
Note that we also define such that the work comes into the system when it is positive. Therefore, the first law of thermodynamics holds in the form
\begin{align}
\label{Wstandard}
W^{\rm st} = {W_1^{\rm st}}+{W_2^{\rm st}} = -(\omega_{\bathh} -\omega_{\bathc})(\rhoee^\bathh-\rhoee^\bathc) = -({Q_\bathh^{\rm st}}+{Q_\bathc^{\rm st}}).
\end{align}

For the standard Otto cycle, the inverse temperatures $\beta_\alpha$ and the energy levels $\omega_\alpha$ are assumed to be positive. We also assume that $\beta_\bathh<\beta_\bathc$. There are three operating modes: engine, refrigerator, and accelerator. The classification of these regimes based on the frequency ratio $\omega_\bathc/\omega_\bathh$ is summarized in Table~\ref{table:standardregimes}. 

\begin{table}[ht]
\centering
\caption{Operational regimes of the standard Otto cycle.}
\label{table:standardregimes}
\begin{tabular}{|l|c|c|c|c|}
\hline
\textbf{Regime} & \textbf{Parameter Range} & \textbf{Work $W$} & \textbf{Heat $Q_c$} & \textbf{Heat $Q_h$} \\ \hline
Accelerator & $\frac{\beta_\bathh}{\beta_\bathc}  <1 < \frac{\omega_\bathc}{\omega_\bathh}$ & $> 0$ & $< 0$ & $> 0$ \\ \hline
$\omega_\bathc = \omega_\bathh$ & $\frac{\beta_\bathh}{\beta_\bathc} <1 = \frac{\omega_\bathc}{\omega_\bathh}$ & $0$ & $< 0$ & $> 0$ \\ \hline
Engine & $\frac{\beta_\bathh}{\beta_\bathc} < \frac{\omega_\bathc}{\omega_\bathh} < 1$ & $< 0$ & $< 0$  & $> 0$ \\ \hline
$\beta_\bathh \omega_\bathc = \beta_\bathh \omega_\bathh$ & $\frac{\omega_\bathc}{\omega_\bathh} = \frac{\beta_\bathh}{\beta_\bathc}<1$ & $0$ & $0$ & $0$ \\ \hline
Refrigerator & $\frac{\omega_\bathc}{\omega_\bathh} < \frac{\beta_\bathh}{\beta_\bathc}<1$ & $> 0$& $> 0$  & $< 0$  \\ \hline
\end{tabular}
\end{table}

The engine produces work to the external environment by transferring heat from the high-temperature bath to the low-temperature bath. 
The refrigerator absorbs heat from the low-temperature bath and releases it to the high-temperature bath by consuming work from the external environment. 
The accelerator accelerates the heat transfer from the high-temperature to the low-temperature bath by consuming work from the external environment. Usually, only the engine and the refrigerator are considered as useful thermal modes, while the accelerator is seen as a dissipative mode. 

The engine regime acts as a thermodynamic bridge between the refrigerator regime and the accelerator regime. There is no direct boundary between the refrigerator and the accelerator because transitioning from pumping heat against a gradient to accelerating heat along a gradient requires passing through the intermediate spontaneous work-producing phase. Hence, the transitions between these modes are governed by two critical boundaries that correspond to zero work and zero heat exchange, respectively. 

One of the boundaries is the zero-heat-exchange boundary between the engine and the refrigerator, given by
    \begin{align}
    \label{standardbound1}
    \frac{\beta_\bathc}{\beta_\bathh} = \frac{\omega_\bathh}{\omega_\bathc}.
    \end{align}
This boundary marks the state where the populations at the end of both isochoric strokes are identical. It separates the engine regime, which produces work for the environment and transfers heat from the hot bath to the cold one, from the refrigerator regime, which transfers heat from the cold bath to the hot one and extracts the work from the environment.

The other is the zero-work boundary between the engine and the accelerator, given by
    \begin{align}
    \label{standardbound2}
    \omega_\bathh = \omega_\bathc.
    \end{align}
This boundary marks the point where the energy level shifts during the expansion and compression strokes perfectly cancel each other out, resulting in zero net work. It separates the engine regime, which produces work for the environment and transfers heat from the hot bath to the cold one, from the accelerator regime, which extracts the work and transfers the heat from the hot bath to the cold one.

On both boundaries, the system cannot operate as a thermal machine, since the energy absorbed from the hot bath equals the energy released to the cold bath and no work is produced.

Let us finally define indicators for the two useful thermal modes. 
The equilibrium efficiency in the engine regime is given by
\begin{align}
\label{standardefficiency}
\eta_{\rm st} = 1 - \frac{\omega_{\bathc}}{\omega_{\bathh}},
\end{align}
wheras the equilibrium coefficient of performance (COP) in the refrigerator regime is given by 
\begin{align}
\label{standardCOP}
\xi_{\rm st} = \frac{\omega_{\bathc}}{\omega_{\bathh} - \omega_{\bathc}}.
\end{align}

\subsection{Dressed-Spectrum Otto Cycle}
\label{Sec1-2}

Some may suspect that the influence of the internal coupling originates merely from the modification of the energy spectrum of the system Hamiltonian. 
In order to clarify the difference, in this subsection, we dare to consider a dressed-spectrum system, as illustrated in Fig.~\ref{Systems}~\subref{DressedSystem}, that works as a standard Otto cycle with the diagonalized Hamiltonian in each isochoric process $\alpha=\bathc,\bathh$. 

\begin{figure*}[h!]
\centering
	\includegraphics[width=0.5\textwidth]{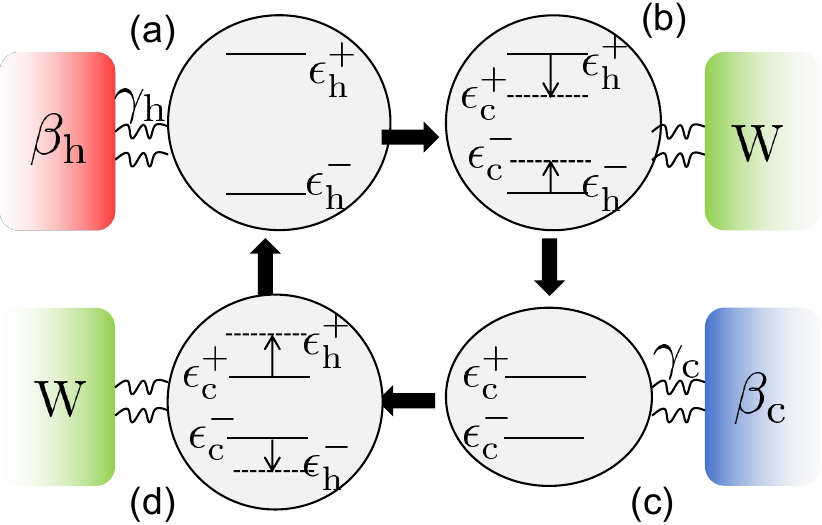}
	\caption{Schematic view of the dressed-spectrum Otto cycle considering the internal coupling as the update of Hamiltonian spectrum. The internal system is a two-level system with energy gap alternating between $\tilde{\omega}_{\bathh}$ and $\tilde{\omega}_{\bathc}$, where $\tilde{\omega}_{\alpha} = \epsilon_\alpha^+ - \epsilon_\alpha^-$ as defined in Eq.~\eqref{effectiveenergygap}. In the isochoric processes (a) and (c), it interacts with heat baths with the inverse temperature $\beta_{\bathh}$ and $\beta_{\bathc}$, respectively. In the adiabatic processes (b) and (d), it interacts with the work storage denoted by $W$. We assume that the interaction time in the adiabatic processes is negligible.}
	\label{dressedcycle}
\end{figure*}

After introducing the internal coupling, we update the Hamiltonian of the internal system from Eq.~(\ref{HSnocp}) to 
\begin{align}
\label{HS}
H_S^\alpha = \begin{pmatrix}
0 &g_\alpha \\
g_\alpha &\omega_\alpha
\end{pmatrix},
\end{align}
where $\alpha = \bathh, \bathc$ corresponds to the hot and cold isochoric strokes, respectively.
Meanwhile, we keep the bath and interaction Hamiltonians in Eqs.~(\ref{HB}) and (\ref{HI}) as they are, assuming that the external environment and the system-bath interaction are unaffected by the internal coupling $g_\alpha$. 
The Hamiltonian in Eq.~\eqref{HS} can be diagonalized as
\begin{align}
\label{tildeHS}
&\tilde{H}_S^\alpha = U_\alpha^\dagger H_S^\alpha U_\alpha = \begin{pmatrix}
\epsilon_\alpha^- & 0\\
0 & \epsilon_\alpha^+
\end{pmatrix},
\end{align}
where the eigenvalues are 
\begin{align}
\label{dressedenergy}
\epsilon_\alpha^\pm = \frac{1}{2}\omega_\alpha \pm \Delta_\alpha
\end{align}
with
\begin{align}
\label{Delta}
 \Delta_\alpha = \sqrt{g_\alpha^2+\frac{\omega_\alpha^2}{4}}.
\end{align}
The effective energy gap is therefore given by
\begin{align}
\label{effectiveenergygap}
\tilde{\omega}_\alpha = \epsilon_\alpha^+ - \epsilon_\alpha^- = \frac{1}{2}\sqrt{g_\alpha^2+\frac{\omega_\alpha^2}{4}} = 2\Delta_\alpha.
\end{align}
The corresponding eigenvector for each eigenvalue is
\begin{align}
\label{P1}
\ket{P_\alpha^+} = \begin{pmatrix}
\cos{\theta_\alpha} & \sin{\theta_\alpha}
\end{pmatrix}^T,\\
\label{P2}
\ket{P_\alpha^-} = \begin{pmatrix}
-\sin{\theta_\alpha} & \cos{\theta_\alpha}
\end{pmatrix}^T
\end{align}
with
\begin{align}
\label{theta}
\theta_\alpha = \arctan{\frac{\epsilon_\alpha^+}{g_\alpha}}.
\end{align}
The diagonalizing unitary transformation $U_\alpha$ in Eq.~(\ref{tildeHS}) is thus given by
\begin{align}
\label{Ualpha}
U_\alpha = \begin{pmatrix} \ket{P_\alpha^-} & \ket{P_\alpha^+} \end{pmatrix}.
\end{align}

In the dressed-spectrum Otto cycle, as shown in Fig.~\ref{dressedcycle}, we first diagonalize the Hamiltonian of the system and then make it run similarly to the standard Otto cycle based on the diagonalized Hamiltonian. Without considering the non-commutativity in the dressed-spectrum Otto cycle, the work-production process remains adiabatic, and we can use the same notation $\rho_\alpha$ as in Eq.~\eqref{rho} to rewrite the energy flows. 
The system state is $\rho_\bathh$ after the isochoric process (a), while $\rho_\bathc$ is the state after process (c). The heat absorbed by the system from each bath $\alpha = \bathh, \bathc$ during the corresponding isochoric process is given by
\begin{align}
\label{Qhdressed}
{Q_\bathh^{\rm ds}} &= \Tr[\tilde{H}_S^\bathh (\rho_\bathh - \rho_\bathc)] 
 = \epsilon_\bathh^- (\rhogg^\bathh-\rhogg^\bathc) + \epsilon_\bathh^+ (\rhoee^\bathh-\rhoee^\bathc)
 =  \tilde{\omega}_\bathh (\rhoee^\bathh-\rhoee^\bathc),\\
\label{Qcdressed}	
{Q_\bathc^{\rm ds}} &= \Tr[\tilde{H}_S^\bathc (\rho_\bathc - \rho_\bathh)] 
 = \epsilon_\bathc^- (\rhogg^\bathc-\rhogg^\bathh) + \epsilon_\bathc^+ (\rhoee^\bathc-\rhoee^\bathh)
  =  \tilde{\omega}_\bathc (\rhoee^\bathc-\rhoee^\bathh),
\end{align}
where the superscript $\rm ds$ denotes quantities related to the dressed-spectrum Otto cycle. The equations are satisfied with $\Tr[\rho_\alpha] = 1$. 
Note that the off-diagonal coherence terms of the system state $\rho^\alpha$ do not contribute at all again. We will stress in Sec.~\ref{Sec1-3} that the coherence terms do contribute in our internally coupled model.

For the adiabatic processes (b) and (d) in Fig.~\ref{dressedcycle}, the work extracted from the environment into the internal system is given by
\begin{align}
\label{W1dressed}
{W_1^{\rm ds}} &= \Tr[(\tilde{H}_S^\bathc-\tilde{H}_S^\bathh)\rho_\bathh] 
 = (\epsilon_\bathc^- - \epsilon_\bathh^-) \rhogg^\bathh + (\epsilon_\bathc^+ - \epsilon_\bathh^+) \rhoee^\bathh
 = (\tilde{\omega}_\bathc-\tilde{\omega}_\bathh) \rhoee^\bathh,\\
\label{W2dressed}
{W_2^{\rm ds}} &=  \Tr[(H_S^\bathh-H_S^\bathc)\rho_\bathc] 
= (\epsilon_\bathh^- - \epsilon_\bathc^-) \rhogg^\bathc + (\epsilon_\bathh^+ - \epsilon_\bathc^+) \rhoee^\bathc
= (\tilde{\omega}_\bathh-\tilde{\omega}_\bathc) \rhoee^\bathc,
\end{align}
which are derived from $\Tr[\rho_\alpha] = 1$. 
Therefore, the first law of thermodynamics holds in the form
\begin{align}
\label{Wdressed}
W^{\rm ds} = {W_1^{\rm ds}}+{W_2^{\rm ds}} = -({Q_\bathh^{\rm ds}}+{Q_\bathc^{\rm ds}}).
\end{align}
Note again that positive energy flows are defined as heat absorption and work extraction from the environment into the internal system. 

The dressed-spectrum Otto cycle operates similarly to the standard Otto cycle in three distinct functional modes: the engine, the refrigerator, and the accelerator. The transitions between these modes are governed by two critical boundaries, representing the zero work and heat exchange, given by
    \begin{align}
    \label{dressedbound1}
    \frac{\beta_\bathc}{\beta_\bathh} = \frac{\tilde{\omega}_\bathh}{\tilde{\omega}_\bathc}.
    \end{align}
The other is the zero work boundary between the engine and the accelerator, given by
    \begin{align}
    \label{dressedbound2}
    \tilde{\omega}_\bathh = \tilde{\omega}_\bathc.
    \end{align}
These boundaries are determined solely by the static parameters of the system, consisting of the effective energy gaps $\tilde{\omega}_\alpha$ and the bath temperatures $\beta_\alpha$, as illustrated in Table~\ref{table:dressedregimes}.
\begin{table}[ht]
\centering
\caption{Operational regimes of the dressed-spectrum Otto cycle.}
\label{table:dressedregimes}
\begin{tabular}{|l|c|c|c|c|}
\hline
\textbf{Regime} & \textbf{Parameter Range} & \textbf{Work $W$} & \textbf{Heat $Q_c$} & \textbf{Heat $Q_h$} \\ \hline
Accelerator & $\frac{\beta_\bathh}{\beta_\bathc}<1 < \frac{\tilde{\omega}_\bathc}{\tilde{\omega}_\bathh}$ & $> 0$ & $< 0$ & $> 0$ \\ \hline
$\tilde{\omega}_\bathh = \tilde{\omega}_\bathc$  & $\frac{\beta_\bathh}{\beta_\bathc}<1 = \frac{\tilde{\omega}_\bathc}{\tilde{\omega}_\bathh}$ & $0$ & $< 0$ & $> 0$ \\ \hline
Engine & $\frac{\beta_\bathh}{\beta_\bathc} < \frac{\tilde{\omega}_\bathc}{\tilde{\omega}_\bathh} < 1$ & $< 0$ & $< 0$  & $> 0$ \\ \hline
$\frac{\beta_\bathc}{\beta_\bathh} = \frac{\tilde{\omega}_\bathh}{\tilde{\omega}_\bathc}$ & $\frac{\tilde{\omega}_\bathc}{\tilde{\omega}_\bathh} = \frac{\beta_\bathh}{\beta_\bathc}<1$ & $0$ & $0$ & $0$ \\ \hline
Refrigerator & $\frac{\tilde{\omega}_\bathc}{\tilde{\omega}_\bathh} < \frac{\beta_\bathh}{\beta_\bathc}<1$ & $> 0$& $> 0$  & $< 0$  \\ \hline
\end{tabular}
\end{table}
The classification of these regimes based on the frequency ratio $\tilde{\omega}_\bathc/\tilde{\omega}_\bathh$ can be summarized simply by replacing $\omega_\bathc/\omega_\bathh$ in Table~\ref{table:standardregimes} with $\tilde{\omega}_\bathc/\tilde{\omega}_\bathh$. 

The corresponding Otto efficiency and coefficient of performance (COP) become
\begin{align}
\label{eta_dressed}
\eta_{\rm ds}
=1-\frac{\tilde{\omega}_\bathc}{\tilde{\omega}_\bathh},
\end{align}
and
\begin{align}
\label{cop_dressed}
\xi_{\rm ds}=\frac{\tilde{\omega}_\bathc}{\tilde{\omega}_\bathh-\tilde{\omega}_\bathc}.
\end{align}

For comparison among these three types of Otto cycle in Sec.~\ref{Sec1-4}, we rewrite them in the form as
\begin{align}
\eta_{\rm ds}
=1-F_A^{\rm ds}
\frac{\omega_\bathc}{\omega_\bathh},
\end{align}
where
\begin{align}
\label{FA_dressed}
F_A^{\rm ds}=\frac{\omega_\bathh}{\omega_\bathc}\frac{\tilde{\omega}_\bathc}{\tilde{\omega}_\bathh}
=\sqrt{\frac{1+4g_\bathc^2/\omega_\bathc^2}{1+4g_\bathh^2/\omega_\bathh^2}}.
\end{align}
Similarly, the dressed-spectrum COP can be rewritten as
\begin{align}
\xi_{\rm ds}
=\frac{\omega_\bathc}{\omega_\bathh-\omega_\bathc}
F_B^{\rm ds},
\end{align}
with
\begin{align}
\label{FB_dressed}
F_B^{\rm ds}
=\frac{(\omega_\bathh-\omega_\bathc)\tilde{\omega}_\bathc}{
\omega_\bathc(\tilde{\omega}_\bathh-\tilde{\omega}_\bathc)}.
\end{align}
We rewrite the efficiency and COP of the internally coupled Otto cycle in the same form in Sec.~\ref{Sec1-3} and compare them in Sec.~\ref{Sec1-4}.

\subsection{Internally Coupled Otto Cycle}
\label{Sec1-3}

\begin{figure*}[h!]
\centering
	\includegraphics[width=0.5\textwidth]{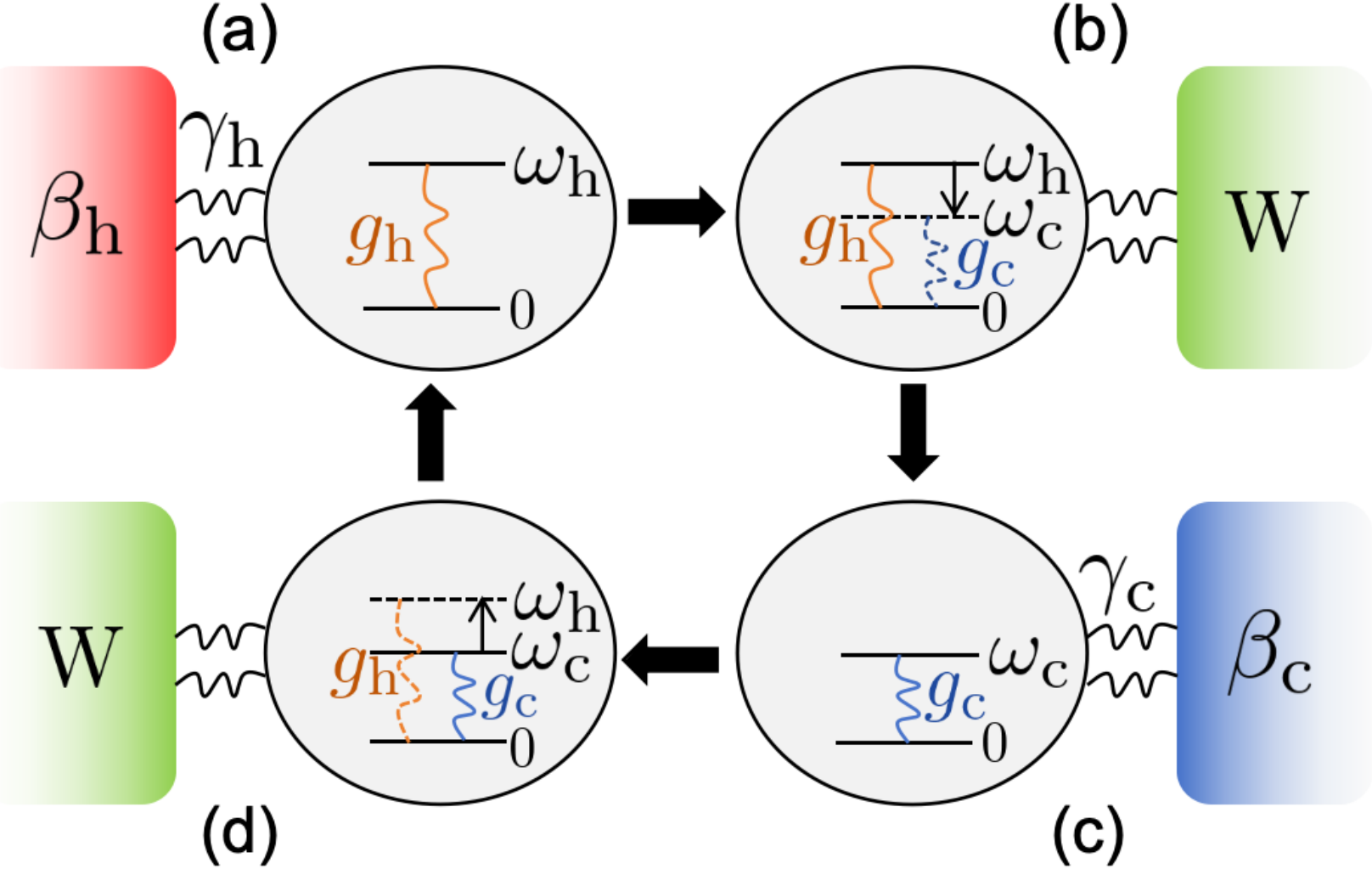}
	\caption{Schematic view of the quantum Otto cycle with the internal coupling between the excited and ground energy levels of the internal two-level system. The internal coupling strength is $g_{\bathh}$ and $g_{\bathc}$ in the isochoric process (a) and (c), respectively. In the unitary work production processes (b) and (d), not only do the excited energy levels alternate between $\omega_{\bathh}$ and $\omega_{\bathc}$, but also the internal coupling strengths alternate between $g_{\bathh}$ and $g_{\bathc}$.}
	\label{cyclecoupling}
\end{figure*}

Then, we consider the internally coupled system, as shown in Fig.~\ref{cyclecoupling}, where not only the excited energy level of the internal system but also the internal coupling strength between the ground and the excited energy levels can change when the internal system interacts with the external work component. Different from the standard and the dressed-spectrum Otto cycles, the work-production process of the internally coupled Otto cycle is not adiabatic due to the non-commutativity $[H_S^\alpha(t),H_S^\alpha(t')]\neq0$. 

The internal Hamiltonian is the same as in Eq.~\eqref{HS}. We also keep the bath and interaction Hamiltonians in Eqs.~(\ref{HB}) and (\ref{HI}) as they are.
After introducing the internal coupling, we define the total heat absorbed from each bath $\alpha = \bathh, \bathc$ during each isochoric process as in
\begin{align}
\label{Qh}
{Q_\bathh^\mathrm{cp}} &= \Tr[H_S^\bathh (\rho_\bathh - \rho_\bathc)] = \Tr[\tilde{H}_S^\bathh (\tilde{\rho}_\bathh - \tilde{\rho}_\bathc)] \\
		& = g_{\bathh} \left [(\rhoeg^\bathh-\rhoeg^\bathc)+(\rhoge^\bathh-\rhoge^\bathc)\right ] +\omega_{\bathh} (\rhoee^\bathh-\rhoee^\bathc),\\
\label{Qc}	
{Q_\bathc^\mathrm{cp}} &= \Tr[H_S^\bathc (\rho^\bathc - \rho^\bathh)] = \Tr[\tilde{H}_S^\bathc (\tilde{\rho}_\bathc - \tilde{\rho}_\bathh)] \\
		& = g_{\bathc} \left [(\rhoeg^\bathc-\rhoeg^\bathh)+(\rhoge^\bathc-\rhoge^\bathh)\right ] +\omega_{\bathc} (\rhoee^\bathc-\rhoee^\bathh),
\end{align}
where the superscript $\rm cp$ denotes quantities related to the present internally coupled cycle, the effective Hamiltonian
$\tilde{H}_S^\alpha$, as in Eq.~\eqref{tildeHS}, is the diagnoalized Hamiltonain of the bare one $H_S^\alpha$ in Eq.~\eqref{HS}, and the system state $\tilde{\rho}_\alpha$ at the eigenbasis of the diagonalized Hamiltonian is given by 
\begin{align}
\tilde{\rho}_\alpha = U_\alpha^\dagger \rho_\alpha U_\alpha.
\end{align}
This is the critical difference from Eqs.~\eqref{Qhdressed} and \eqref{Qcdressed} for the dressed-spectrum cycle. 
The system state $\rho_\alpha$, as in Eq.~\eqref{rho}, is the original map corresponding to the bare Hamiltonian $H_S^\alpha$. Due to the transformation $U_\alpha$, the system state $\tilde{\rho}_\alpha$ is at a different eigenbasis in the isochoric process $\bathh$ and $\bathc$, leading to contributions from the coherence terms $\rhoeg^\alpha$ and $\rhoge^\alpha$.

The work-production process is non-adiabatic due to the non-commutativity. The work extracted from the environment is given by
\begin{align}
\label{W1}
{W_1^\mathrm{cp}}&= \Tr[(H_S^\bathc-H_S^\bathh)\rho_\bathh] \\ 
	&= (\rhoeg^\bathh + \rhoge^\bathh)(g_{\bathc}-g_{\bathh})+(\omega_{\bathc}-\omega_{\bathh})\rhoee^\bathh,\\
\label{W2}
{W_2^\mathrm{cp}}&= \Tr[(H_S^\bathh-H_S^\bathc)\rho_\bathc] \\ 
	&= (\rhoeg^\bathc + \rhoge^\bathc)(g_{\bathh}-g_{\bathc})+(\omega_{\bathh}-\omega_{\bathc})\rhoee^\bathc,
\end{align}
and the total work is given by
\begin{align}
\label{W}
{W^\mathrm{cp}} ={W_1^\mathrm{cp}} + {W_2^\mathrm{cp}} = -(g_{\bathh}-g_{\bathc}) \left [(\rhoeg^\bathh-\rhoeg^\bathc)+(\rhoge^\bathh-\rhoge^\bathc)\right ] -(\omega_{\bathh} -\omega_{\bathc})(\rhoee^\bathh-\rhoee^\bathc) =  -({Q_\bathh^\mathrm{cp}}+{Q_\bathc^\mathrm{cp}}),
\end{align}
which respects the first law of thermodynamics. Note again that positive energy flows are defined as the heat absorption and the work extraction from the environment to the internal system. 

We define that the Otto cycle operates as an engine when the energy flows as ${Q_\bathh^\mathrm{cp}}>0$, ${Q_\bathc^\mathrm{cp}}<0$, and ${W^\mathrm{cp}}<0$. Its efficiency $\eta_\mathrm{cp}=-W/{Q_\bathh^\mathrm{cp}}$ is given by
\begin{align}
\label{efficiency}
\eta_\mathrm{cp} & = \frac{(g_{\bathh}-g_{\bathc}) \left [(\rhoeg^\bathh-\rhoeg^\bathc)+(\rhoge^\bathh-\rhoge^\bathc)\right ] +(\omega_{\bathh} -\omega_{\bathc})(\rhoee^\bathh-\rhoee^\bathc)}{g_{\bathh} \left [(\rhoeg^\bathh-\rhoeg^\bathc)+(\rhoge^\bathh-\rhoge^\bathc)\right ] +\omega_{\bathh} (\rhoee^\bathh-\rhoee^\bathc)}.
\end{align}

Similarly, we define that the quantum Otto cycle runs as a refrigerator, when the energy flows as ${Q_\bathh^\mathrm{cp}}<0$, ${Q_\bathc^\mathrm{cp}}>0$, and $W>0$,. Its coefficient of performance (COP) $\xi_\mathrm{cp} = {Q_\bathc^\mathrm{cp}}/W$ is given by
\begin{align}
\label{COP}
\xi_\mathrm{cp} & = \frac{g_{\bathc} \left [(\rhoeg^\bathc-\rhoeg^\bathh)+(\rhoge^\bathc-\rhoge^\bathh)\right ] +\omega_{\bathc} (\rhoee^\bathc-\rhoee^\bathh)}{(g_{\bathc}-g_{\bathh}) \left [(\rhoeg^\bathh-\rhoeg^\bathc)+(\rhoge^\bathh-\rhoge^\bathc)\right ] +(\omega_{\bathc} -\omega_{\bathh})(\rhoee^\bathh-\rhoee^\bathc)}
\end{align}

Below, we will use the sum of the off-diagonal elements
\begin{align}
\label{functionF}
f(g_\alpha,\beta_\alpha,\omega_\alpha) &= \rhoge^\alpha+\rhoeg^\alpha,
\end{align}
which depends on the internal coupling strength $g_\alpha$, the inverse temperature $\beta_\alpha$, and the energy spacing $\omega_\alpha$ in the isochoric process $\alpha=\bathh,\bathc$. Let us stress here that the internal coupling $g_\alpha$ generates off-diagonal coherence in the system state.

For better comparison with other cycles, we rewrite the efficiency $\eta_\mathrm{cp}$ and COP $\xi_\mathrm{cp} $ of the Otto cycle with internal coupling in Eqs.~\eqref{efficiency} and \eqref{COP} as
\begin{align}
\eta_\mathrm{cp} & = \frac{(g_{\bathh}-g_{\bathc}) \left [(\rhoeg^\bathh-\rhoeg^\bathc)+(\rhoge^\bathh-\rhoge^\bathc)\right ] +(\omega_{\bathh} -\omega_{\bathc})(\rhoee^\bathh-\rhoee^\bathc)}{g_{\bathh} \left [(\rhoeg^\bathh-\rhoeg^\bathc)+(\rhoge^\bathh-\rhoge^\bathc)\right ] +\omega_{\bathh} (\rhoee^\bathh-\rhoee^\bathc)}\nonumber \\
\label{efficiencyFA}
& = 1- F_A^\mathrm{cp}\frac{\omega_{\bathc}}{\omega_{\bathh}},
\end{align}
where
\begin{align}
\label{functionA}
F_A^\mathrm{cp} &= \frac{ \frac{g_{\bathc}}{\omega_{\bathc}} (f(g_{\bathh},\beta_{\bathh},\omega_{\bathh})-f(g_{\bathc},\beta_{\bathc},\omega_{\bathc}))  + (\rhoee^\bathh-\rhoee^\bathc)}{\frac{g_{\bathh}}{\omega_{\bathh}} (f(g_{\bathh},\beta_{\bathh},\omega_{\bathh})-f(g_{\bathc},\beta_{\bathc},\omega_{\bathc}))  + (\rhoee^\bathh-\rhoee^\bathc)},
\end{align}
and
\begin{align}
\xi_\mathrm{cp} & = \frac{g_{\bathc} \left [(\rhoeg^\bathc-\rhoeg^\bathh)+(\rhoge^\bathc-\rhoge^\bathh)\right ] +\omega_{\bathc} (\rhoee^\bathc-\rhoee^\bathh)}{(g_{\bathc}-g_{\bathh}) \left [(\rhoeg^\bathh-\rhoeg^\bathc)+(\rhoge^\bathh-\rhoge^\bathc)\right ] +(\omega_{\bathc} -\omega_{\bathh})(\rhoee^\bathh-\rhoee^\bathc)}\nonumber \\
\label{COPFB}
& = \frac{\omega_{\bathc}}{\omega_{\bathh}-\omega_{\bathc}}F_B^\mathrm{cp},
\end{align}
where
\begin{align}
\label{functionB}
F_B^\mathrm{cp} = \frac{\frac{g_{\bathc}}{\omega_{\bathc}} (f(g_{\bathc},\beta_{\bathc},\omega_{\bathc})-f(g_{\bathh},\beta_{\bathh},\omega_{\bathh}))+(\rhoee^\bathc-\rhoee^\bathh)}{\frac{g_{\bathh}-g_{\bathc}}{\omega_{\bathh}-\omega_{\bathc}}(f(g_{\bathc},\beta_{\bathc},\omega_{\bathc})-f(g_{\bathh},\beta_{\bathh},\omega_{\bathh}))+(\rhoee^\bathc-\rhoee^\bathh)}.
\end{align}

\subsection{Comparison of the Internally Coupled Otto Cycle with the Other Two}
\label{Sec1-4}

Compared to the standard Otto efficiency $\eta_{\rm st}=1-\omega_{\bathc}/\omega_{\bathh}$, which depends only on the bare energy levels, after introducing the internal coupling, the efficiency $\eta_\mathrm{cp}$ of the internally coupled Otto cycle can be either greater $F_A^\mathrm{cp} <1$ or less $F_A^\mathrm{cp} >1$, depending on the six parameters $\{\omega_{\bathh}, \omega_{\bathc}, g_{\bathh}, g_{\bathc}, \beta_{\bathh}, \beta_{\bathc}\}$, as summarized in Table~\ref{table:compareefficiency}. 
When $f(g_{\bathh},\beta_{\bathh},\omega_{\bathh})>f(g_{\bathc},\beta_{\bathc},\omega_{\bathc})$ and $g_{\bathh}/\omega_{\bathh}>g_{\bathc}/\omega_{\bathc}$, or when $f(g_{\bathh},\beta_{\bathh},\omega_{\bathh})<f(g_{\bathc},\beta_{\bathc},\omega_{\bathc})$ and $g_{\bathh}/\omega_{\bathh}<g_{\bathc}/\omega_{\bathc}$, the efficiency of the internally coupled system can surpass that of the standard Otto engine without the internal coupling. For a system with fixed bath temperatures $\beta_\alpha$ and energy levels $\omega_\alpha$, we can always adjust the internal coupling strengths $g_\alpha$ to achieve a higher efficiency, as shown in Table~\ref{tablecompare}~\subref{table:compareefficiency}.

Similarly, the comparison between the COP $\xi_\mathrm{cp}$ of the internally coupled system and the standard Otto COP $\xi_{\rm st} =  \omega_{\bathc}/(\omega_{\bathh}-\omega_{\bathc})$ is summarized in Table~\ref{tablecompare}~\subref{table:compareCOP}. We can also adjust the internal coupling strengths $g_\alpha$ to achieve a higher COP.

\begin{table}[h]
\centering
\subfloat[Comparison of efficiency]{
\begin{tabular}{c !{\vrule width 1pt} c | c | c }
\hline
 & $f(g_{\bathh},\beta_{\bathh},\omega_{\bathh})<f(g_{\bathc},\beta_{\bathc},\omega_{\bathc})$ & $f(g_{\bathh},\beta_{\bathh},\omega_{\bathh})=f(g_{\bathc},\beta_{\bathc},\omega_{\bathc})$ & $f(g_{\bathh},\beta_{\bathh},\omega_{\bathh})>f(g_{\bathc},\beta_{\bathc},\omega_{\bathc})$ \\
\noalign{\hrule height 1pt}
$g_{\bathh}/\omega_{\bathh}<g_{\bathc}/\omega_{\bathc}$ & $\eta_\mathrm{cp}>\eta_{\rm st}$ & $\eta_\mathrm{cp}=\eta_{\rm st}$ & $\eta_\mathrm{cp}<\eta_{\rm st}$ \\
$g_{\bathh}/\omega_{\bathh}=g_{\bathc}/\omega_{\bathc}$ & $\eta_\mathrm{cp}=\eta_{\rm st}$ & $\eta_\mathrm{cp}=\eta_{\rm st}$ & $\eta_\mathrm{cp}=\eta_{\rm st}$ \\
$g_{\bathh}/\omega_{\bathh}>g_{\bathc}/\omega_{\bathc}$ & $\eta_\mathrm{cp}<\eta_{\rm st}$ & $\eta_\mathrm{cp}=\eta_{\rm st}$ & $\eta_\mathrm{cp}>\eta_{\rm st}$ \\
\hline
\end{tabular}\label{table:compareefficiency}}

\subfloat[Comparison of coefficient of performance (COP)]{\begin{tabular}{c !{\vrule width 1pt} c | c | c }
\hline
 & $f(g_{\bathh},\beta_{\bathh},\omega_{\bathh})<f(g_{\bathc},\beta_{\bathc},\omega_{\bathc})$ & $f(g_{\bathh},\beta_{\bathh},\omega_{\bathh})=f(g_{\bathc},\beta_{\bathc},\omega_{\bathc})$ & $f(g_{\bathh},\beta_{\bathh},\omega_{\bathh})>f(g_{\bathc},\beta_{\bathc},\omega_{\bathc})$ \\
\noalign{\hrule height 1pt}
\scriptsize$(g_{\bathh}-g_{\bathc})/(\omega_{\bathh}-\omega_{\bathc})<g_{\bathc}/\omega_{\bathc}$& $\xi_\mathrm{cp}>\xi_{\rm st}$ & $\xi_\mathrm{cp}=\xi_{\rm st}$ & $\xi_\mathrm{cp}<\xi_{\rm st}$ \\
\scriptsize$(g_{\bathh}-g_{\bathc})/(\omega_{\bathh}-\omega_{\bathc})=g_{\bathc}/\omega_{\bathc}$ & $\xi_\mathrm{cp}=\xi_{\rm st}$ & $\xi_\mathrm{cp}=\xi_{\rm st}$ & $\xi_\mathrm{cp}=\xi_{\rm st}$ \\
\scriptsize$(g_{\bathh}-g_{\bathc})/(\omega_{\bathh}-\omega_{\bathc})>g_{\bathc}/\omega_{\bathc}$ & $\xi_\mathrm{cp}<\xi_{\rm st}$ & $\xi_\mathrm{cp}=\xi_{\rm st}$ & $\xi_\mathrm{cp}>\xi_{\rm st}$ \\
\hline
\end{tabular}\label{table:compareCOP}}
\caption{Comparison of (a)~efficiency $\eta$ and (b)~coefficient of performance (COP) $\xi$ between the standard Otto cycle denoted by the superscript $\rm st$ and the internally coupled cycle denoted by the superscript $\rm cp$, given by $f(g_\alpha,\beta_\alpha,\omega_\alpha)$ and the relations between $g_{\bathh}/\omega_{\bathh}$ and $g_{\bathc}/\omega_{\bathc}$.}
\label{tablecompare}
\end{table}

At the standard Otto bounds, as in Eqs.~\eqref{standardbound1} and \eqref{standardbound2}, the standard Otto cycle without or neglecting the internal coupling cannot run as a thermal mode. However, the internally coupled system can operate in diverse thermal modes. Hence, the presence of the internal coupling should extend the operational regime of the quantum Otto cycle. 

Some may suspect that the influence of internal coupling arises solely from the modification of the energy spectrum of the system Hamiltonian. Here, we compare the internally coupled Otto cycle explained in Sec~\ref{Sec1-3} to the dressed-spectrum Otto cycle explained in Sec.~\ref{Sec1-2}. 
We here claim that the internal coupling influences the performance of the Otto cycle not only by updating the Hamiltonian, but also by introducing the off-diagonal coherence due to the transformation $U_\alpha$ in Eq.~\eqref{Ualpha}. 
By comparing the factors $F_A^{\rm cp}$ in Eq.~(\ref{functionA}) and $F_B^{\rm cp}$ in Eq.~(\ref{functionB}) of the internal coupled Otto cycle to the factors $F_A^{\rm ds}$ in Eq.~(\ref{FA_dressed}) and $F_B^{\rm ds}$ in Eq.~(\ref{FB_dressed}) of the dressed-spectrum Otto cycle, we can examine that they are inequivalent.

Under the dressed-spectrum interpretation, the influence of the internal coupling is entirely encoded in the modified effective energy gaps $\tilde{\omega}_\alpha$. 
Different from the factors $F_A^\mathrm{cp}$ in Eq.~(\ref{functionA}) and $F_B^\mathrm{cp}$ in Eq.~(\ref{functionB}), which contain the coherence terms $\rho_{eg}^\alpha$ and $\rho_{ge}^\alpha$, the factors $F_A^{\rm ds}$ in Eq.~(\ref{FA_dressed}) and $F_B^{\rm ds}$ in Eq.~(\ref{FB_dressed}) depend only on the dressed spectrum of the Hamiltonian without the presence of the coherence terms $\rho_{eg}^\alpha$ and $\rho_{ge}^\alpha$. Hence, the performance of the internal coupled Otto cycle does not merely depend on the dressed-spectrum update of the Hamiltonian spectrum. 
In this sense, the eigenbasis transformation $U_\alpha$ in Eq.~\eqref{Ualpha} also plays a key role in introducing off-diagonal coherence. 

For the efficiency of the engine, if $F_A^\mathrm{cp}>F_A^{\rm ds}$, we can obtain $\eta_\mathrm{cp}<\eta_{\rm ds}$, and otherwise, we have $\eta_\mathrm{cp}>\eta_{\rm ds}$, as illustrated in Table~\ref{tablecomparedressed}~\subref{table:compareefficiencydressed}. Similarly, for the COP of the refrigerator, if $F_B^\mathrm{cp}>F_B^{\rm ds}$, we can obtain $\xi_\mathrm{cp}>\xi_{\rm ds}$, and otherwise, we have $\xi_\mathrm{cp}<\xi_{\rm ds}$, as illustrated in Table~\ref{tablecomparedressed}~\subref{table:compareCOPdressed}. 

\begin{table}[h]
\centering
\subfloat[Comparison of efficiency]{
\begin{tabular}{c | c | c }
\hline
$F_A^\mathrm{cp}>F_A^{\rm ds}$ & $F_A^\mathrm{cp}=F_A^{\rm ds}$ & $F_A^\mathrm{cp}<F_A^{\rm ds}$ \\
\noalign{\hrule height 1pt}
$\eta_\mathrm{cp}<\eta_{\rm ds}$ & $\eta_\mathrm{cp}=\eta_{\rm ds}$ & $\eta_\mathrm{cp}>\eta_{\rm ds}$ \\
\hline
\end{tabular}\label{table:compareefficiencydressed}}

\subfloat[Comparison of coefficient of performance (COP)]{
\begin{tabular}{c | c | c }
\hline
$F_B^\mathrm{cp}>F_B^{\rm ds}$ & $F_B^\mathrm{cp}=F_B^{\rm ds}$ & $F_B^\mathrm{cp}<F_B^{\rm ds}$ \\
\noalign{\hrule height 1pt}
$\xi_\mathrm{cp}<\xi_{\rm ds}$ & $\xi_\mathrm{cp}=\eta_{\rm ds}$ & $\xi_\mathrm{cp}>\xi_{\rm ds}$ \\
\hline
\end{tabular}\label{table:compareCOPdressed}}
\caption{Comparison of (a)~efficiency $\eta$ and (b)~coefficient of performance (COP) $\xi$ between the dressed Otto cycle denoted by the superscript $\rm ds$ and the internally coupled cycle denoted by the superscript $\rm cp$, given by the factors $F_A$ and $F_B$.}
\label{tablecomparedressed}
\end{table}

\section{Gibbs-State Limit Cycle}
\label{Sec2}
In this section, we compare the internally coupled Otto cycle in detail with the standard and dressed-spectrum cycles under the framework of a thermalized steady cycle. As shown in Fig.~\ref{Diagrams}~\subref{GSLCDiagram}, we assume that the system rapidly equilibrates with the heat baths and remains in thermal equilibrium throughout each isochoric stroke. Because the system reaches the thermalized Gibbs state immediately, we refer to this as the Gibbs-state limit cycle (GSLC). For brevity, we omit specific scripts indicating the GSLC, but all quantities discussed in this section pertain to this limit cycle.

We demonstrate that internal coupling plays a dual role in the Otto cycle. Actively, it updates the Hamiltonian spectrum, thereby broadening the operational regimes and enhancing performance. Passively, the non-commutativity of the Hamiltonians introduces an eigenbasis mismatch. This mismatch acts as quantum internal friction, which constrains the operational regimes and degrades performance. Finally, we confirm that utilizing internal coupling as a thermodynamic resource to surpass the standard GSLC incurs an unavoidable cost: when this internal friction vanishes, the Hamiltonians commute, and all three GSLC models entirely coincide with each other.

\subsection{Internally Coupled Gibbs-State Limit Cycle}
\label{Sec2-1}

In this subsection, we derive the GSLC of the internally coupled system in detail and then compare it to those of the standard and dressed-spectrum systems.  As established in Secs.~\ref{Sec1-1} and \ref{Sec1-2}, although the energy exchanges in the standard and dressed-spectrum cycles depend on the excited-state population $\rhoee$, their thermal performance is strictly independent of the system state. Consequently, the generic results from these previous sections apply directly to their respective GSLCs. In contrast, as noted in Sec.~\ref{Sec1-3}, the thermal performance of the internally coupled Otto cycle depends intimately on both the energy levels and the system state, which in this context is fixed as the Gibbs state.

To determine the Gibbs state in each isochoric process, we diagonalize the system Hamiltonian from $H_S^\alpha$ in Eq.~\eqref{HS} to $\tilde{H}_S^\alpha$ in Eq.~\eqref{tildeHS}. The effective basis in each process differs, and we express all physical quantities in the entire Otto cycle in the original basis. We switch from the original to the effective basis at each isochoric process to calculate the system state, and then switch back from the effective to the original basis to calculate the physical quantities. 

In the effective basis $\ket{P_\alpha^\pm}$ in each isochoric process $\alpha=\bathh,\bathc$, the Gibbs state is given by
\begin{align}
\label{GibbsState1}
&{\rho_\alpha^\mathrm{cp}} = \frac{1}{Z_\alpha}(e^{-\beta_\alpha \epsilon_\alpha^+}\ket{P_\alpha^+}\bra{P_\alpha^+} + e^{-\beta_\alpha \epsilon_\alpha^-} \ket{P_\alpha^-}\bra{P_\alpha^-}),
\end{align}
where the inverse temperature $\beta_\alpha = 1/(kT_\alpha)$, and the partition function is
\begin{align}
Z_\alpha &= e^{-\beta_\alpha \epsilon_\alpha^-}+ e^{-\beta_\alpha \epsilon_\alpha^+} = 2 e^{-\frac{\beta_\alpha \omega_\alpha}{2}} \cosh(\beta_\alpha \Delta_\alpha)
\end{align}
with $\Delta_\alpha=\sqrt{4 g_\alpha^2+\omega_\alpha^2}/2$ as in Eq.~\eqref{Delta}.
We put the Boltzmann constant $k$ to unity hereafter.

Without the internal coupling, the efficiency and COP can be written in terms of the bare energy levels $\omega_\alpha$. When the internal coupling $g_\alpha$ is introduced, the Gibbs states are defined by the effective energy levels $\tilde{\omega}_\alpha$ in Eq.~\eqref{effectiveenergygap} and the diagonalizing eigenbases in Eqs.~(\ref{P1}) and (\ref{P2}).
To maintain consistency, we hereafter express all physical quantities in the original eigenbasis $\begin{pmatrix} \ket{\mathrm{g}}& \ket{\mathrm{e}} \end{pmatrix}$. The Gibbs state in Eq.~(\ref{GibbsState1}) is then expressed as
\begin{align}
\label{GibbsState2}
{\rho_\alpha^\mathrm{cp}} &= \frac{1}{Z_\alpha}\left [e^{-\beta \epsilon_\alpha^+}\begin{pmatrix}
\cos{\theta_\alpha}^2 & \cos{\theta_\alpha} \sin{\theta_\alpha} \\
\cos{\theta_\alpha} \sin{\theta_\alpha} & \sin{\theta_\alpha}^2
\end{pmatrix}+ e^{-\beta \epsilon_\alpha^-} \begin{pmatrix}
\sin{\theta_\alpha}^2 & -\cos{\theta_\alpha} \sin{\theta_\alpha} \\
-\cos{\theta_\alpha} \sin{\theta_\alpha} & \cos{\theta_\alpha}^2
\end{pmatrix}\right ]\\
&=\frac{1}{4\Delta_\alpha}\begin{pmatrix}
2\Delta_\alpha+\omega_\alpha \tanh(\beta_\alpha \Delta_\alpha)&-2 g_\alpha \tanh(\beta_\alpha \Delta_\alpha) \\
-2 g_\alpha \tanh(\beta_\alpha \Delta_\alpha) & 2\Delta_\alpha-\omega_\alpha \tanh(\beta_\alpha \Delta_\alpha)
\end{pmatrix}.
\end{align}

In the forms of the Gibbs states, there is not only the effective energy gaps caused by the dressed-spectrum update of Hamiltonian spectrum, but also the coherence caused by the internal coupling. These terms can be represented by the function $f(g_\alpha,\beta_\alpha,\omega_\alpha)$ defined in Eq.~(\ref{functionF}), which depends on three parameters. For the Gibbs states~(\ref{GibbsState2}), this function is specifically given by
\begin{align}
\label{falphaGSLC}
f(g_\alpha,\beta_\alpha,\omega_\alpha)
= -\frac{g_\alpha}{\Delta_\alpha}\tanh(\beta_\alpha \Delta_\alpha).
\end{align}

The heat absorptions and work extraction from the external environment to the internal system, as in Eqs.~(\ref{Qh})--(\ref{W}), can be rewritten for the GSLC as
\begin{align}
\label{QhGibbs}
{Q_\bathh^\mathrm{cp}} =& g_{\bathh}  \left [f(g_{\bathh},\beta_{\bathh},\omega_{\bathh})- f(g_{\bathc},\beta_{\bathc},\omega_{\bathc})\right ] + \omega_{\bathh}  (\rhoee^\bathh-\rhoee^\bathc) \nonumber \\
=&\frac{4 g_\bathc g_\bathh+\omega_\bathc \omega_\bathh}{4\Delta_\bathc}\tanh(\beta_\bathc \Delta_\bathc)-\Delta_\bathh \tanh(\beta_{\bathh} \Delta_\bathh),\\
\label{QcGibbs}
{Q_\bathc^\mathrm{cp}} =& -g_{\bathc}  \left [f(g_{\bathh},\beta_{\bathh},\omega_{\bathh})- f(g_{\bathc},\beta_{\bathc},\omega_{\bathc})\right ] - \omega_{\bathc}  (\rhoee^\bathh-\rhoee^\bathc)\nonumber  \\
=&\frac{4g_\bathc g_\bathh+\omega_\bathc\omega_\bathh}{4\Delta_\bathh}
\tanh(\beta_\bathh\Delta_\bathh)
-\Delta_\bathc \tanh(\beta_\bathc\Delta_\bathc), \\
\label{WGibbs}
{W^\mathrm{cp}} =&-(g_{\bathh}-g_{\bathc}) \left [f(g_{\bathh},\beta_{\bathh},\omega_{\bathh})- f(g_{\bathc},\beta_{\bathc},\omega_{\bathc})\right ] -(\omega_{\bathh} -\omega_{\bathc})(\rhoee^\bathh-\rhoee^\bathc)\nonumber  \\
=&\frac{\left[4g_\bathc(g_\bathc-g_\bathh)+\omega_\bathc(\omega_\bathc-\omega_\bathh)\right]}{4\Delta_\bathc}\tanh(\beta_\bathc\Delta_\bathc)\nonumber  \\
&+\frac{\left[4g_\bathh(g_\bathh-g_\bathc)+\omega_\bathh(\omega_\bathh-\omega_\bathc)\right]}{4\Delta_\bathh}\tanh(\beta_\bathh\Delta_\bathh) .
\end{align}
The heat and work exchanges thus depend on six parameters: ${\omega_{\bathh}, \omega_{\bathc}, g_{\bathh}, g_{\bathc}, \beta_{\bathh}, \beta_{\bathc}}$. Without loss of generality, we can fix the energy unit to $\omega_{\bathc}$ and the temperature unit to $\beta_{\bathc}$, leaving four independent parameters.

For thermal performance, under the thermalized condition, we focus on the engine's efficiency and the refrigerator's coefficient of performance (COP), while power is meaningless due to rapid thermalization in the GSLC. When the internally coupled GSLC operates as an engine, its efficiency in Eq.~\eqref{efficiency} is rewritten as
\begin{align}
\label{etaGibbs}
\eta_{\rm cp} =& \frac{-W^\mathrm{cp}}{Q_\bathh^\mathrm{cp}} \nonumber \\
=& \frac{ \frac{\left[4g_\bathc(g_\bathh-g_\bathc)+\omega_\bathc(\omega_\bathh-\omega_\bathc)\right]}{4\Delta_\bathc}\tanh(\beta_\bathc\Delta_\bathc) + \frac{\left[4g_\bathh(g_\bathc-g_\bathh)+\omega_\bathh(\omega_\bathc-\omega_\bathh)\right]}{4\Delta_\bathh}\tanh(\beta_\bathh\Delta_\bathh) }{ \frac{4 g_\bathc g_\bathh+\omega_\bathc \omega_\bathh}{4\Delta_\bathc}\tanh(\beta_\bathc \Delta_\bathc)-\Delta_\bathh \tanh(\beta_{\bathh} \Delta_\bathh) }. 
\end{align}
When it operates as a refrigerator, the COP defined in Eq.~\eqref{COP} is rewritten as
\begin{align}
\label{xiGibbs}
\xi_{\rm cp} =& \frac{Q_\bathc^\mathrm{cp}}{W^\mathrm{cp}} \nonumber \\
=& \frac{ \frac{4g_\bathc g_\bathh+\omega_\bathc\omega_\bathh}{4\Delta_\bathh} \tanh(\beta_\bathh\Delta_\bathh) -\Delta_\bathc \tanh(\beta_\bathc\Delta_\bathc) }{ \frac{\left[4g_\bathc(g_\bathc-g_\bathh)+\omega_\bathc(\omega_\bathc-\omega_\bathh)\right]}{4\Delta_\bathc}\tanh(\beta_\bathc\Delta_\bathc) + \frac{\left[4g_\bathh(g_\bathh-g_\bathc)+\omega_\bathh(\omega_\bathh-\omega_\bathc)\right]}{4\Delta_\bathh}\tanh(\beta_\bathh\Delta_\bathh) }.
\end{align}

\subsection{Operation Regimes}
\label{Sec2-2}

Then, in this subsection, we compare the operational regimes of the internally coupled GSLC with those of the standard and dressed-spectrum GSLCs. Note that the operating regimes of the standard and dressed-spectrum Otto cycles depend only on their energy gaps and the bath temperatures, as illustrated in Tables~\ref{table:standardregimes} and \ref{table:dressedregimes} with boundaries as in Eqs.~\eqref{standardbound1}--\eqref{standardbound2} and \eqref{dressedbound1}--\eqref{dressedbound2}. Through these comparisons, we find that the internal coupling plays two distinct roles under the thermalized condition: broadening the operating regimes by updating the Hamiltonian spectrum and narrowing the useful thermal-mode regimes due to eigenbasis mismatch caused by the non-commutativity of the Hamiltonian. 

\begin{figure*}[h!]
\centering
	\subfloat[]{\includegraphics[width=0.33\textwidth]{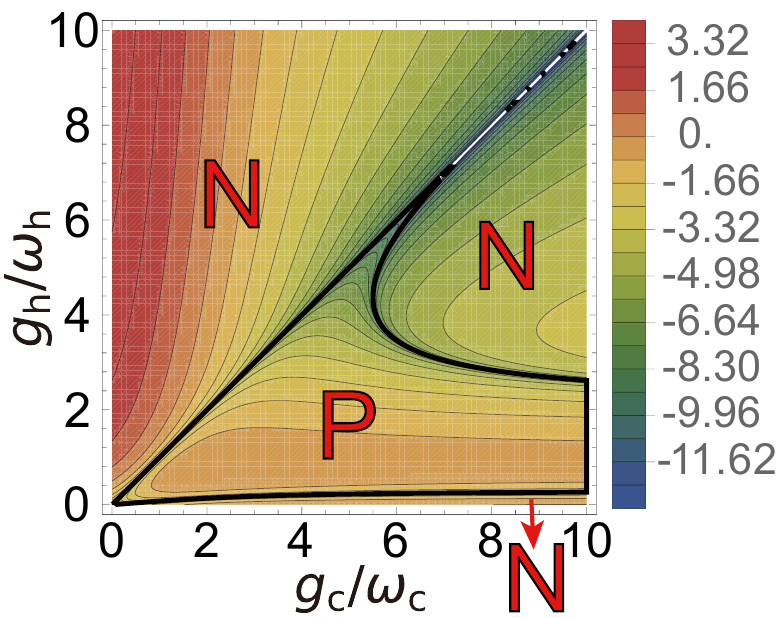}\label{GibbsQhLog}}
	\subfloat[]{\includegraphics[width=0.33\textwidth]{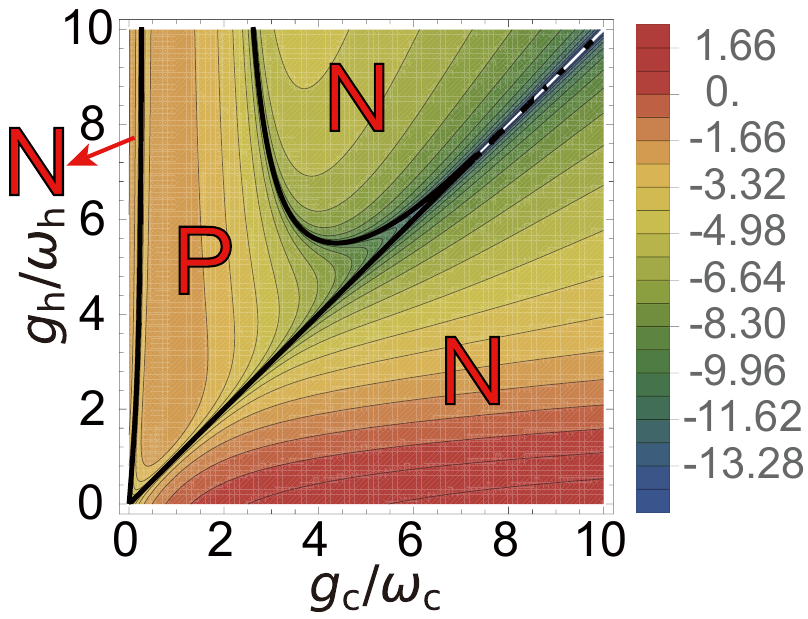}\label{GibbsQcLog}}
	\subfloat[]{\includegraphics[width=0.33\textwidth]{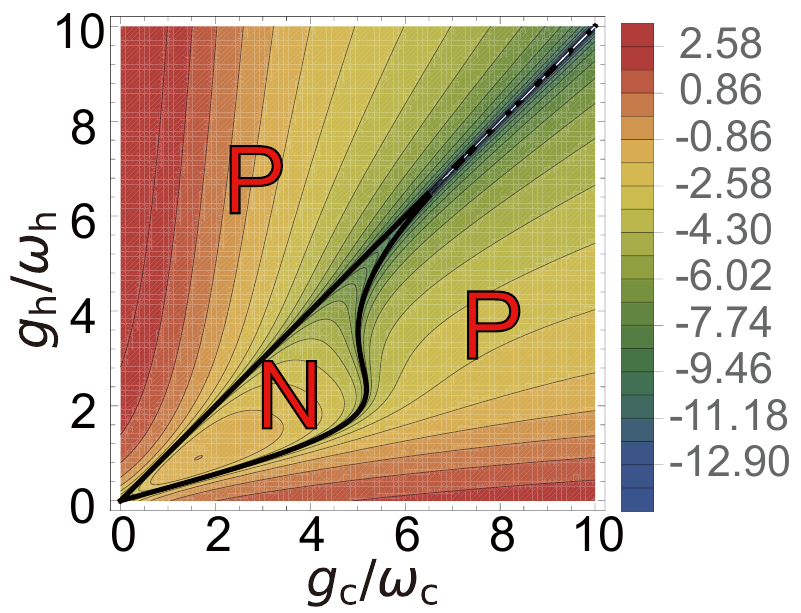}\label{GibbsWLog}}
	\caption{Logarithmic plots of the energy flows (a)~$\log|{Q_\bathh^\mathrm{cp}}|$, (b)~$\log|{Q_\bathc^\mathrm{cp}}|$, and (c)~$\log|{W^\mathrm{cp}}|$ depending on  $g_\alpha/\omega_\alpha$ evaluated in the Gibbs-state limit cycle (GSLC) for the internally coupled system. The black boundaries indicate the zero energy flow. The symbol P indicates the areas of the positive energy flow, in which the system absorbs energy from the environment. The symbol N indicates the areas of the negative energy flows, in which the system releases energy to the environment. We set the parameter values to $\omega_{\bathh}=5$, $\omega_{\bathc}=1$, $\beta_{\bathh}=0.2$, $\beta_{\bathc}=1$, and hence $\omega_{\bathh}/\omega_{\bathc} = \beta_{\bathc}/\beta_{\bathh} = 1/5$.}
	\label{GibbsEnergy}
\end{figure*}

We set the bath temperatures $\beta_\alpha$ and the system energy levels $\omega_\alpha$ using the boundary conditions of the standard Otto cycle, as in Eqs.~\eqref{standardbound1}, and the energy exchanges are zero at this bound. Hence, the standard GSLC cannot operate as a thermal machine here. However, as shown in Fig.~\ref{GibbsEnergy}, after introducing the internal coupling, the energy exchanges of the internally coupled GSLC are nonzero, allowing it to operate in several thermal modes.

\begin{figure*}[h!]
\centering
	\subfloat[]{\makebox[0.33\textwidth][c]{\includegraphics[scale=0.32]{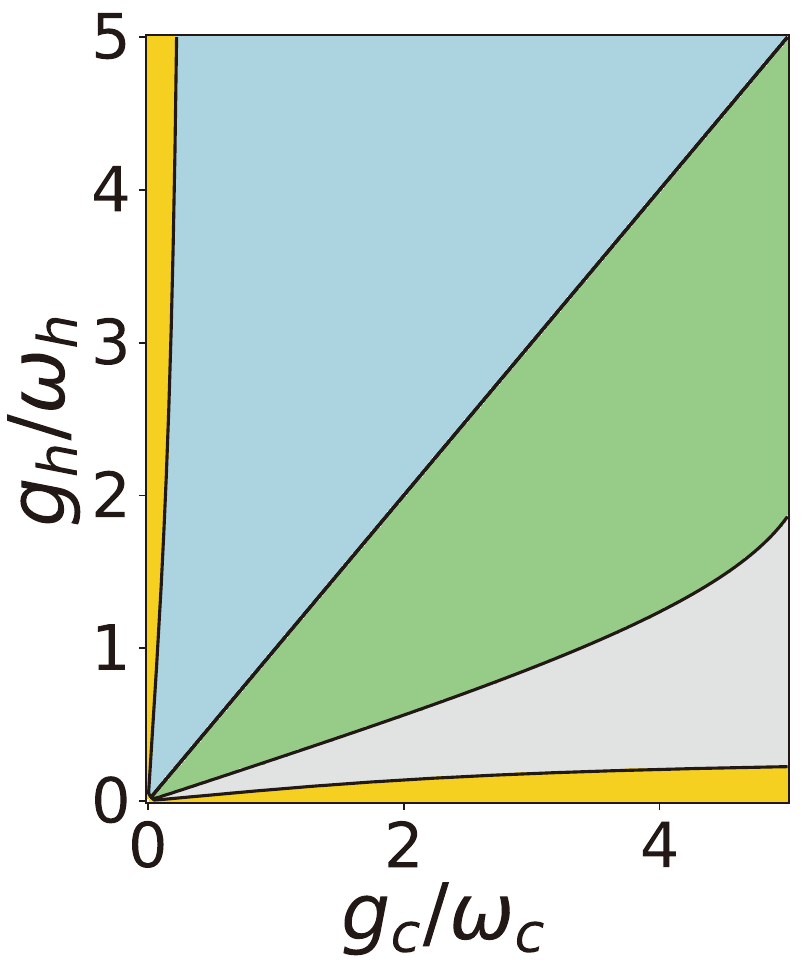}}\label{fig:GSLCregimes}}
	\subfloat[]{\makebox[0.33\textwidth][c]{\includegraphics[scale=0.32]{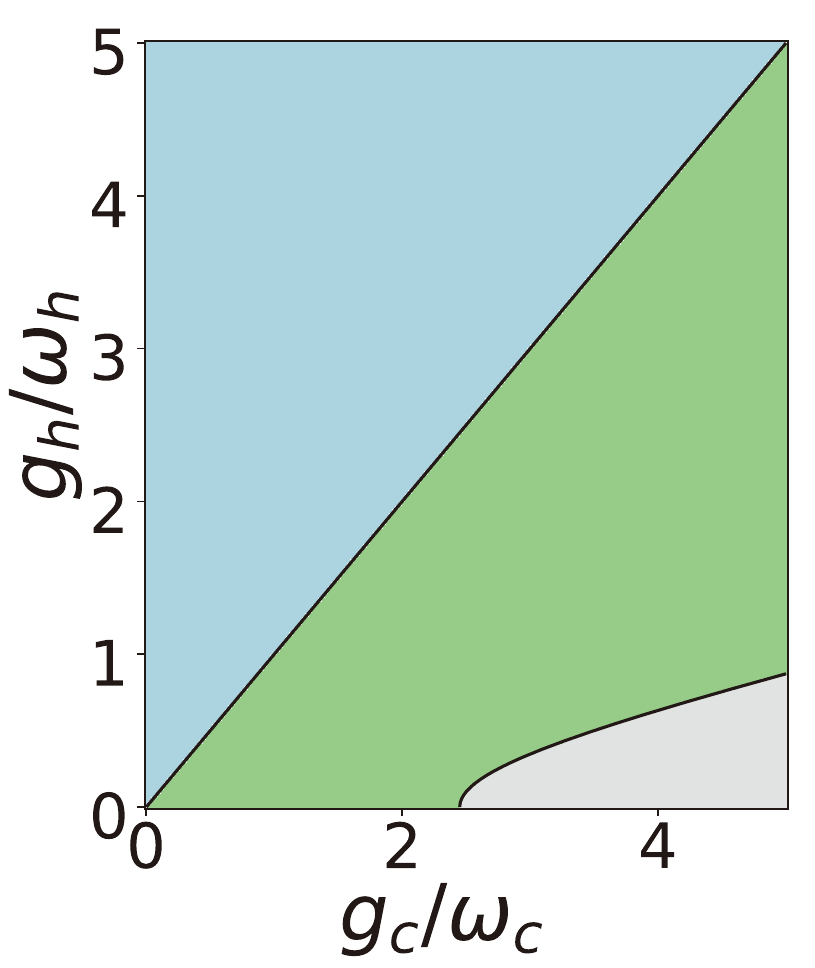}}\label{fig:dressedregimes}}
	\subfloat{\makebox[0.33\textwidth][c]{\raisebox{1.2cm}{\includegraphics[scale=0.4]{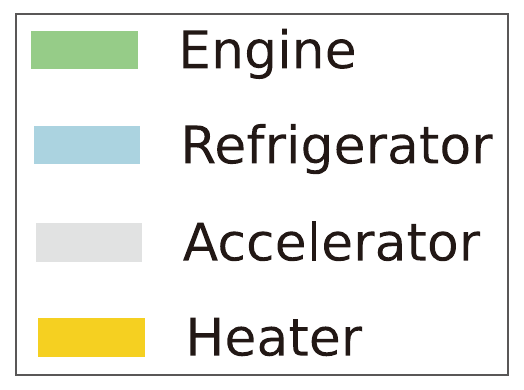}}}}
	\caption{Operation regimes of the Gibbs-state limit cycle (GSLC) for (a)~the internally coupled system and (b)~the dressed-spectrum system. The green area indicates the Otto cycle running in an engine, the blue area indicates the refrigerator, the gray area indicates the accelerator, and the yellow area indicates the heater. We set the parameter values to $\omega_{\bathh}=5$, $\omega_{\bathc}=1$, $\beta_{\bathh}=0.2$, $\beta_{\bathc}=1$, and hence $\omega_{\bathh}/\omega_{\bathc} = \beta_{\bathc}/\beta_{\bathh} = 1/5$.}
	\label{fig:Gibbsregimes}
\end{figure*}

As shown in Fig.~\ref{fig:Gibbsregimes}, a change in the system Hamiltonian spectrum allows the internally coupled GSLC to operate as several thermal modes. We consider only the engine and the refrigerator to be functional thermal machines, while the accelerator and heater modes represent dissipative states in which the machines' functional capacity is overwhelmed by internal friction. 

The operating regimes of the dressed-spectrum GSLC, as shown in Fig.~\ref{fig:Gibbsregimes}~\subref{fig:dressedregimes}, only depend on the effective energy gaps $\tilde{\omega}_\alpha$ and the bath temperatures $\beta_\alpha$, leading to three modes and two boundaries, as illustrated in Table~\ref{table:dressedregimes}.
Differently, the internally coupled GSLC is not equivalent to the dressed-spectrum one, showing four thermal modes and three boundaries, as shown in Fig.~\ref{fig:Gibbsregimes}~\subref{fig:GSLCregimes}. 
The three characteristic lines $W=0$, $Q_\bathh=0$, and $Q_\bathc = 0$ uniquely partition the parameter space into four operational phases, as summarized in Table~\ref{table:GSLCregimes}:
\begin{table}[ht]
\centering
\caption{Operational regimes of the GSLC and their transition boundaries.}
\label{table:GSLCregimes}
\renewcommand{\arraystretch}{1.3} 
\begin{tabular}{|l|c|c|c|l|}
\hline
\textbf{Phase / Boundary} & \textbf{Work $W$} & \textbf{Heat ${Q_\bathc}$} & \textbf{Heat ${Q_\bathh}$} & \textbf{Physical Interpretation} \\ \hline
\textbf{Accelerator} & $>0$ & $<0$ & $>0$ & Work assists spontaneous heat flow. \\ \hline
\rowcolor{gray!10} 
\boldmath{${Q_\bathh} = 0$} & --- & --- & --- & \textit{Accelerator $\leftrightarrow$ Heater} \\ \hline
\textbf{Heater} & $>0$ & $<0$ & $<0$ & Work dissipated as friction into both baths. \\ \hline
\rowcolor{gray!10}
\boldmath{$W = 0$} & --- & --- & --- & \textit{Heater $\leftrightarrow$ Engine} \\ \hline
\textbf{Engine} & $<0$ & $<0$ & $>0$ & Thermal energy converted into useful work. \\ \hline
\rowcolor{gray!10}
\boldmath{${Q_\bathc} = 0$} & --- & --- & --- & \textit{Engine $\leftrightarrow$ Refrigerator} \\ \hline
\textbf{Refrigerator} & $>0$ & $>0$ & $<0$ & Work pumps heat from cold to hot. \\ \hline
\end{tabular}
\end{table}

As shown in Fig.~\ref{fig:Gibbsregimes}, the three boundaries (${W^\mathrm{cp}}=0, {Q_\bathc^\mathrm{cp}}=0$, and ${Q_\bathh^\mathrm{cp}}=0$) of the internally coupled GSLC do not coincide with the two conditions~\eqref{dressedbound1} and \eqref{dressedbound2} of the dressed-spectrum one, except for a few special cases. 

We analyze the conditions under which the dressed-spectrum boundaries coincide with the internally coupled ones based on the eigenbasis mismatch of the system Hamiltonians $H_S^\bathh$ and $H_S^\bathc$. 
In Appendix~\ref{A4}, we analyze each boundary of the internally coupled GSLC, as illustrated in Table~\ref{table:GSLCregimes}, and compare it with the dressed-spectrum boundaries, as in Eqs.~\eqref{dressedbound1} and \eqref{dressedbound2}. We confirm that when the eigenbases match with each other, as the Hamiltonians are commutable in the cycle, the operating regimes of the internally coupled GSLC coincide with those of the dressed-spectrum GSLC. 

By clarifying the strict geometric connection between the off-diagonal coherence $f(g_\alpha,\beta_\alpha,\omega_\alpha)$ in Eq.~\eqref{falphaGSLC} and the eigenbasis-mismatch parameter $\Lambda$ in Eq.~\eqref{Lambdacos}, we reveal that the extra dissipative thermal mode and different operating regimes arise from an eigenbasis mismatch rather than off-diagonal coherence $f(g_\alpha,\beta_\alpha,\omega_\alpha)$ itself. 

\begin{figure*}[h!]
\centering
	\includegraphics[width=0.7\textwidth]{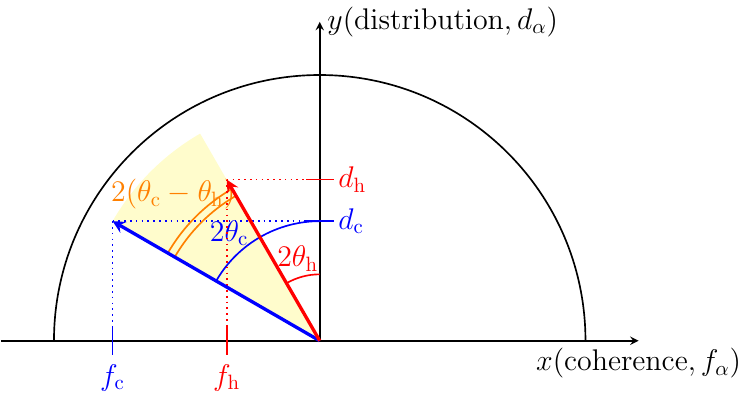}
	\caption{The internally coupled Gibbs state $\rho_\alpha^\mathrm{cp}$ decomposed as a polarization vector on the $xz$-plane of the Bloch sphere. The off-diagonal coherence $f_\alpha$ is defined in Eq.~\eqref{falphaGSLC}. The population difference between the excited and ground states $d_\alpha$ is defined in Eq.~\eqref{dalpha_definition}.}
	\label{BlochSphere}
\end{figure*}

By invoking the standard Pauli matrices $\sigma_x$ and $\sigma_z$, the internally coupled Gibbs state $\rho_\alpha^\mathrm{cp}$ expressed in the original bare basis can be elegantly decomposed as a polarization vector on the $xz$-plane of the Bloch sphere, as shown in Fig.~\ref{BlochSphere}. 
\begin{equation}
\label{eq:Bloch_decomposition}
\rho_\alpha^\mathrm{cp} = \frac{1}{2} \left( \mathbb{I} + f_\alpha \sigma_x - d_\alpha \sigma_z \right),
\end{equation}
where the off-diagonal coherence defined in Eq.~\eqref{falphaGSLC} satisfies
\begin{align}
f_\alpha = f(g_\alpha,\beta_\alpha,\omega_\alpha) = \text{Tr}(\rho_\alpha^\mathrm{cp} \sigma_x).
\end{align}
To ensure mathematical self-consistency with the matrix form in Eq.~\eqref{GibbsState2}, we introduce another parameter 
\begin{align}
\label{dalpha_definition}
d_\alpha = -\text{Tr}(\rho_\alpha^\mathrm{cp} \sigma_z) = \rho_{ee}^\alpha- \rho_{gg}^\alpha,
\end{align}
which represents the stationary population difference between the excited and ground states. 
By comparing Eq.~\eqref{eq:Bloch_decomposition} with the explicit elements of Eq.~\eqref{GibbsState2}, these two geometric components can be rigorously parameterized by the thermodynamic polarization factor $\tanh(\beta_\alpha \Delta_\alpha)$ and the geometric angle $2\theta_\alpha$ as
\begin{align}
\label{eq:f_alpha_geometric}
f_\alpha &= -\sin(2\theta_\alpha) \tanh(\beta_\alpha \Delta_\alpha), \\
\label{eq:d_alpha_geometric}
d_\alpha &= \cos(2\theta_\alpha) \tanh(\beta_\alpha \Delta_\alpha),
\end{align}
where the angle $2\theta_\alpha$ precisely characterizes the spatial orientation of the thermalized state vector on the Bloch sphere, corresponding to twice the Hilbert-space diagonalization angle $\theta_\alpha$ in Eq.~\eqref{theta}. The norm of this Bloch vector is bounded by 
\begin{align}
\sqrt{f_\alpha^2 + d_\alpha^2} = \tanh(\beta_\alpha \Delta_\alpha),
\end{align}
which represents the thermal purity dictated by the respective bath.

During the work-production process, the abrupt switch between the two internally coupled Hamiltonians $H_S^\bathh$ and $H_S^\bathc$ induces a spatial rotation of the quantization axis on the Bloch sphere by a mismatch angle $2(\theta_\bathh - \theta_\bathc)$, as shown in Fig.~\ref{BlochSphere}. The overlap between these mismatched eigenbases is quantified by the scalar factor 
\begin{align}
\label{Lambdacos}
 \Lambda & = \cos[2(\theta_\bathh - \theta_\bathc)] = \frac{4g_\bathc g_\bathh+\omega_\bathc\omega_\bathh}{4\Delta_\bathc\Delta_\bathh}.
\end{align}

By utilizing basic trigonometric identities and substituting Eqs.~\eqref{eq:f_alpha_geometric} and \eqref{eq:d_alpha_geometric}, the mismatch factor $\Lambda$ can be remarkably formulated as the normalized inner product of the hot and cold steady-state Bloch vectors, given by
\begin{align}
\label{eq:Lambda_dot_product}
\Lambda &= \cos(2\theta_\bathh)\cos(2\theta_\bathc) + \sin(2\theta_\bathh)\sin(2\theta_\bathc) \nonumber \\
        &= \frac{d_\bathh d_\bathc + f_\bathh f_\bathc}{\tanh(\beta_\bathh \Delta_\bathh) \tanh(\beta_\bathc \Delta_\bathc)}.
\end{align}
Equation~\eqref{eq:Lambda_dot_product} reveals that the eigenbasis mismatch $\Lambda$ is fundamentally rooted in the alignment of the Gibbs states in the isochoric processes.

We reveal how the eigenbasis mismatch parameter $\Lambda$ is related to the dynamic quantum friction. 
In the generalized framework of finite-time quantum thermodynamics~\cite{PhysRevLett.113.260601, PhysRevX.5.031044}, the irreversible work is commonly referred to as quantum inner friction and quantitatively determined by the non-adiabatic transition probabilities that occur when a quantum state is projected across non-commuting Hamiltonians. For our internally coupled system, we can define the inner friction using the parameter $\Lambda$ as
\begin{align}
\label{PfrictionGSLC}
P_{\rm fric} = \sin^2[2(\theta_\bathh - \theta_\bathc)] = 1 - \Lambda^2.
\end{align}
This transition probability indicates the fraction of energy that is not converted into useful work but instead irreversibly dissipated as waste heat during the isochoric stroke. 
Using $f_\alpha$ in Eq.~\eqref{eq:f_alpha_geometric} and $d_\alpha$ in Eq.~\eqref{eq:d_alpha_geometric}, the magnitude of this dynamic quantum friction term $\sin[2(\theta_\bathh - \theta_\bathc)]$ can be elegantly quantified by the cross product of the stationary Bloch vectors:
\begin{align}
\label{eq:friction_cross_product}
\sin[2(\theta_\bathh - \theta_\bathc)] &= \sin(2\theta_\bathh)\cos(2\theta_\bathc) - \cos(2\theta_\bathh)\sin(2\theta_\bathc) \nonumber \\
                                      &= \frac{d_\bathh f_\bathc - f_\bathh d_\bathc}{\tanh(\beta_\bathh \Delta_\bathh) \tanh(\beta_\bathc \Delta_\bathc)}.
\end{align}
Equations~\eqref{Lambdacos} and \eqref{eq:friction_cross_product} indicate that the emergence of dynamic quantum friction does not merely stem from the solo presence of the coherence $f_\alpha$ within a single stroke. Instead, it is strictly governed by the misalignment $d_\bathh f_\bathc \neq f_\bathh d_\bathc$ of the quantum states across different isochoric strokes. When the internal coupling is specifically tuned such that the mixing angles are perfectly aligned $\theta_\bathh = \theta_\bathc$, the cross product vanishes identically. Under this frictionless condition $\Lambda = 1$, the dynamic quantum friction drops to zero $P_{\rm fric}=0$, and the internally coupled GSLC becomes equivalent to the dressed-spectrum one.

As shown in Fig.~\ref{fig:Gibbsregimes}~\subref{fig:GSLCregimes}, the emergence of the heater regime and the expansion of the accelerator regime at the expense of the engine domain illustrate the dissipative role of the eigenbasis mismatch $\Lambda \neq 1$ caused by the non-commutativity of the Hamiltonians $[H_S^\bathh,H_S^\bathc]\neq 0$. We consider the refrigerator and the engine serve as useful thermal machines, and the heater and the accelerator act merely as dissipative modes that consume work without yielding beneficial thermodynamic outputs. Consequently, compared to the dressed-spectrum GSLC, the internally coupled one exhibits a wider variety of thermal modes but has narrower operational regimes for useful thermal modes, due to non-commutativity.

\subsection{Efficiency and coefficient of performance (COP)}
\label{Sec2-3}

In this subsection, we confirm the bi-roles of internal coupling by comparing the efficiency and coefficient of performance (COP) of the internally coupled GSLC with that of the standard and dressed-spectrum GSLCs. 
In Sec.~\ref{Sec2-3-1}, we first compare the Gibbs-state limit cycle (GSLC) with the standard GSLC and analytically derive the conditions under which the internally coupled GSLC's efficiency and COP surpass those of the standard GSLC. Then, in Sec.~\ref{Sec2-3-2}, we find that the performance of the GSLC is worse than that of the dressed-spectrum GSLC, due to the non-commutativity $\Lambda \neq 1$ of the system Hamiltonians. Note that the efficiency and COP in the GSLC for the standard and dressed-spectrum systems are the same as the generic discussion in Secs.~\ref{Sec1-1} and \ref{Sec1-2}, which are independent of the system state.

\subsubsection{Enhanced Performance Compared to Standard Otto cycle}
\label{Sec2-3-1}

As illustrated in Tables~\ref{tablecompare}, the coherence factors $f(g_\alpha, \beta_\alpha, \omega_\alpha)$ in distinct isochoric processes $\alpha=\bathh,\bathc$ play a key role in the relation between the standard and the internally coupled limit cycles. In this section, by analyzing these quantities in a Gibbs-state limit cycle (GSLC), we examine the condition that the internally coupled efficiency surpasses the standard one.

To rigorously prove the constraints on the parameter space without relying on geometric or graphical interpretations, we define a positive variable $x_\alpha \equiv -f(g_\alpha, \beta_\alpha, \omega_\alpha)$ for $\alpha \in \{\bathh, \bathc\}$. Since the parameters $g_\alpha, \beta_\alpha, \Delta_\alpha $ are real-positive values, we have $x_\alpha = (g_\alpha/\Delta_\alpha)\tanh(\beta_\alpha\Delta_\alpha) > 0$. 
Using this definition, we can rewrite the mathematical condition for the engine efficiency of the internally coupled GSLC to surpass that of the standard one $\eta_\mathrm{cp} > \eta_{\rm st}$ as
\begin{equation}
     \left(\frac{g_\bathh}{\omega_\bathh} - \frac{g_\bathc}{\omega_\bathc}\right)(x_\bathc - x_\bathh) > 0.
\end{equation}
This inequality permits two possible branches:
\begin{itemize}
    \item \textbf{Branch A:} $g_\bathh/\omega_\bathh > g_\bathc/\omega_\bathc$ and $x_\bathc > x_\bathh$.
    \item \textbf{Branch B:} $g_\bathh/\omega_\bathh < g_\bathc/\omega_\bathc$ and $x_\bathc < x_\bathh$.
\end{itemize}

We use proof by contradiction to eliminate Branch B. For the system to operate as a heat engine, it must absorb heat from the hot bath, imposing the absolute physical requirement ${Q_\bathh^\mathrm{cp}} > 0$. Assuming that Branch B holds, as in $x_\bathh > x_\bathc > 0$, we substitute this condition into $Q_\bathh^\mathrm{cp}>0$ with $Q_\bathh^\mathrm{cp}$ as defined in Eq.~\eqref{Qh}. Then, we obtain
\begin{equation}
    \left(g_\bathh + \frac{\omega_\bathc\omega_\bathh}{4g_\bathc}\right) x_\bathc > \left(g_\bathh + \frac{\omega_\bathh^2}{4g_\bathh}\right) x_\bathh.
\end{equation}
Since $x_\bathh > x_\bathc > 0$, the inequality dictates that the coefficient of $x_\bathc$ must be strictly greater than the coefficient of $x_\bathh$
\begin{equation}
    g_\bathh + \frac{\omega_\bathc\omega_\bathh}{4g_\bathc} > g_\bathh + \frac{\omega_\bathh^2}{4g_\bathh},
\end{equation}
which yields
\begin{equation}
    \frac{\omega_\bathc\omega_\bathh}{4g_\bathc} > \frac{\omega_\bathh^2}{4g_\bathh} \implies \frac{\omega_\bathc}{g_\bathc} > \frac{\omega_\bathh}{g_\bathh} \implies \frac{g_\bathh}{\omega_\bathh} > \frac{g_\bathc}{\omega_\bathc}.
\end{equation}
This directly contradicts the initial assumption of Branch B. Thus, Branch B is physically forbidden. A heat engine surpassing the standard Otto limit can only exist under the constraints of Branch A.

As summarized in Table~\ref{tablecompare}~\subref{table:compareefficiency}, after introducing internal coupling, a higher efficiency is obtained under the conditions
\begin{align}
\label{higherefficiency}
g_{\bathh}/\omega_{\bathh}>g_{\bathc}/\omega_{\bathc} \quad \text{and} \quad f(g_{\bathh},\beta_{\bathh},\omega_{\bathh})>f(g_{\bathc},\beta_{\bathc},\omega_{\bathc}),
\end{align}
with the operational constraints $W<0$, ${Q_\bathc^\mathrm{cp}}<0$, and ${Q_\bathh^\mathrm{cp}}>0$. Under fixed temperatures and energy levels, the coupling strengths can thus be tuned to achieve the optimal efficiency.

Similarly, the condition for the refrigerator coefficient of performance (COP) of the internally coupled GSLC to surpass the standard one $ \xi_\mathrm{cp} > \xi_{\rm st}$ can be rewritten as
\begin{equation}
    \left( \frac{g_\bathc}{\omega_\bathc} - \frac{g_\bathh-g_\bathc}{\omega_\bathh-\omega_\bathc} \right)(x_\bathh - x_\bathc) > 0,
\end{equation}
resulting in two mathematical branches:
\begin{itemize}
    \item \textbf{Branch C:} $g_\bathc/\omega_\bathc > (g_\bathh-g_\bathc)/(\omega_\bathh-\omega_\bathc)$ and $x_\bathh > x_\bathc$.
    \item \textbf{Branch D:} $g_\bathc/\omega_\bathc < (g_\bathh-g_\bathc)/(\omega_\bathh-\omega_\bathc)$ and $x_\bathh < x_\bathc$.
\end{itemize}

We here disprove Branch D similarly, using proof by contradiction. The system must act as a refrigerator, which strictly requires absorbing heat from the cold bath as ${Q_\bathc^{\rm cp}} > 0$. Given $x_\bathc > x_\bathh > 0$ from Branch D, substituting this into ${Q_\bathc^{\rm cp}} > 0$ with $Q_\bathc^{\rm cp}$ as defined in Eq.~(\ref{Qc}), we obtain
\begin{equation}
    \left(g_\bathc + \frac{\omega_\bathc\omega_\bathh}{4g_\bathh}\right) x_\bathh > \left(g_\bathc + \frac{\omega_\bathc^2}{4g_\bathc}\right) x_\bathc.
\end{equation}
By the same algebraic logic, since $x_\bathc > x_\bathh > 0$, the coefficient of $x_\bathh$ must be strictly greater than that of $x_\bathc$
\begin{equation}
    g_\bathc + \frac{\omega_\bathc\omega_\bathh}{4g_\bathh} > g_\bathc + \frac{\omega_\bathc^2}{4g_\bathc},
\end{equation}
which leads to
\begin{align}
    \frac{\omega_\bathc\omega_\bathh}{4g_\bathh} > \frac{\omega_\bathc^2}{4g_\bathc} \implies \frac{\omega_\bathh}{g_\bathh} > \frac{\omega_\bathc}{g_\bathc} \implies \frac{g_\bathc}{\omega_\bathc} > \frac{g_\bathh}{\omega_\bathh}.
\end{align}
Algebraically, the inequality $g_\bathc/\omega_\bathc > g_\bathh/\omega_\bathh$ is rigorously equivalent to $g_\bathc/\omega_\bathc > (g_\bathh-g_\bathc)/(\omega_\bathh-\omega_\bathc)$. This outcome directly contradicts the initial assumption of Branch D. Therefore, Branch D cannot physically satisfy ${Q_\bathc^\mathrm{cp}} > 0$, proving that an enhanced refrigerator is strictly bound to the parameter space of Branch C.

Following the illustration in Table~\ref{tablecompare}~\subref{table:compareCOP}, after introducing the internal coupling, a higher COP is achieved under the conditions
\begin{align}
\label{higherCOP}
f(g_{\bathh},\beta_{\bathh},\omega_{\bathh}) < f(g_{\bathc},\beta_{\bathc},\omega_{\bathc}) \quad \text{and} \quad
\frac{g_{\bathc}}{\omega_{\bathc}} > \frac{g_{\bathh}-g_{\bathc}}{\omega_{\bathh}-\omega_{\bathc}},
\end{align}
together with ${Q_\bathc}>0$, ${Q_\bathh}<0$, and $W>0$. 
By adjusting the coupling strengths at fixed temperatures and energy levels, the quantum Otto refrigerator can thus achieve enhanced COP performance.

\begin{figure*}[h!]
\centering
	\subfloat[]{\includegraphics[width=0.46\textwidth]{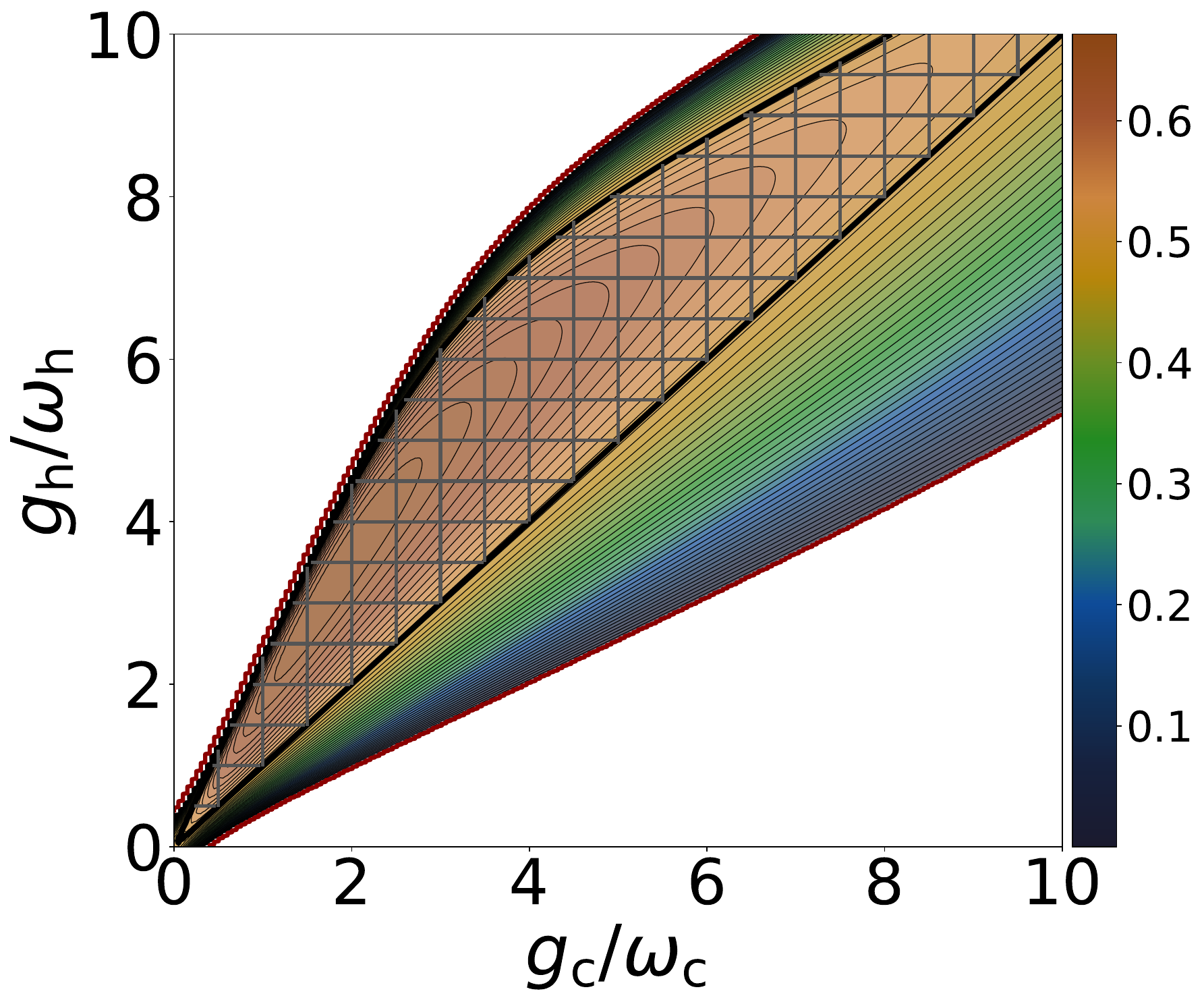}\label{GibbsEfficiency}}
	\hspace{3mm}
	\subfloat[]{\includegraphics[width=0.46\textwidth]{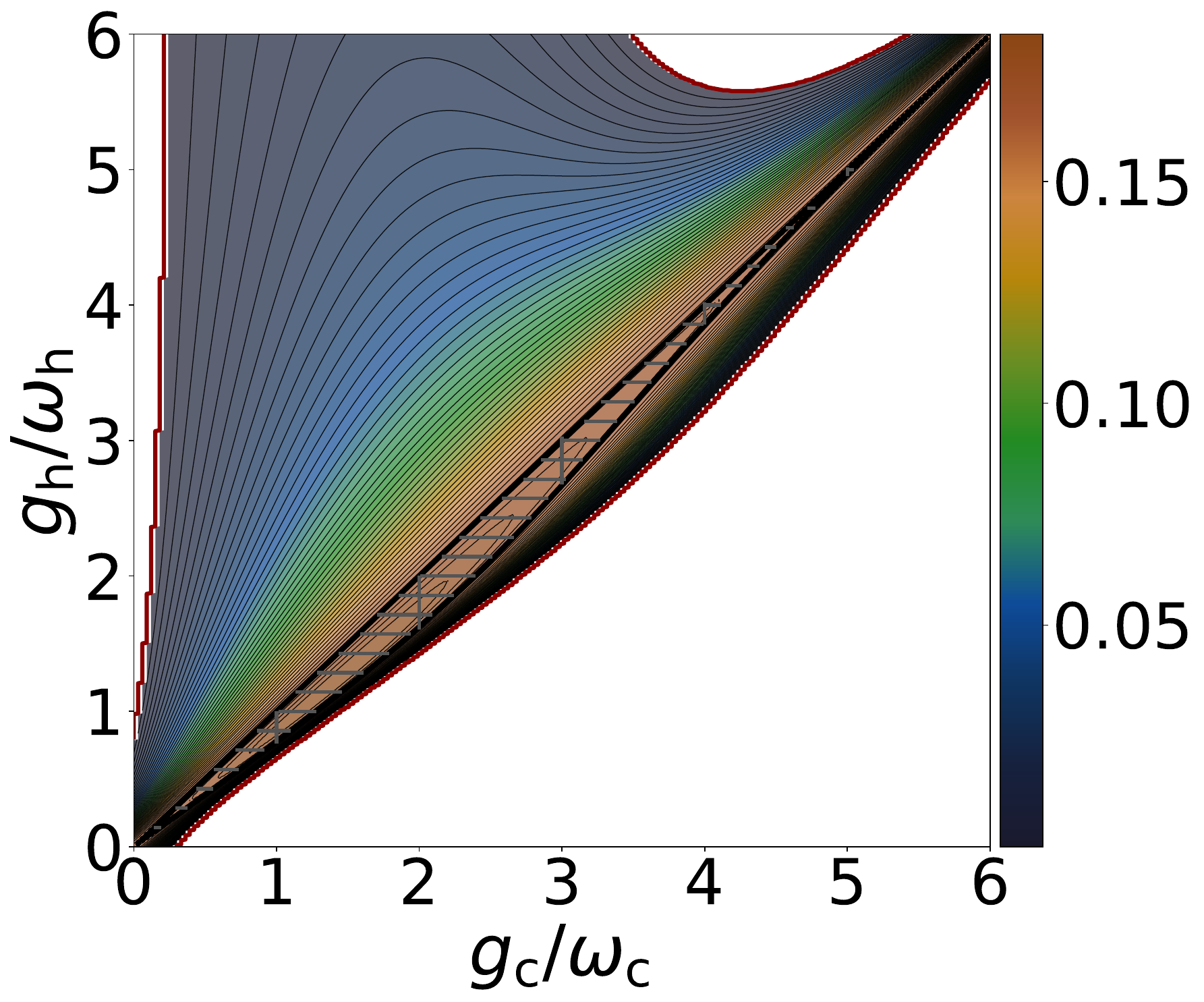}\label{GibbsCOP}}
	\caption{Dependence of (a)~efficiency $\eta_\mathrm{cp}$ of the Otto engine with $\omega_{\bathh}=2$ and (b)~COP $\xi_\mathrm{cp}$ of the Otto cooler with $\omega_{\bathh}=7$ in the Gibbs-state limit cycle (GSLC) on the ratio $g_\alpha/\omega_\alpha$. The black boundary indicates the efficiency $\eta_{\rm st}$ and COP $\xi_{\rm st}$ for the Otto cycle without the internal coupling. The meshing area indicates that the efficiency and the COP of the internally coupled GSLC exceed those of the standard GSLC, with $\eta_\mathrm{cp}>\eta_{\rm st}$ and $\xi_\mathrm{cp}>\xi_{\rm st}$. We set the energy unit and temperature unit to $\omega_{\bathc}=1$ and $\beta_{\bathc}=1$, and the inverse temperature of the hot bath to $\beta_{\bathh}=0.2$.}
	\label{GibbsPerformance}
\end{figure*}

In Fig.~\ref{GibbsPerformance}, we compared the efficiency and the COP of the internally GSLC with the standard one numerically. The numerical results confirm the relations~\eqref{higherefficiency} and \eqref{higherCOP} derived above, in which the performance of the GSLC can surpass the standard Otto bound. 
Hence, with proper coupling strength, the internal coupling can improve the thermal performance of the Otto cycle.

\subsubsection{Reduced Performance Compared to Dressed-Spectrum Otto cycle}
\label{Sec2-3-2}

In Sec.~\ref{Sec2-2}, we introduced a parameter $\Lambda$ to analyze the boundary coincidence between the internally coupled and dressed-spectrum GSLCs and reveal the reason for the extra dissipative thermal mode by connecting the off-diagonal coherence $f(g_\alpha,\beta_\alpha,\omega_\alpha)$ in Eq.~\eqref{falphaGSLC}, the eigenbasis-mismatch parameter $\Lambda$ in Eq.~\eqref{Lambdacos}, and the inner friction term $P_{\rm fric}$ in Eq.~\eqref{PfrictionGSLC}.

In this section, we continue using the parameter $\Lambda$ in Eq.~\eqref{Lambdacos} to compare the efficiency and the coefficient of performance (COP) of the internally coupled and dressed-spectrum GSLCs, revealing the dissipative role of the eigenbasis mismatch caused by the non-commutativity of the system Hamiltonian.

\begin{figure*}[h!]
\centering
	\subfloat[]{\makebox[0.47\textwidth][c]{\includegraphics[scale=0.4]{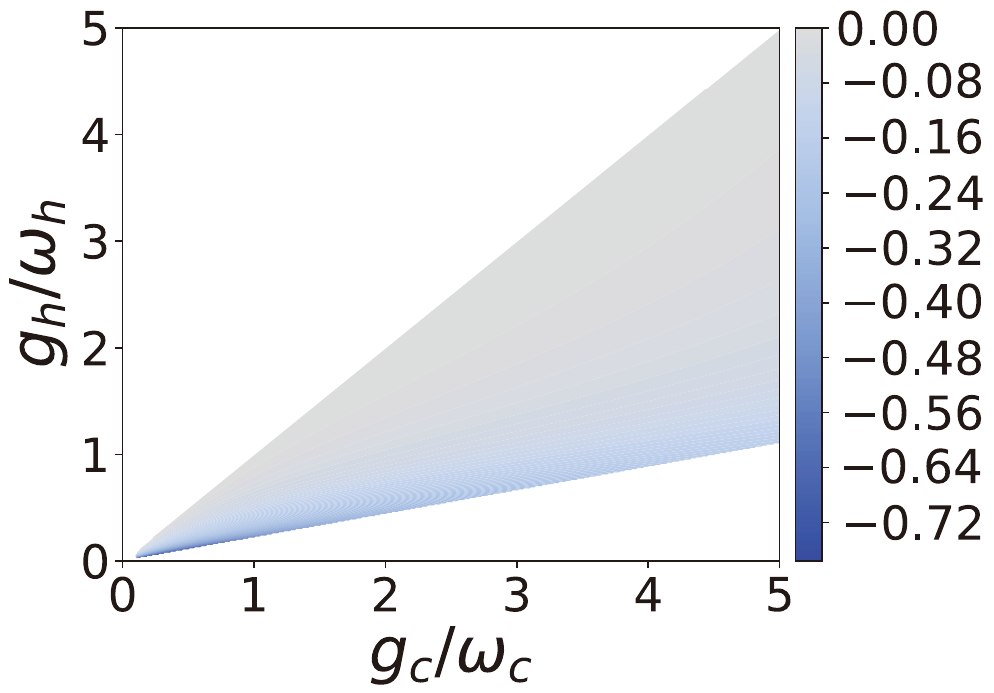}}\label{diff_efficiency}}
	\subfloat[]{\makebox[0.47\textwidth][c]{\includegraphics[scale=0.4]{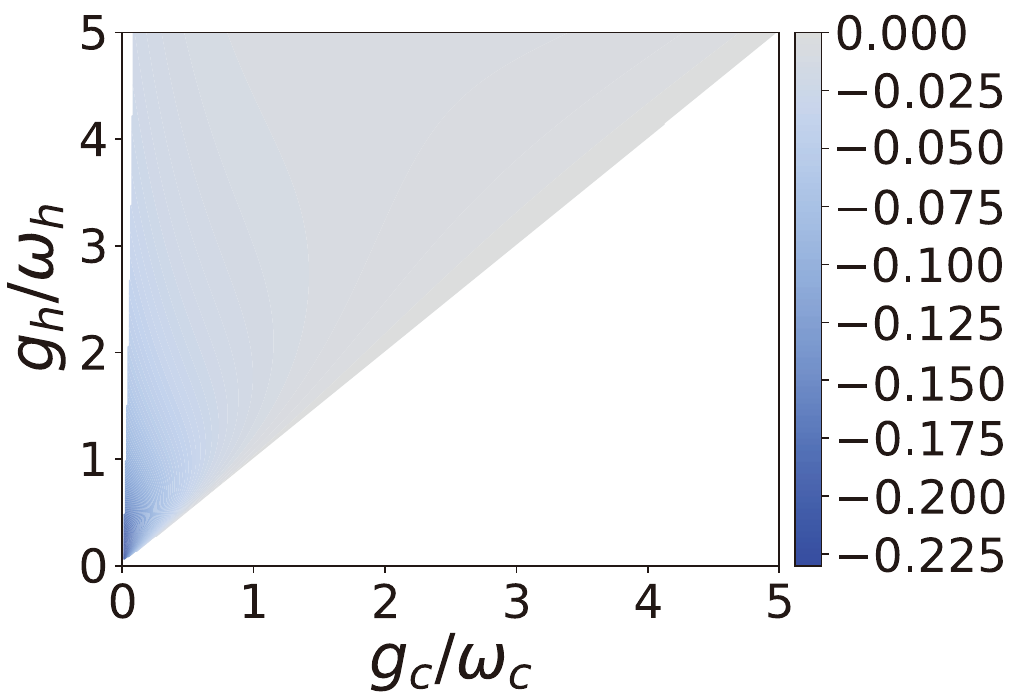}}\label{diff_COP}}
	\caption{Difference of (a)~ the efficiency $\eta_\mathrm{cp}-\eta_{\rm ds}$ and (b)~the coefficient of performance (COP) $\xi_\mathrm{cp}-\xi_{\rm ds}$ between the internal coupled Gibbs-state limit cycle (GSLC) and dressed-spectrum limits. We set the parameter values to $\omega_{\bathh}=5$, $\omega_{\bathc}=1$, $\beta_{\bathh}=0.2$, $\beta_{\bathc}=1$, and hence $\omega_{\bathh}/\omega_{\bathc} = \beta_{\bathc}/\beta_{\bathh} = 1/5$.}
	\label{diff_performance}
\end{figure*}

As shown in Fig.~\ref{diff_performance}, we find that the efficiency of the engine and the coefficient of performance (COP) of the refrigerator of the internally coupled GSLC are smaller than those of the dressed-spectrum one. 
To further clarify this, we again rewrite the efficiency and the COP of the internally coupled GSLC using the parameter $\Lambda$ as defined in Eq.~\eqref{Lambdacos}. As discussed in Sec.~\ref{Sec2-2}, the parameter $\Lambda$ indicates the alignment of the Gibbs states in different isochoric processes. It can be rewritten as a cosine function of the angle difference $2(\theta_\bathh-\theta_\bathc)$. The internal friction is defined as the sine function of the angle difference, as in Eq.~\eqref{eq:friction_cross_product}, and hence 
\begin{align}
\label{LambdaBound}.
|\Lambda|\le1.
\end{align} 
The fraction of energy, which is not converted into useful work but instead irreversibly dissipated as waste heat during the isochoric stroke, is indicated by $1-\Lambda^2$, as in Eq.~\eqref{PfrictionGSLC}.

The GSLC operates as an engine when
\begin{align}
{Q_\bathh^\mathrm{cp}}>0,\qquad {Q_\bathc^\mathrm{cp}}<0,\qquad {W^\mathrm{cp}}<0.
\end{align}
Using the definitions of $Q_\bathh^\mathrm{cp}$ and $Q_\bathc^\mathrm{cp}$ in Eqs.~\eqref{QhGibbs} and \eqref{QcGibbs}, we can reduce these conditions to
\begin{align}
\Lambda\tanh\left(\beta_\bathc\Delta_\bathc\right)>
\tanh\left(\beta_\bathh\Delta_\bathh\right).
\end{align}
The engine's efficiency, as in Eq.~\eqref{etaGibbs}, can be simplified using the parameter $\Lambda$ as
\begin{align}
\eta_\mathrm{cp}
&=-\frac{{W^\mathrm{cp}}}{{Q_\bathh^\mathrm{cp}}} \\ \nonumber
&= \frac{ \frac{\left[4g_\bathc(g_\bathh-g_\bathc)+\omega_\bathc(\omega_\bathh-\omega_\bathc)\right]}{4\Delta_\bathc}\tanh(\beta_\bathc\Delta_\bathc) + \frac{\left[4g_\bathh(g_\bathc-g_\bathh)+\omega_\bathh(\omega_\bathc-\omega_\bathh)\right]}{4\Delta_\bathh}\tanh(\beta_\bathh\Delta_\bathh) }{ \frac{4 g_\bathc g_\bathh+\omega_\bathc \omega_\bathh}{4\Delta_\bathc}\tanh(\beta_\bathc \Delta_\bathc)-\Delta_\bathh \tanh(\beta_{\bathh} \Delta_\bathh) }  \\ \nonumber
&=1-\frac{\Delta_\bathc\left[\tanh\left(\beta_\bathc\Delta_\bathc\right)-
\Lambda\tanh\left(\beta_\bathh\Delta_\bathh\right)\right]}{\Delta_\bathh
\left[\Lambda\tanh\left(\beta_\bathc\Delta_\bathc\right)-\tanh\left(
\beta_\bathh\Delta_\bathh\right)\right]}.
\end{align}
Defining
\begin{align}
x=\tanh\left(\beta_\bathc\Delta_\bathc\right),
\qquad
y=\tanh\left(\beta_\bathh\Delta_\bathh\right),
\end{align}
we have the efficiency in the forms
\begin{align}
\eta_\mathrm{cp}=
1-\frac{\Delta_\bathc}{\Delta_\bathh}
\frac{x-\Lambda y}{\Lambda x-y}.
\end{align}
For the dressed-spectrum Otto cycle, the efficiency is
\begin{align}
\eta_{\rm ds}
=1-\frac{\Delta_\bathc}{\Delta_\bathh}.
\end{align}
Subtracting the two expressions yields
\begin{align}
\eta_\mathrm{cp}-\eta_{\rm ds}
&=\frac{\Delta_\bathc}{\Delta_\bathh}
\left[1-\frac{x-\Lambda y}{\Lambda x-y}\right]\\
&=\frac{\Delta_\bathc}{\Delta_\bathh}
\frac{(\Lambda-1)(x+y)}{\Lambda x-y}.
\end{align}

The engine condition requires
\begin{align}
\Lambda x-y>0,
\end{align}
while
\begin{align}
x+y>0,
\qquad
\Lambda\le1,
\end{align}
and hence, we obtain
\begin{align}
\label{efficiency_bound}
\eta_\mathrm{cp}\leq\eta_{\rm ds}.
\end{align}
The equality holds only when
\begin{align}
\Lambda=1,
\end{align}
in which the dressed eigenbases of the two Hamiltonians coincide with each other.

Similarly, the refrigerator regime is determined by
\begin{align}
\tanh\left(\beta_\bathc\Delta_\bathc\right)>
\Lambda\tanh\left(\beta_\bathh\Delta_\bathh\right),
\end{align}
and the coefficient of performance is
\begin{align}
\xi_\mathrm{cp}=&\frac{{Q_\bathc^\mathrm{cp}}}{{W^\mathrm{cp}}} \\ \nonumber
 = & \frac{ \frac{4g_\bathc g_\bathh+\omega_\bathc\omega_\bathh}{4\Delta_\bathh} \tanh(\beta_\bathh\Delta_\bathh) -\Delta_\bathc \tanh(\beta_\bathc\Delta_\bathc) }{ \frac{\left[4g_\bathc(g_\bathc-g_\bathh)+\omega_\bathc(\omega_\bathc-\omega_\bathh)\right]}{4\Delta_\bathc}\tanh(\beta_\bathc\Delta_\bathc) + \frac{\left[4g_\bathh(g_\bathh-g_\bathc)+\omega_\bathh(\omega_\bathh-\omega_\bathc)\right]}{4\Delta_\bathh}\tanh(\beta_\bathh\Delta_\bathh) }\\ \nonumber
=&\frac{\Delta_\bathc\left[\Lambda y-x\right]}{
\Delta_\bathh\left[x-\Lambda y\right]
-\Delta_\bathc\left[\Lambda y-x\right]}.
\end{align}
After straightforward algebra, we obtain
\begin{align}
\xi_\mathrm{cp}
=\frac{\Delta_\bathc}{\Delta_\bathh-\Delta_\bathc}
\frac{\Lambda y-x}{x-\Lambda y+\dfrac{\Delta_\bathc}{\Delta_\bathh}(\Lambda y-x)}.
\end{align}
The coefficient of performance (COP) of the dressed-spectrum refrigerator is given by
\begin{align}
\xi_{\rm ds}=\frac{\Delta_\bathc}{\Delta_\bathh-\Delta_\bathc}.
\end{align}
Using the refrigerator condition
\begin{align}
x>\Lambda y 
\end{align}
together with $\Lambda\le1$, we find
\begin{align}
\label{COP_bound}
\xi_\mathrm{cp}\leq\xi_{\rm ds}.
\end{align}
The equality holds when $\Lambda = 1$. 

Hence, the efficiency and the COP of the internally coupled GSLC are bounded by those of the dressed-spectrum GSLC, since the non-commutativity works as the quantum internal friction, which passively influences the thermal performance. Only when the Hamiltonians are commutative in different strokes does the internally coupled Otto cycle reach the upper bound given by the dressed-spectrum one.

\subsubsection{The Zero-Friction Limit}
\label{Sec2-3-3}

To systematically evaluate the thermodynamic advantage of the internal coupling, we compare the three distinct efficiencies, (i)~the standard Otto efficiency $\eta_{\rm st}$ in Eq.~\eqref{standardefficiency}, (ii)~the dressed-spectrum efficiency $\eta_{\rm ds}$ in Eq.~\eqref{eta_dressed}, representing the theoretical upper bound when internal coupling reshapes the energy spectrum but operates without any geometric friction, and (iii)~the internally coupled efficiency $\eta_{\mathrm{cp}}$ in Eq.~\eqref{efficiency}, which is the actual performance subjected to the internal friction caused by non-commutativity.

As discussed above, introducing internal coupling effectively alters the compression ratio of the engine. By adjusting the internal coupling strength $g_\alpha$ and the bare energy level $\omega_\alpha$, we can achieve higher efficiency for the internally coupled system than for the standard one, $\eta_{\mathrm{cp}} > \eta_{\mathrm{st}}$. However, this energetic advantage is inevitably accompanied by a geometric cost, given by the eigenbasis mismatch $\Lambda \neq 1$ caused by the non-commutativity $[H_{\bathh}, H_{\bathc}] \neq 0$, which induces an internal friction area and strictly enforces $\eta_{\mathrm{cp}} < \eta_{\mathrm{ds}}$.`

It is natural to optimize the system by setting the geometric mismatch angle to zero, thereby completely eliminating this internal friction and making the internally coupled efficiency achieve the upper bound given by the dressed-spectrum efficiency $\eta_{\rm cp} = \eta_{\mathrm{ds}}$. However, we mathematically demonstrate that eliminating the internal friction completely erases the energetic advantage, collapsing the efficiencies back to the standard limit $\eta_{\rm cp} = \eta_{\mathrm{ds}} = = \eta_{\mathrm{st}}$.

The condition for vanishing internal friction implies a perfect alignment $\Lambda = 1$ of the eigenbases between the hot and cold strokes. Mathematically, the geometric mismatch angle $2(\theta_\bathh - \theta_\bathc)$ in Eq.~\eqref{Lambdacos} must be zero, which implies
\begin{equation}
   \theta_\bathh = \theta_\bathc.
\end{equation}
The ratio of the effective energy gaps $\tilde{\omega}_\alpha$ in Eq.~\eqref{effectiveenergygap} is constrained by the bare energy gaps $\omega_\alpha$
\begin{equation}
    \frac{\tilde{\omega}_\bathc}{\tilde{\omega}_\bathh} = \frac{\omega_\bathc}{\omega_\bathh}.
\end{equation}
Substituting this relation into the definition of the dressed-spectrum efficiency and the standard efficiency, we find:
\begin{equation}
    \eta_{\mathrm{ds}} = 1 - \frac{\tilde{\omega}_\bathc}{\tilde{\omega}_\bathh} = 1 - \frac{\omega_\bathc}{\omega_\bathh} = \eta_{\rm st}.
\end{equation}
Furthermore, since $2(\theta_\bathh - \theta_\bathc)-0$, the geometric friction factor vanishes completely $P_{\rm fric} = 0$, allowing the actual efficiency to reach the theoretical upper bound $\eta_{\mathrm{cp}} = \eta_{\mathrm{ds}}$. Consequently, we reach a strict equivalence 
\begin{equation}
    \eta_{\mathrm{cp}} = \eta_{\mathrm{ds}} = \eta_{\mathrm{st}}.
\end{equation}
When there is no internal friction, a similar equivalence 
\begin{equation}
    \xi_{\mathrm{cp}} = \xi_{\mathrm{ds}} = \xi_{\mathrm{st}}.
\end{equation}
can also be confirmed for the COP when the systems run as a refrigerator.

This result reveals a fundamental constraint in internally coupled quantum engines: the energetic advantage of asymmetric basis rotation is mathematically entangled with the dissipation arising from the eigenbasis mismatch. Eliminating the internal friction necessitates an isotropic scaling of the energy spectrum, which merely preserves the original performance and entirely neutralizes the coupling-induced thermodynamic enhancement. Therefore, the internal friction should not be merely viewed as an engineering flaw, but rather as an unavoidable physical cost exacted for operating beyond the standard thermodynamic limit.

\section{Non-Equilibrating Limit Cycle (NELC)}
\label{Sec3}

In the previous section~\ref{Sec2}, we considered the case in which the three systems are thermalized to the Gibbs state promptly in each isochoric process and reach the Gibbs-state limit cycle (GSLC). By comparing internally coupled GSLC with the standard and dressed-spectrum cases, we found the dual roles of the internal coupling under equilibrium conditions. 

However, rapid thermalization is often impossible. In practice, the dynamics in a finite-time cycle are significant. 
In this section, we analyze the quantum Otto cycle with finite interaction time in the isochoric process, as illustrated in Fig.~\ref{Diagrams}~\subref{NELCDiagram}, namely the non-equilibrating limit cycle (NELC). 
With infinite interaction time in the isochoric process, as illustrated in Fig.~\ref{Diagrams}~\subref{ELCDiagram}, the Otto cycle equilibrates and converges to the equilibrium limit cycle (ELC), which exhibits performance similar to that of the Gibbs-state limit cycle (GSLC). Here, we clarify that the term ``limit cycle'' used here is the unique periodic steady state satisfying $\rho(t + T_{\mathrm{cycle}}) = \rho(t)$ due to the contractivity of the quantum dynamical map. 
Unlike the limit cycles in classical non-linear dynamics, which emerge from non-linear feedback loops~\cite{Ginoux2012, JENKINS2013167}, once the limit cycle here is established, the engine operates as a closed-loop switching between two non-equilibrium Floquet steady states.

In both the ELC with infinite interaction time and GSLC with trivial evolution time, the power is meaningless. However, the power is significant in practice, which makes sense only in the NELC.  The operating regimes and the thermal performance of the standard and dressed-spectrum NELCs, as discussed in Secs.~\ref{Sec1-1} and \ref{Sec1-2}, are independent of the interaction time, whereas the power depends on the interaction time since the energy change is time-dependent. 
This is different from the power-efficiency trade-off law in classical mechanics. A classical thermal machine achieves the equilibrium efficiency as the maximum when the thermal operation is slow and long enough, but its power is zero in this situation. However, in Sec.~\ref{Sec3-2}, we confirm that the efficiency and COP of the dressed-spectrum NELC are time-independent, which means the finite-time interaction is not the only reason for the power-efficiency trade-off law. 
Instead, in this section, by exploiting the non-commutativity of the internally coupled NELC, we obtain the power-efficiency trade-off law, revealing that internal friction is essential to it. 

Additionally, we detail the validation of the global approach to the GKSL master equation in Appendix~\ref{A1}. The global approach used in the internally coupled NELC decreases the decay rates $\tilde{\gamma}_\alpha^\pm$ in Eq.~\eqref{tildegamma} by introducing a coefficient $\cos^(2\theta_\alpha)$, compared to the local decay rates $\gamma_\alpha^\pm$ in Eq.~\eqref{gamma} used in the dressed-spectrum NELC. The term $\cos(2\theta_\alpha)$ is tightly related to the off-diagonal coherence caused by the eigenbasis transformation $U_\alpha$ in Eq.~\eqref{Ualpha}, instead of the eigenbasis mismatch $\Lambda = \cos[2(\theta_\bathc-\theta_\bathh)]$ caused by the non-commutativity of the Hamiltonians. Hence, the internal coupling introduced the off-diagonal coherence of the system state, resulting in the power of the internally coupled NELC always being lower than that of the dressed-spectrum NELC. 
With these discussions, we find that non-commutativity influences the efficiency and COP, while the off-diagonal coherence itself influences the power.

\subsection{Internally Coupled NELC and ELC}
\label{Sec3-1}

We assume that the interaction time is negligible during the work-production processes (b) and (d) in Fig.~\ref{cyclecoupling}, so that the system Hamiltonian changes rapidly without altering the internal system state. 
The non-equilibrium process is inevitable during the isochoric processes (a) and (c) in Fig.~\ref{cyclecoupling}. We describe the non-equilibrium process by the Gorini--Kossakowski--Sudarshan--Lindblad (GKSL) master equation~\cite{breuer2002theory, Hofer_2017}.

When the interaction time is short, the system cannot entirely equilibrate.
As proven in Appendix~\ref{A2}, the stroboscopic evolution over one Otto cycle defines a primitive, completely positive and trace-preserving (CPTP) map. According to the quantum Perron--Frobenius theorem, the dynamics finally converge to a unique periodic limit cycle. Within each cycle, if each isochoric process is short, the system alternates between two distinct states $\rho_{\mathrm{NELC}}^{h}$ and $\rho_{\mathrm{NELC}}^{c}$, and the limit cycle is a non-equilibrating limit cycle (NELC), as illustrated in Fig.~\ref{Diagrams}\subref{NELCDiagram}.

If we increase the interaction time to infinity in the isochoric processes, the limit cycle becomes an equilibrium limit cycle (ELC), as shown in Fig.~\ref{Diagrams}~\subref{NELCDiagram}. In the ELC, the state equilibrates at the end of each isochoric process and should exhibit properties similar to those of the GSLC. 
In Appendix~\ref{A1}, we derive the GKSL master equation using both local and global approaches. As detailed in Appendix~\ref{A1-1}, the local master equation is the standard GKSL master equation, which does not include the influence of the internal coupling. The equilibrium state~(\ref{localR0}) derived from the local master equation is not the thermalized Gibbs state. On the other hand, as detailed in Appendix~\ref{A1-2}, the global master equation, which is updated to account for internal coupling, produces the correct equilibrium Gibbs state. Therefore, in the analysis, we utilize the global master equation rather than the local one.

As derivation in Appendix~\ref{A1-2}, we utilize the global GKSL master equation to derive the time evolution of the internally coupled Otto cycle, given by
\begin{align}
\label{globalLiouvillian}
\frac{d\tilde{\rho}_\alpha}{dt} = \tilde{\mathcal{L}}_\alpha \tilde{\rho}_\alpha = -i [\tilde{H}_S^\alpha, \tilde{\rho}_\alpha] +\tilde{\gamma}_\alpha^- \mathcal{D}(\cos(2\theta)\tilde{\sigma}_S^+) + \tilde{\gamma}_\alpha^+ \mathcal{D}(\cos(2\theta)\tilde{\sigma}_S^-) + \tilde{\gamma}_\alpha(0) \mathcal{D}(-\sin(2\theta)\tilde{\sigma}_S^z)
\end{align}
with the dissipator
\begin{align}
\label{dissipator}
\mathcal{D}(\hat{O}) = \hat{O}\rho \hat{O}^\dagger -\frac{1}{2}\{\hat{O}^\dagger\hat{O},\rho\}.
\end{align}
Under the diagonalized system basis \[\tilde{\rho}_\alpha = \begin{pmatrix} \tilderhogg^\alpha &\tilderhoge^\alpha \\\tilderhoeg^\alpha &\tilderhoee^\alpha \end{pmatrix},\] as in Eq.~(\ref{tildeHS}), there is no energy gap for the transition \[\tilde{\sigma}_S^z = \begin{pmatrix}1 &0\\ 0 & -1 \end{pmatrix},\] while the excitation frequency $\Delta \omega = \tilde{\omega}_\alpha$ is associated with \[\tilde{\sigma}_S^+ = \begin{pmatrix}0 & 0 \\ 1 & 0\end{pmatrix},\] and the de-excitation frequency $\Delta \omega = -\tilde{\omega}_\alpha$ is associated with \[\tilde{\sigma}_S^- = \begin{pmatrix}0 & 1 \\ 0 & 0\end{pmatrix}.\] 
In the analysis of the Markovian dynamical system, we assume an Ohmic environment with the spectral density $J(\omega) \propto \omega$, we have $\tilde{\gamma}_\alpha(0) = 0$, and
\begin{align}
\label{tildegamma}
&\tilde{\gamma}_\alpha^\pm = \gamma(\pm\tilde{\omega}_\alpha,\beta_\alpha) 
\end{align}
with the bath correlation function
\begin{align}
\label{gamma}
\gamma(\omega,\beta) =
\begin{cases}
J(\omega)\, n(\beta,\omega), & \omega < 0,\\
J(\omega)\,[n(\beta,\omega)+1], & \omega > 0,
\end{cases}
\end{align}
with $J(\omega)$ being the spectral density and $n(\beta,\omega)$ the Bose-Einstein distribution:
\begin{align}
&J(\omega) = \Gamma \omega e^{-\omega^s/\omega_{ct}^{1-s}},\\
\label{BEdistribution}
&n(\beta,\omega) = \frac{1}{e^{\beta \omega}-1},
\end{align}
where $\omega_\mathrm{ct}$ is the cut-off frequency and $\Gamma$ indicates the coupling strength of the spectral density. The exponent $s$ characterizes the low-frequency behavior of the bath spectral density and determines whether the environment is sub-Ohmic ($s<1$), Ohmic ($s=1$), or super-Ohmic ($s>1$).
Hence, the global Liouvillian superoperator can be expressed as
\begin{align}
\label{globalmatrix}
\tilde{\mathcal{L}}_\alpha = \begin{pmatrix}
-\tilde{\Gamma}_\alpha^- &0 &0 &\tilde{\Gamma}_\alpha^+\\
0 &-\frac{1}{2}(\tilde{\Gamma}_\alpha^+ + \tilde{\Gamma}_\alpha^-)+i \tilde{\omega}_\alpha &0 &0 \\
0 &0 &-\frac{1}{2}(\tilde{\Gamma}_\alpha^+ + \tilde{\Gamma}_\alpha^-)-i \tilde{\omega}_\alpha &0\\
\tilde{\Gamma}_\alpha^- &0 &0 &-\tilde{\Gamma}_\alpha^+,
\end{pmatrix}
\end{align}
with
\begin{align}
\label{globalGamma}
\tilde{\Gamma}_\alpha^\pm = \cos^2(2 \theta_\alpha) \tilde{\gamma}_\alpha^\pm,
\end{align}
where the basis is $ \begin{pmatrix} \tilderhogg & \tilderhoge &\tilderhoeg & \tilderhoee \end{pmatrix}^T$. 

The spectral properties of the global Liouvillian superoperator $\tilde{\mathcal{L}}_\alpha$ can be obtained analytically. The eigenvalues $\tilde{\lambda}_i^\alpha$ are
\begin{align}
\label{globallambda0}
&\tilde{\lambda}_0^\alpha = 0,\\
\label{globallambda1}
&\tilde{\lambda}_1^\alpha = -\frac{1}{2} (\tilde{\Gamma}_\alpha^- + \tilde{\Gamma}_\alpha^+) + i \tilde{\omega}_\alpha,\\
\label{globallambda2}
&\tilde{\lambda}_2^\alpha = -\frac{1}{2} (\tilde{\Gamma}_\alpha^- + \tilde{\Gamma}_\alpha^+) - i \tilde{\omega}_\alpha,\\
\label{globallambda3}
&\tilde{\lambda}_3^\alpha = -(\tilde{\Gamma}_\alpha^- + \tilde{\Gamma}_\alpha^+).
\end{align}
The zero eigenvalue $\tilde{\lambda}_0^\alpha$ corresponds to a stationary state. The other three eigenvalues are negative, leading to the decay of the non-equilibrium terms. 

In the limit of infinite interaction time, the system reaches equilibrium during each isochoric process, leading to an equilibrium limit cycle (ELC), as illustrated in Fig.~\ref{Diagrams}\subref{ELCDiagram}, in which the influence of non-equilibrium terms becomes negligible after a long time. In each isochoric process of the ELC, the system state is the equilibrium steady state in the end, which is the zero-eigenvalue state $\tilde{\rho}_{0}^\alpha$~(\ref{globalR0}) of the global Liouvillian~(\ref{globalLiouvillian}), corresponding to the zero eigenvalue~(\ref{globallambda0}):
\begin{align}
\label{globalELC}
\tilde{\rho}_\mathrm{ELC}^\alpha =& \tilde{\rho}_{0}^\alpha =  \frac{1}{\tilde{\Gamma}_\alpha^- + \tilde{\Gamma}_\alpha^+}\begin{pmatrix}
\tilde{\Gamma}_\alpha^+ &0 \\ 0 &\tilde{\Gamma}_\alpha^-
\end{pmatrix}.
\end{align}
The relation between the system states in the original basis $\rho_\text{ELC}^\alpha$ and the diagonalized basis $\tilde{\rho}_\text{ELC}^\alpha$ of the Liouvillian~(\ref{globalLiouvillian}) in each isohoric process $\alpha=\bathh,\bathc$ is given by
\begin{align}
\rho_\text{ELC}^\alpha = U_\alpha \tilde{\rho}_\text{ELC}^\alpha U_\alpha^\dagger.
\end{align}

We can rewrite the equilibrium heat exchanges of the ELC as
\begin{align}
\label{QhELC}
{Q_\bathh} &= \Tr[H_S^\bathh (\rho_\text{ELC}^\bathh - \rho_\text{ELC}^\bathc)],\\
\label{QcELC}	
{Q_\bathc} &= \Tr[H_S^\bathc (\rho_\text{ELC}^\bathc - \rho_\text{ELC}^\bathh)].
\end{align}
In each work-production process, the work extracted from the environment is given by
\begin{align}
\label{W1ELC}
{W_1}&= \Tr[(H_S^\bathc-H_S^\bathh)\rho_\text{ELC}^\bathh], \\ 
\label{W2ELC}
{W_2}&= \Tr[(H_S^\bathh-H_S^\bathc)\rho_\text{ELC}^\bathc],
\end{align}
which still obeys the first law of thermodynamics. 
Due to the infinite time, the power is zero, and the thermal performance of the ELC is only governed by the efficiency and the COP. Both the Gibbs-state limit cycle (GSLC) and the ELC describe the equilibrium and thermalized Otto cycle, and they should reveal the same thermal performance.

Let us now find the non-equilibrium limit cycle (NELC) shown in Fig.~\ref{Diagrams}\subref{NELCDiagram}.
In each cycle, the internal system develops over time $t_\bathh$ from a non-equilibrium state $\tilde{\rho}_\mathrm{NELC}^\bathc$ to another non-equilibrium state $\tilde{\rho}_\mathrm{NELC}^\bathh$ in the isochoric process (a), and develops back from $\tilde{\rho}_\mathrm{NELC}^\bathh$ to $\tilde{\rho}_\mathrm{NELC}^\bathc$ in the isochoric process (c). 
The time evolution in each isochoric process is governed by the global GKSL master equation with the Liouvillian superoperator $\tilde{\mathcal{L}}^\alpha$ given in Eq.~(\ref{globalLiouvillian}) for the isochoric processes (a) and (c).

In addition to the time evolution in each isochoric process, we need to make basis transformations.
The global master equation is derived in the diagonalizing basis of the system Hamiltonian, as detailed in Appendix~\ref{A1-2}. 
Hence, we perform additional eigenbasis transformations to go back and forth between the eigenbasis of the Hamiltonians for the two isochoric processes.

The proper combination of the time evolutions and the basis transformations constitutes the operator for one cycle.
Indeed, we prove in Appendix~\ref{A2} that this cycle with the finite interaction times $t_\bathh$ and $t_\bathc$ converges to a limit cycle in the end.

Note that each non-equilibirium state $\tilde{\rho}_\mathrm{NELC}^\alpha$ with $\alpha=\bathh,\bathc$ is represented in the eigenbasis of $H_S^\alpha$, as in
\begin{align}
\tilde{\rho}_\mathrm{NELC}^\alpha=U_\alpha^\dag \rho_\mathrm{NELC}^\alpha U_\alpha,
\end{align}
where $\rho_\mathrm{NELC}^\alpha$ denotes the state in the original basis of the system Hamiltonian in Eq.~\eqref{HS}.
Therefore, to represent the time evolution from $\tilde{\rho}_\mathrm{NELC}^\bathc$ to $\tilde{\rho}_\mathrm{NELC}^\bathh$, we first go back from the eigenbasis of $H_S^\bathc$ to the original basis, and then further go to the eigenbasis of $H_S^\bathh$, before finally applying the time evolution generated by $\tilde{\mathcal{L}}^\bathh$.
The same goes to the time evolution from $\tilde{\rho}_\mathrm{NELC}^\bathh$ back to $\tilde{\rho}_\mathrm{NELC}^\bathc$.
We thus have
\begin{align}
\label{globalsteadycycleH}
\tilde{\rho}_\mathrm{NELC}^\bathh &= e^{\tilde{\mathcal{L}}^\bathh t_\bathh} (U^\dagger_\bathh U_\bathc \tilde{\rho}_\mathrm{NELC}^\bathc U_\bathc^\dagger U_\bathh),\\
\label{globalsteadycycleC}
\tilde{\rho}_\mathrm{NELC}^\bathc &= e^{\tilde{\mathcal{L}}^\bathc t_\bathc} (U^\dagger_\bathc U_\bathh \tilde{\rho}_\mathrm{NELC}^\bathh U_\bathh^\dagger U_\bathc),
\end{align}
and hence the operator for one non-equilibrium cycle is given by
\begin{align}
\label{onecycleoperator}
\mathcal{M}(\tilde{\rho})=e^{\tilde{\mathcal{L}}^\bathc t_\bathc} (U^\dagger_\bathc U_\bathh e^{\tilde{\mathcal{L}}^\bathh t_\bathh} (U^\dagger_\bathh U_\bathc \tilde{\rho} U_\bathc^\dagger U_\bathh) U_\bathh^\dagger U_\bathc).
\end{align}

We now analyze Eqs~\eqref{globalsteadycycleH}--\eqref{globalsteadycycleC} by expanding each non-equilibrium state $\tilde{\rho}_\mathrm{NELC}^\alpha$ in the eigenbasis of the superoperator $\tilde{\mathcal{L}}^\alpha$, as in
\begin{align}
\tilde{\rho}^\alpha_\mathrm{NELC} = \sum_{i=0}^3 \tilde{c}_i^\alpha \tilde{\rho}_i^\alpha,
\end{align}
with the expansion coefficients given by
\begin{align}
\tilde{c}_i^\alpha=\Tr[\tilde{\sigma}_i^\alpha \tilde{\rho}^\alpha_\mathrm{NELC}],
\end{align}
where $\tilde{\sigma}_i^\alpha$ and $\tilde{\rho}_i^\alpha$ are, as in Eqs.~(\ref{globalL0})--(\ref{globalR3}), the left and right eigenmatrices, respectively, of the global Liouvillian superoperator $\mathcal{L}^\alpha$, corresponding to the $i$th eigenvalue $\tilde{\lambda}_i^\alpha$ in Eqs.~(\ref{globallambda0})--(\ref{globallambda3}).
%
%
%
We apply this expansion to each of Eqs.~\eqref{globalsteadycycleH}--\eqref{globalsteadycycleC}, obtaining
\begin{align}
\label{globalsteadycycleH1}
\tilde{\rho}_\mathrm{NELC}^\bathh &= \sum_{i=0}^3 \tilde{c}_i^\bathh \tilde{\rho}_i^\bathh e^{\tilde{\lambda}_i^\bathh t_\bathh},\\
\label{globalsteadycycleC1}
\tilde{\rho}_\mathrm{NELC}^\bathc &= \sum_{j=0}^3 \tilde{c}_j^\bathc \tilde{\rho}_j^\bathc e^{\tilde{\lambda}_j^\bathc t_\bathc},
\end{align}
where
\begin{align}
\label{globalsteadycycleH2}
\tilde{c}_i^\bathh &= \mathrm{Tr}[\tilde{\sigma}_i^\bathh U^\dagger_\bathh U_\bathc \tilde{\rho}_\mathrm{NELC}^\bathc U_\bathc^\dagger U_\bathh] ,\\
\label{globalsteadycycleC2}
\tilde{c}_j^\bathc &= \mathrm{Tr}[\tilde{\sigma}_j^\bathc U^\dagger_\bathc U_\bathh \tilde{\rho}_\mathrm{NELC}^\bathh U_\bathh^\dagger U_\bathc].
 \end{align}

Inserting Eq.~\eqref{globalsteadycycleC1} into Eq.~\eqref{globalsteadycycleH2} as well as Eq.~\eqref{globalsteadycycleH1} into Eq.~\eqref{globalsteadycycleC2}, we find
\begin{align}
\label{globalsteadycycleH3}
\tilde{c}_i^\bathh &= \sum_{k=0}^3 d_{ik}  \tilde{c}_k^\bathc e^{\tilde{\lambda}_k^\bathc t_\bathc},\\
\label{globalsteadycycleC3}
\tilde{c}_j^\bathc &= \sum_{\ell=0}^3 d_{j\ell}  \tilde{c}_\ell^\bathh   e^{\tilde{\lambda}_\ell^\bathh t_\bathh},
\end{align}
where
\begin{align}
\label{globalsteadycycleH4}
d_{ik}=\mathrm{Tr}[\tilde{\sigma}_i^\bathh U^\dagger_\bathh U_\bathc \tilde{\rho}_k^\bathc U_\bathc^\dagger U_\bathh] ,\\
\label{globalsteadycycleC4}
d_{j\ell}= \mathrm{Tr}[\tilde{\sigma}_j^\bathc U^\dagger_\bathc U_\bathh \tilde{\rho}_\ell^\bathh U_\bathh^\dagger U_\bathc].
 \end{align}
Finally inserting Eq.~\eqref{globalsteadycycleH3} into Eq.~\eqref{globalsteadycycleC3} to make it represent one whole cycle, we arrive at
\begin{align}
\label{globalsteadycycleC5}
\tilde{c}_j^\bathc=\sum_{k=0}^3 M_{jk} \tilde{c}_k^\bathc,
\end{align}
where
\begin{align}
\label{globalsteadycycleM}
M_{jk}=\sum_{\ell=0}^3 d_{j\ell}d_{\ell k}e^{\tilde{\lambda}_\ell^\bathh t_\bathh+\tilde{\lambda}_k^\bathc t_\bathc}.
\end{align}
%
Introducing the coefficient vector 
\begin{align}
\vec{\tilde{c}}^\alpha = \begin{pmatrix} \tilde{c}_0^\alpha & \tilde{c}_1^\alpha & \tilde{c}_2^\alpha & \tilde{c}_3^\alpha \end{pmatrix}^T,
\end{align}
we can express Eq.~\eqref{globalsteadycycleC5} in the matrix form
\begin{align}
\label{vectildec1}
(\mathbb{I}_4-M)\vec{\tilde{c}}^\bathc = 0,
\end{align}
where $\mathbb{I}_4$ is the $4\times4$ identity matrix and the matrix $M$ is defined by the matrix elements given in Eq.~\eqref{globalsteadycycleM}.
This is the eigenvalue equation for the matrix $M$ with the unity eigenvalue.
As we mentioned above, we prove in Appendix~\ref{A2} that this is the only unity eigenvalue and the absolute values of the other eigenvalues are less than unity.
Thus, the eigenvector $\vec{\tilde{c}}^\bathc$ corresponding to the unity eigenvalue survives as the non-equilibrium state for the cold isochoric process in the NELC, as in Eq.~\eqref{globalsteadycycleC1}.
We can also find the corresponding eigenvector $\vec{\tilde{c}}^\bathc$ by intersing $\vec{\tilde{c}}^\bathc$ into Eq.~\eqref{globalsteadycycleH3};
it represents the non-equilibrium state for the hot isochoric process in the NELC, as in Eq.~\eqref{globalsteadycycleH1}.
%
%
We can thus determine the non-equilibrium states in the NELC once we fix the parameters $\{\beta_\alpha, \omega_\alpha, g_\alpha\}$ ($\alpha=\bathh,\bathc$), making the NELC depend only on the interaction times $t_\bathh$ and $t_\bathc$.

It is important to emphasize that the NELC does not correspond to a time-independent stationary state. 
Instead, it represents a periodic steady-state trajectory within each cycle, as shown in Fig.~\ref{Diagrams}\subref{NELCDiagram}. 
In general, these two states are distinct:
\begin{equation}
\tilde{\rho}_{\mathrm{NELC}}^\bathh \neq \tilde{\rho}_{\mathrm{NELC}}^\bathc.
\end{equation}
The system thus converges to a unique periodic orbit rather than a single static state, enabling sustained energy exchange with the hot and cold reservoirs.

In Eqs.~(\ref{Qh})--(\ref{W}), the total energy flows during the isochoric processes are given by the system Hamiltonian and the states in the original basis. For the NELC, we can rewrite them as the functions of the interaction times $t_\bathh$ and $t_\bathc$, given by
\begin{align}
\label{QhNELC}
{Q_\bathh}(t_\bathh,t_\bathc) &= \Tr[H_S^\bathh (\rho_\text{NELC}^\bathh - \rho_\text{NELC}^\bathc)] = \Tr[\tilde{H}_S^\bathh (\tilde{\rho}_\text{NELC}^\bathh - \tilde{\rho}_\text{NELC}^\bathc)],\\
\label{QcNELC}	
{Q_\bathc}(t_\bathh,t_\bathc) &= \Tr[H_S^\bathc (\rho_\text{NELC}^\bathc - \rho_\text{NELC}^\bathh)] = \Tr[\tilde{H}_S^\bathc (\tilde{\rho}_\text{NELC}^\bathc - \tilde{\rho}_\text{NELC}^\bathh)].
\end{align}
In each work-production process, the total work extracted from the environment is given by
\begin{align}
\label{W1NELC}
{W_1}(t_\bathh,t_\bathc)&= \Tr[(H_S^\bathc-H_S^\bathh)\rho_\text{NELC}^\bathh], \\ 
\label{W2NELC}
{W_2}(t_\bathh,t_\bathc)&= \Tr[(H_S^\bathh-H_S^\bathc)\rho_\text{NELC}^\bathc],
\end{align}
and the total work is given by
\begin{align}
\label{WNELC}
W(t_\bathh,t_\bathc) ={W_1}(t_\bathh,t_\bathc) + {W_2}(t_\bathh,t_\bathc) =  -\left [ {Q_\bathh}(t_\bathh,t_\bathc)+{Q_\bathc}(t_\bathh,t_\bathc)\right ],
\end{align}
which still obeys the first law of thermodynamics. Note that the relation between the system state in the original basis, $\rho_{\rm NELC}^\alpha$, and the one in the basis that diagonalizes the Liouvillian, $\tilde{\rho}_{\rm NELC}^\alpha$, is given by
\begin{align}
\rho_\text{NELC}^\alpha = U_\alpha \tilde{\rho}_\text{NELC}^\alpha U_\alpha^\dagger.
\end{align}

As discussed in Sec.~\ref{Sec2-1}, the internally coupled system can operate as four thermal modes with three boundaries. We focus on the phases of the engine and the refrigerator. For the engine, not only the efficiency but also the power is significant~\cite{PhysRevResearch.6.023172}.
\begin{align}
\label{efficiencyNELC}
\eta_\text{NELC}(t_\bathh,t_\bathc) = -\frac{W(t_\bathh,t_\bathc)}{{Q_\bathh}(t_\bathh,t_\bathc)},\\
\label{powerNELC}
P_\text{NELC}(t_\bathh,t_\bathc) = -\frac{W(t_\bathh,t_\bathc)}{t_\bathh+t_\bathc}.
\end{align}
The performance of the refrigerator is still given by the coefficient of performance (COP)
\begin{align}
\label{COPNELC}
\xi_\text{NELC}(t_\bathh,t_\bathc) = \frac{{Q_\bathc}(t_\bathh,t_\bathc)}{W(t_\bathh,t_\bathc)}.
\end{align}

For a symmetric non-equilibrating limit cycle (NELC), we assume that the interaction time in each isochoric process is equal $t_\bathh = t_\bathc = \tau$. Hence, after fixing other parameters $\{g_\alpha, \omega_\alpha, \beta_\alpha\}$, both the internally coupled and the dressed-spectrum NELC only depend on the interaction time $\tau$.

\subsection{Power-Efficiency Trade-Off Law}
\label{Sec3-2}

Similarly to the discussion of the thermalized cycles in Sec.~\ref{Sec2}, we also choose the zero-energy boundary of the standard Otto cycle between the phases of the engine and the refrigerator, as in Eq.~\eqref{standardbound1}. The operating regimes of the standard Otto cycle depend only on the bare energy gap $\omega_\alpha$ and the bath temperature $\beta_\alpha$, which are independent of the system state evolution. Consequently, at the zero-energy boundary, there is no heat exchange in the standard NELC, and it cannot operate in any thermal mode. However, as previously discussed in Sec.~\ref{Sec2-1}, both the dressed-spectrum and the internally coupled Otto cycle can operate in diverse thermal modes due to the Hamiltonian spectrum update introduced by the internal coupling. 

\begin{figure*}[h!]
	\centering
	\subfloat[]{\includegraphics[width=0.47\textwidth]{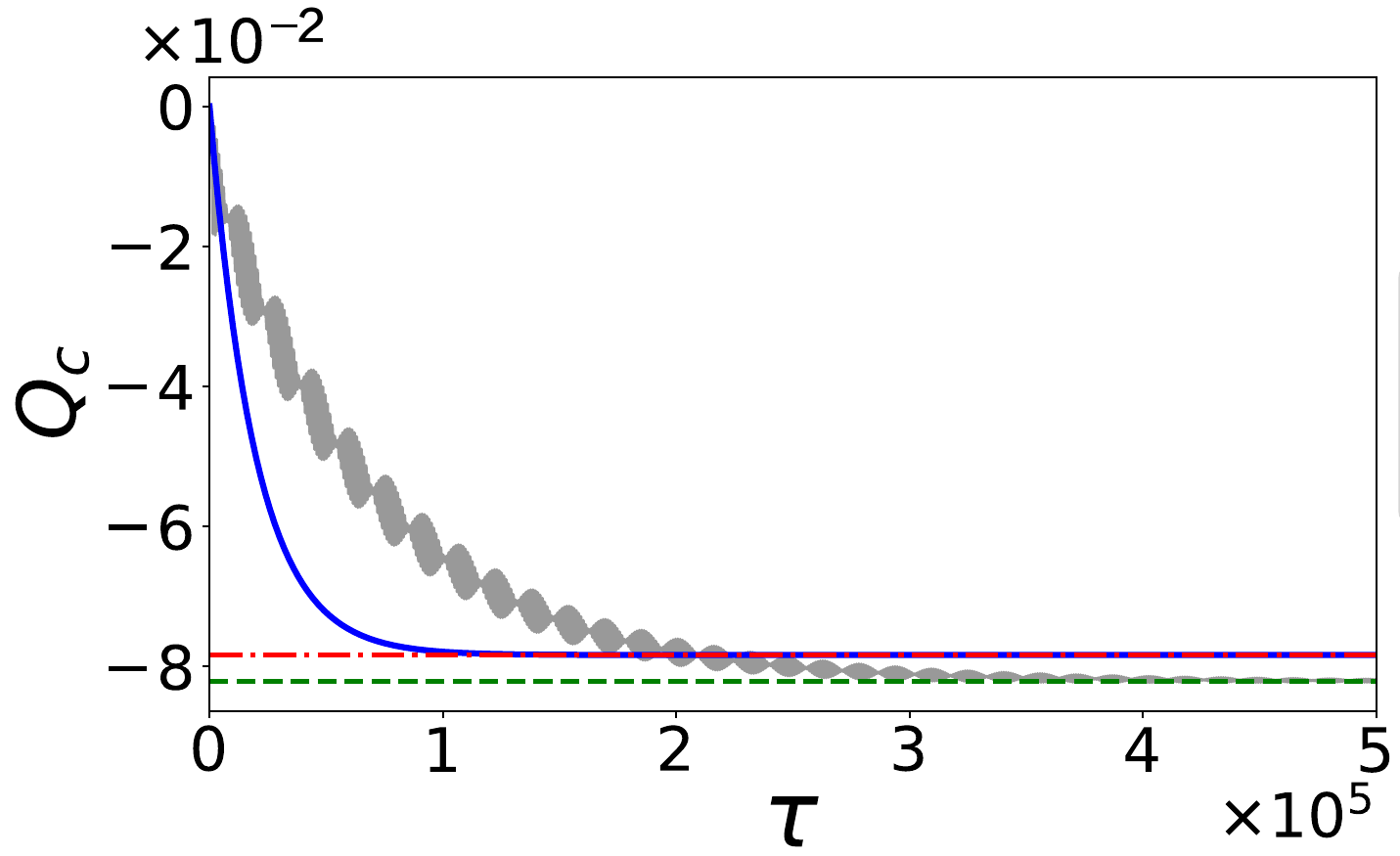}\label{Qc_t}}
	\subfloat[]{\includegraphics[width=0.47\textwidth]{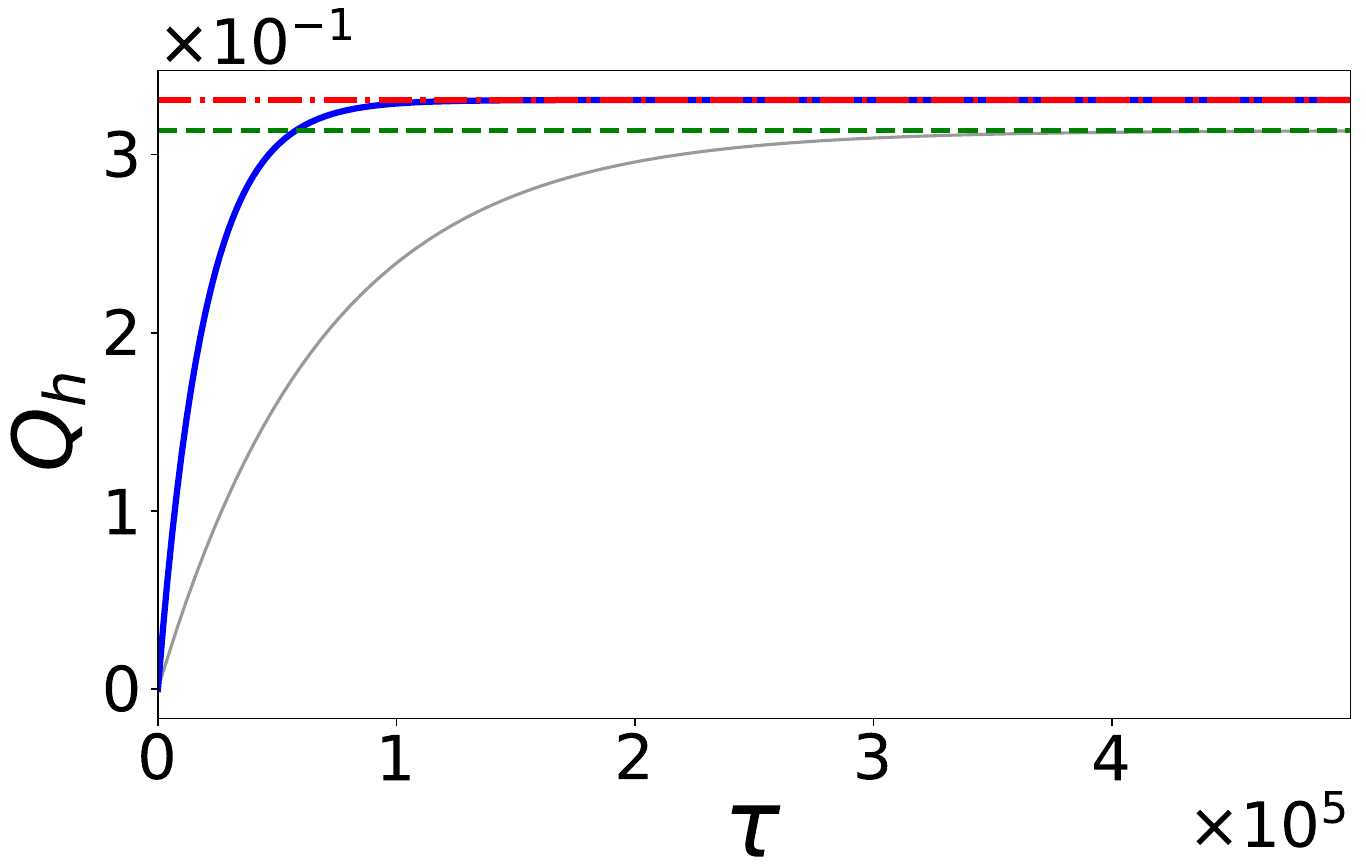}\label{Qh_t}} \par
	\subfloat[]{\includegraphics[width=0.47\textwidth]{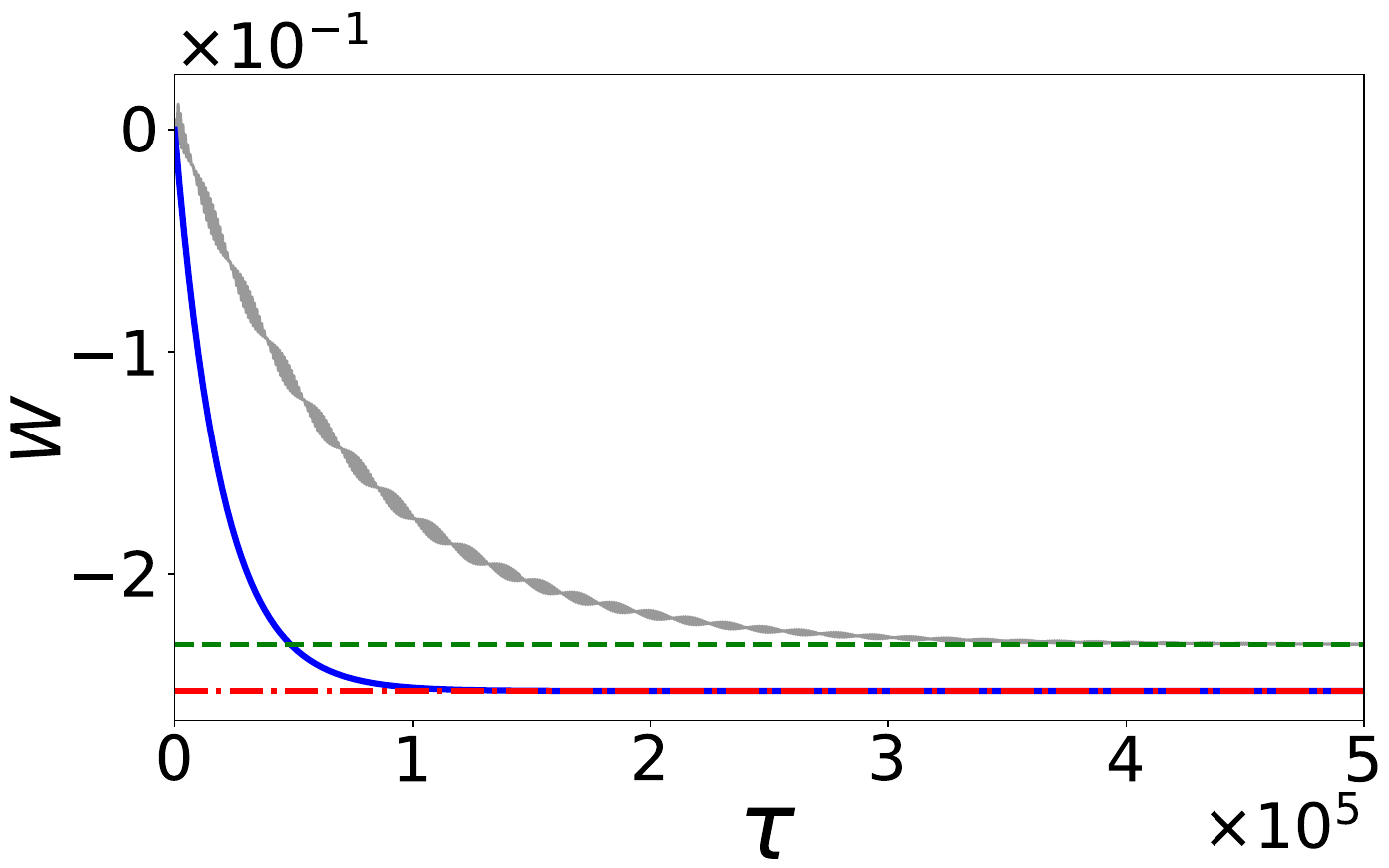}\label{W_t}} 
	\subfloat{\makebox[0.47\textwidth][c]{\raisebox{1.2cm}{\includegraphics[scale=0.5]{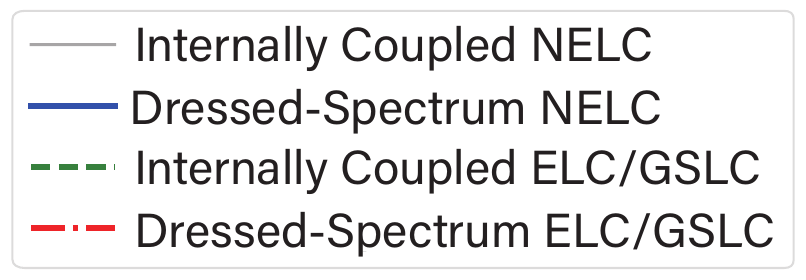}}}}
	\caption{Time dependence of the energy flows (a)~${Q_\bathh^\mathrm{cp}}$, (b)~${Q_\bathc^\mathrm{cp}}$, and (c)~$W$. The gray lines indicate the non-equilibrating limit cycle (NELC) of the internally coupled system, while the blue lines indicate those of the dressed-spectrum system, which neglects the eigenbasis mismatch. After long-time evolution, the NELCs approach the equilibrium limit cycle (ELC). The green dashed lines indicate the ELC of the internally coupled system, which is the same as the Gibbs-state limit cycle (GSLC), while the red dashed lines indicate the ELC of the dressed-spectrum system. We set the parameter values to $\omega_{\bathc}=1$, $\omega_{\bathh} = 5$, $\beta_{\bathc} = 1$, $\beta_{\bathh} = 0.2$, $g_{\bathc} = 1$, $g_{\bathh} = 4$.}
	\label{QW_t}
\end{figure*}

As shown in Fig.~\ref{QW_t}, there is energy exchange between the internal system and the environment in each stroke for both dressed and internally coupled NELC, confirming that the internal coupling extends the operation regimes of the Otto cycle. 
As the interaction time increases, the NELC approaches the ELC, which should show the same thermal performance as a Gibbs-state limit cycle (GSLC) in Sec.~\ref{Sec2}, validating the global GKSL master equation, as discussed in Appendix~\ref{A1}.

However, the internal coupling not only updates the Hamiltonian spectrum but also introduces an eigenbasis mismatch due to the non-commutativity of the system Hamiltonian. 
To rigorously demonstrate the role of the non-commutativity, we compare the internally coupled NELC with the dressed-spectrum one. 
As detailed in Appendix~\ref{A3}, the derivation of the dressed-spectrum NELC can be simplified as $U_\alpha = \mathbb{I}$ with the transformation $U_\alpha$ defined in Eq.~\eqref{Ualpha}. In the absence of the basis mismatch, the dressed system acts as an effective two-level system with an energy gap $\tilde{\omega}_\alpha$, and its energy eigenbasis $\{\ketg, \kete\}$ remains invariant across both the hot and cold baths ($\alpha = \bathh, \bathc$). 
The difference between the dressed-spectrum NELC and the internally coupled one reveals the role of the eigenbasis mismatch arising from non-commutativity. As shown in Fig.~\ref{QW_t}, the energy exchanges between these two NELCs are always different, which should lead to different thermal performances. 

\begin{figure*}[h!]
	\centering
	\subfloat[efficiency]{\includegraphics[width=0.47\textwidth]{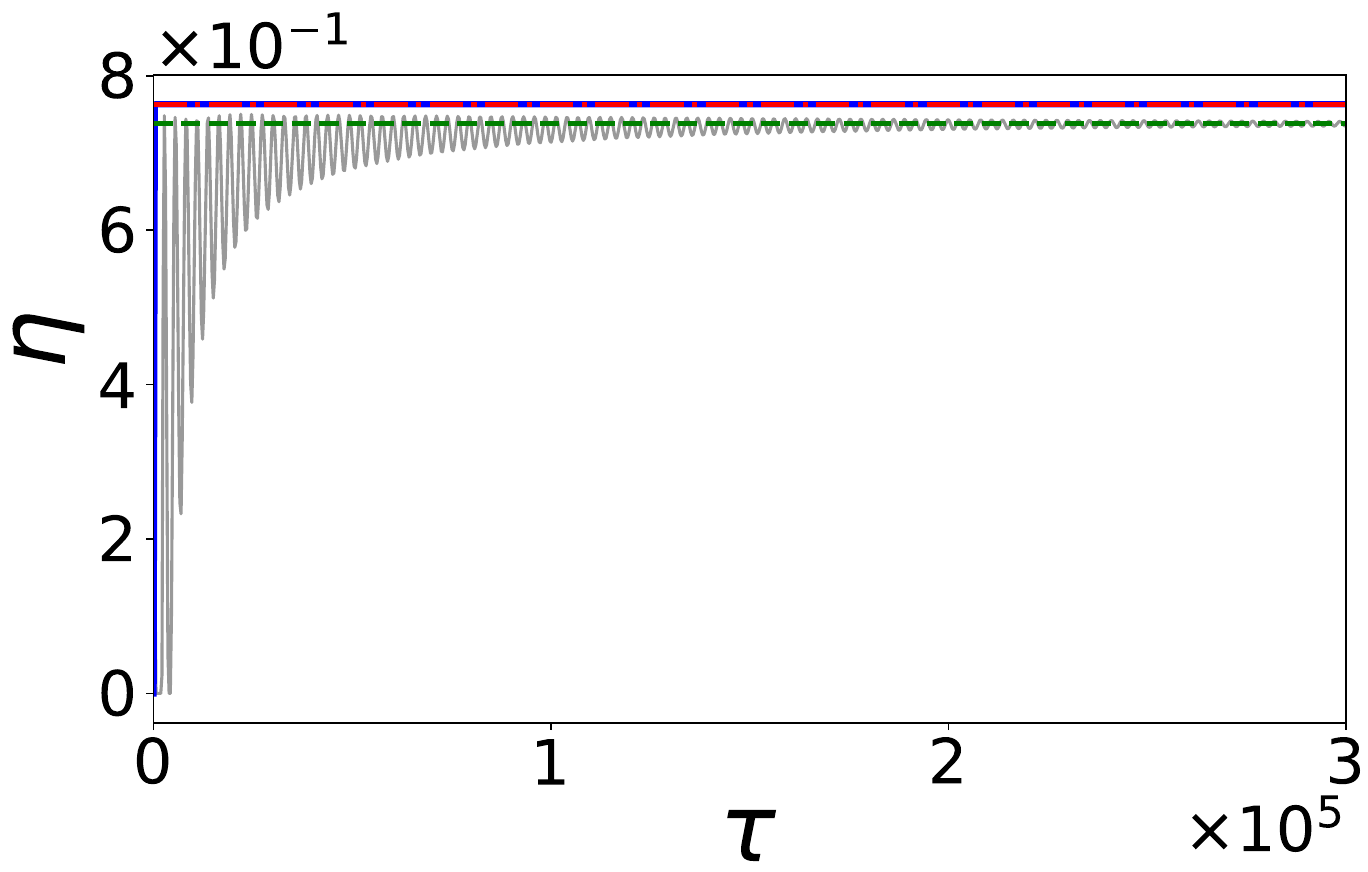}\label{efficiency_t}} 
	\subfloat[COP]{\includegraphics[width=0.5\textwidth]{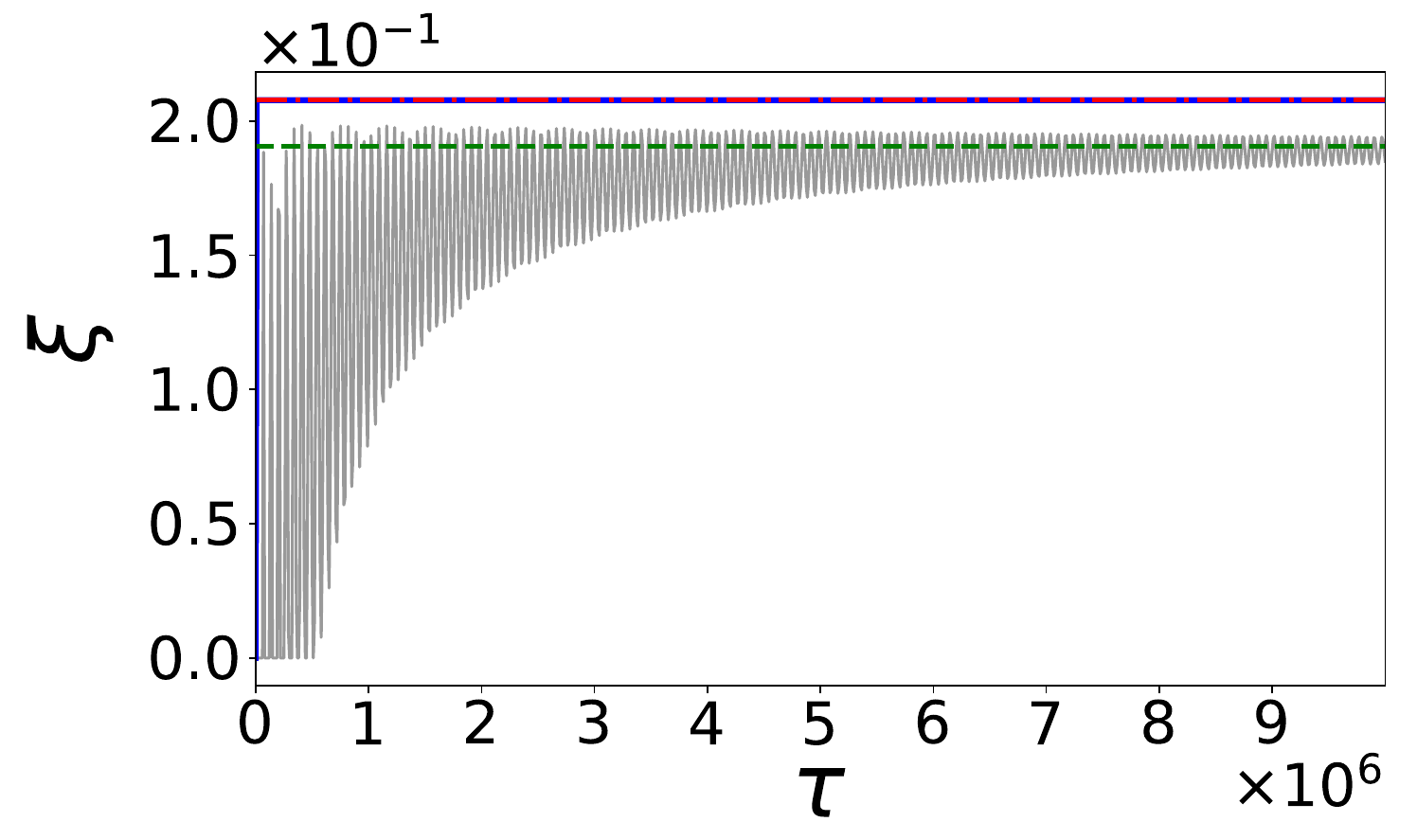}\label{COP_t}}
	
	\subfloat{\makebox[0.47\textwidth][c]{\raisebox{1.2cm}{\includegraphics[scale=0.5]{QWlegend.pdf}}}}
	\caption{Time dependence of (a)~the efficiency $\eta$ and (b)~the coefficient of performance (COP) $xi$. The gray lines indicate the non-equilibrating limit cycle (NELC) of the internally coupled system, while the blue lines indicate those of the dressed-spectrum system, which neglects the eigenbasis mismatch. After long-time evolution, the NELCs approach the equilibrium limit cycle (ELC). The green dashed lines indicate the ELC of the internally coupled system, which is the same as the Gibbs-state limit cycle (GSLC), while the red dashed lines indicate the ELC of the dressed-spectrum system, which is the dressed Otto bounds in Sec.~\ref{Sec1-2}. We set the parameter values to $\omega_{\bathc}=1$, $\omega_{\bathh} = 5$, $\beta_{\bathc} = 1$, $\beta_{\bathh} = 0.2$, $g_{\bathc} = 1$. The systems run as an engine with $g_{\bathh} = 4$ in subfig.~(a), and as a refrigerator with $g_{\bathh} = 6$ in subfig.~(b).}
	\label{performance_t}
\end{figure*}

Different energy exchange coefficients affect the thermal performance of the dressed spectrum and the internally coupled NELCs. 
As shown in Fig.~\ref {performance_t} for the dressed-spectrum NELC (blue line) and the ELC (red dashed line), we find that finite-time operation is not the only reason for the power-efficiency trade-off, and that non-commutativity plays a significant role. 

In classical thermodynamics, finite-time operation universally induces a power-efficiency trade-off relation due to inherent macroscopic mechanical or fluidic friction~\cite{PhysRevLett.117.190601, Koyuk_2019, PhysRevLett.120.190602}, so that the efficiency is maximized under thermalized conditions with extremely long interaction times, whereas power increases in the short term but decreases over longer timescales. 
A few previous studies on quantum thermal machines have also identified the power-efficiency trade-off law but have concluded that it is due to finite-time operations or the coherence itself~\cite{PhysRevE.106.024137, PhysRevLett.127.190604}. 

However, the comparison between the dressed spectrum and internally coupled NELCs reveals a fundamentally different paradigm in the quantum regime: finite-time operation is not a sufficient condition for the trade-off relation.
Instead, the trade-off strictly requires the presence of internal quantum friction, which mathematically originates from the eigenbasis mismatch given by the non-commutativity $[H_S^\bathh, H_S^\bathc] \neq 0$. 

We can analytically confirm that the power-efficiency trade-off law disappears with finite interaction time in the dressed-spectrum NELC. 
For the dressed-spectrum NELC, as detailed in Appendix~\ref{A3}, we can rewrite the density matrix $\rho_{\rm ds}^\alpha$~\eqref{dressedNELCstatefinal} as a function of $\tau$, which strictly remains in the diagonal population subspace of the effective Hamiltonian $\tilde{H}_S^\alpha$ in Eq.~\eqref{tildeHS}, given by
\begin{equation}
\label{dressed_population_state}
\rho_{\rm ds}^\alpha(\tau) = \Pg^\alpha(\tau) \ketg\brag  + \Pe^\alpha(\tau) \kete\brae,
\end{equation}
with the probability normalization condition $\Pg^\alpha(\tau) + \Pe^\alpha(\tau) = 1$. 
Regardless of the finite interaction time $\tau$ and the complexity of the non-equilibrium coefficient $c_3^\alpha(\tau)$ derived in Eqs.~\eqref{dressedc3hfinal} and \eqref{dressedc3cfinal}, the system state can be entirely parameterized by its populations.

During the isochoric processes, the heat exchanged with the hot and cold baths can be evaluated directly from the change in the system's internal energy. For the hot bath, the absorbed heat is
\begin{align}
{Q_\bathh^{\rm ds}}(\tau) &= \mathrm{Tr}\left[ \left(\rho_{\rm ds}^\bathh(\tau) - \rho_{\rm ds}^\bathc(t)\right) \tilde{H}^\bathh \right] \nonumber \\
&= \left[ \Pg^\bathh(t) - \Pg^\bathc(\tau) \right]\epsilon_\bathh^- + \left[ \Pe^\bathh(\tau) - \Pe^\bathc(t) \right]\epsilon_\bathh^+.
\end{align}
If we define the variation in the excited state population between the two isochoric strokes as $\Delta p(\tau) = \Pe^\bathh(\tau) - \Pe^\bathc(\tau)$, the variation in the ground state population strictly follows $\Pg^\bathh(\tau) - \Pg^\bathc(\tau) = -\Delta p(\tau)$. Substituting these relations into the heat transfer equation yields
\begin{align}
\label{Qh_dressed_proof}
{Q_\bathh^{\rm ds}}(\tau) &= -\Delta p(\tau) \epsilon_\bathh^- + \Delta p(\tau) \epsilon_\bathh^+ \nonumber \\
&= \Delta p(\tau) (\epsilon_\bathh^+ - \epsilon_\bathh^-) \nonumber \\
&= \Delta p(\tau) \tilde{\omega}_\bathh,
\end{align}
where the effective energy levels $\epsilon_\alpha^\pm$ are defined in Eq.~\eqref{dressedenergy}, and the effective energy gap $\tilde{\omega}_\alpha$ is defined in Eq.~\eqref{effectiveenergygap}. 
Similarly, the heat released to the cold bath is given by
\begin{align}
\label{Qc_dressed_proof}
{Q_\bathc^{\rm ds}}(\tau) &= \mathrm{Tr}\left[ \left(\rho_{\rm ds}^\bathc(\tau) - \rho_{\rm ds}^\bathh(\tau)\right) \tilde{H}^\bathc \right] \nonumber \\
&= -\Delta p(\tau) \tilde{\omega}_\bathc.
\end{align}
By taking the ratio of the heat exchanged with the two baths, we observe a remarkable cancellation:
\begin{equation}
\label{Qratio_proof}
\frac{{Q_\bathc^{\rm ds}}(\tau)}{{Q_\bathh^{\rm ds}}(\tau)} = \frac{-\Delta p(\tau) \tilde{\omega}_\bathc}{\Delta p(\tau) \tilde{\omega}_\bathh} = -\frac{\tilde{\omega}_\bathc}{\tilde{\omega}_\bathh}.
\end{equation}
All the complex time-dependent dynamics, including the interaction time $t$, the decay rates $\lambda_3^\alpha$, and the transient non-equilibrium components stored in $c_3^\alpha$, are entirely absorbed into the collective scalar factor $\Delta p(\tau)$ and cancel out perfectly. 

Therefore, the thermodynamic efficiency of the dressed-spectrum NELC is strictly time-independent and is universally locked to the dressed-spectrum GSLC and ELC, given by
\begin{equation}
\label{eta_dressed_proof}
\eta_{\rm ds}(\tau) = 1 + \frac{{Q_\bathc^{\rm ds}}(\tau)}{{Q_\bathh^{\rm ds}}(\tau)} = 1 - \frac{\tilde{\omega}_\bathc}{\tilde{\omega}_\bathh} = \eta_{\rm ds}.
\end{equation}
Similarly, when the dressed-spectrum NELC operates as a refrigerator, the coefficient of performance (COP) exhibits an identical time-independent characteristic.
In the refrigeration cycle, the heat extracted from the cold bath ${Q_\bathc^{\rm ds}}(\tau)$ and the heat released to the hot bath ${Q_\bathh^{\rm ds}}(\tau)$ maintain the same algebraic forms as derived in Eqs.~\eqref{Qh_dressed_proof} and \eqref{Qc_dressed_proof}. However, to drive the heat transfer from the cold bath to the hot bath against the temperature gradient, the internal system must extract work from the external environment.

According to the first law of thermodynamics, the total work exchanged per cycle is
\begin{align}
\label{W_dressed_proof}
W^{\rm ds}(\tau) &= -({Q_\bathh^{\rm ds}}(\tau) + {Q_\bathc^{\rm ds}}(\tau)) \nonumber \\
&= -\Delta P(\tau) \tilde{\omega}_\bathh + \Delta P(\tau) \tilde{\omega}_\bathc \nonumber \\
&= \Delta P(\tau) (\tilde{\omega}_\bathc - \tilde{\omega}_\bathh).
\end{align}
The coefficient of performance (COP) $\xi(\tau)$ for the refrigerator is rewritten as
\begin{equation}
\label{COP_def}
\xi_{\rm ds}(\tau) = \frac{{Q_\bathc^{\rm ds}}(\tau)}{W^{\rm ds}(\tau)}.
\end{equation}
Substituting the expressions for ${Q_\bathc^{\rm ds}}(\tau)$ in Eq.~\eqref{Qc_dressed_proof} and $W^{\rm ds}(\tau)$ in Eq.~\eqref{W_dressed_proof} into this definition, we obtain:
\begin{align}
\label{COP_dressed_proof}
\xi_{\rm ds}(\tau) &= \frac{-\Delta P(\tau) \tilde{\omega}_\bathc}{-\Delta P(\tau) (\tilde{\omega}_\bathh - \tilde{\omega}_\bathc)} \nonumber \\
&= \frac{\tilde{\omega}_\bathc}{\tilde{\omega}_\bathh - \tilde{\omega}_\bathc} \nonumber \\
&= \xi_{\rm ds}.
\end{align}
Once again, we observe a remarkable cancellation. The collective scalar factor $\Delta p(\tau)$, which encapsulates all the complex transient non-equilibrium dynamics and the finite interaction time $\tau$, cancels out perfectly in the numerator and denominator. 
Similarly, in the standard system without internal coupling, there is no trade-off relation between power and efficiency, even with finite interaction time, showing the difference between classical and quantum thermodynamics. 

Then, we analyze that the eigenbasis mismatch caused by the non-commutativity $[H_S^\bathh, H_S^\bathc] \neq 0$ works as the internal friction and is essential for the power-efficiency trade-off law in a quantum thermal machine. 
Here, we rewrite the basis mismatch of the internally coupled NELC as a matrix 
\begin{align}
\label{mismatchU}
U = U_\bathh^\dagger U_\bathc = \begin{bmatrix}
\cos(\Theta) & -\sin(\Theta) \\
\sin(\Theta) & \cos(\Theta)
\end{bmatrix},
\end{align}
using the angle difference 
\begin{align}
\label{mismatchTheta}
\Theta = \theta_\bathh-\theta_\bathc.
\end{align}
The angle $\theta_\alpha$ in Eq.~\eqref{theta} shows the basis mismatch between the effective Hamiltonian $\tilde{H}_S^\alpha$ in Eq.~\eqref{tildeHS} and the original Hamiltonian $H_S^\alpha$ in Eq.~\eqref{HS}. The matrix $U$ in Eq.~\eqref{mismatchU} acts as a coherent rotation at the boundary of each isochoric stroke. 

We can easily find that the coefficient of non-commutativity here, given by the angle difference in Eq.~\ref{mismatchTheta} in different strokes, is directly related to the parameter $\Lambda$, which is defined as the cosine function of the angle difference in Eq.~\ref{Lambdacos}, in the discussion of the Gibbs-state limit cycle (GSLC) in Sec.~\ref{Sec2}. We can also rewrite the basis mismatch matrix $U$ using $\Lambda$:
\begin{align}
&\cos(\Theta) = \pm \sqrt{\frac{1+\Lambda}{2}},
&\sin(\Theta) = \pm \sqrt{\frac{1-\Lambda}{2}}.
\end{align}
When $\Lambda < 1$, corresponding to $\Theta \neq 0$, the rotation $U=\mathbb{I}$ geometrically projects a fraction of the useful population difference into off-diagonal coherences, given by non-zero coefficients $\tilde{c}_1^\alpha, \tilde{c}_2^\alpha$. During the finite-time isochoric evolution, these induced coherences exponentially decay due to the bath-induced decoherence governed by the negative real parts of $\tilde{\lambda}_{1,2}^\alpha$. This irreversible decay permanently destroys the amplitude that was rotated into the coherence subspace, effectively leaking useful populations into waste heat. This coherence-induced dissipation acts as the genuine quantum friction, which inherently lowers the efficiency at short cycle times and yields the power-efficiency trade-off law observed in the internally coupled system.

To further isolate the physical impact of the eigenbasis mismatch from the internal coupling itself, we can consider a specific configuration where the internal coupling is maintained $g_\alpha \neq 0$ but tuned such that $\theta_\bathh = \theta_\bathc$. In this scenario, the eigenbasis mismatch vanishes $U = \mathbb{I}$, and no coherence is generated during the unitary strokes. Without this internal friction, similarly to the dressed-spectrum NELC, we can easily confirm that the efficiency of the internally coupled NELC becomes time-independent and coincides exactly with the optimal efficiency of the dressed-spectrum system. The power-efficiency trade-off law disappears here despite finite-time operation.

Additionally, as shown in Fig.~\ref{performance_t}, we find that the efficiency and the COP of the internally coupled NELC cannot be better than the dressed-spectrum one, since the non-commutativity leads to quantum internal friction. 
The efficiency and COP of the internally coupled NELC increase from zero to the equilibrium values observed in the internally coupled ELC. Hence, in the internally coupled system, the efficiency and the COP of the ELC, which are equal to those of the GSLC, are the upper bounds for the NELC. 
As discussed in Sec.~\ref{Sec2-2}, the efficiency and the COP of the internally coupled GSLC are lower than those of the dressed-spectrum GSLC due to the eigenbasis mismatch caused by the incommutable Hamiltonians. 
The efficiency and the COP of the dressed-spectrum NELC are strictly time-independent and are universally locked to the equilibrium values for any interaction time $\tau$. Thus, the efficiency and the COP of the internally coupled NELC are strictly bounded below the dressed-spectrum case for any finite interaction time. 

\begin{figure*}[h!]
	\centering
	\includegraphics[width=0.47\textwidth]{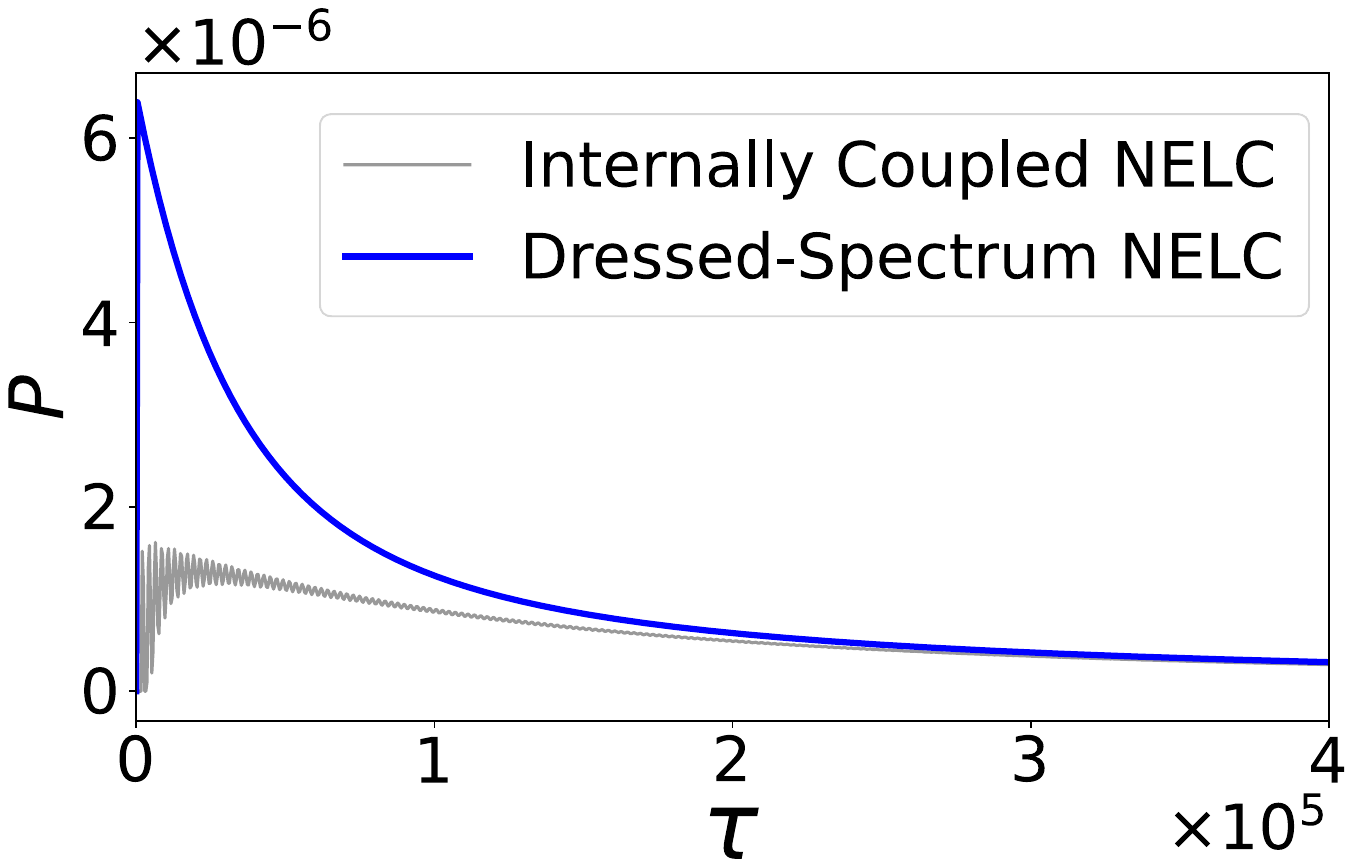}
	\caption{Time dependence of the power $P$. The gray lines indicate the non-equilibrating limit cycle (NELC) of the internally coupled system, while the blue lines indicate those of the dressed-spectrum system, which neglects the eigenbasis mismatch. After long-time evolution, the NELCs approach the equilibrium limit cycle (ELC), and the power approaches zero in both cases. We set the parameter values to $\omega_{\bathc}=1$, $\omega_{\bathh} = 5$, $\beta_{\bathc} = 1$, $\beta_{\bathh} = 0.2$, $g_{\bathc} = 1$, $g_{\bathh} = 4$.}
	\label{power_t}
\end{figure*}

As shown in Fig.~\ref{power_t}, the power of the internally coupled NELC is also lower than that of the dressed-spectrum NELC, but the reason is because of the off-diagonal coherence itself rather than the eigenbasis mismatch. 
Even in the frictionless limit $\Lambda = 1$, the internally coupled system still exhibits a strictly lower power output compared to the dressed-spectrum one. This discrepancy originates from a secondary, purely kinetic effect of the internal coupling. As detailed in the derivation of the global GKSL master equation in Appendix~\ref{A1}, the system-bath interaction operators are projected onto the rotated internal energy eigenbasis. This geometric projection introduces an additional suppression factor of $\cos^2(2\theta_\alpha)$ into the effective transition rates $\tilde{\gamma}_\alpha^\pm$ as in Eq.~\eqref{tildegamma}. 
As a result, the internally coupled system equilibrates more slowly than the dressed-spectrum system. The slower accumulation rate of the extracted work $W$ directly translates to a reduced power output $P$, even when their thermodynamic efficiency $\eta$ is the same and independent of interaction time.

In summary, rather than merely claiming that an internally coupled system is more realistic because it phenomenologically mimics the classical trade-off law, we establish that internal coupling modifies the non-equilibrium dynamics of quantum thermal machines in different ways. On the one hand, the internal coupling updates the Hamiltonian, leading to broader thermal regimes and better thermal performance. On the other hand, internal coupling introduces off-diagonal coherence and non-commutativity of the Hamiltonian, leading to dissipative thermal modes and reducing the thermal performance. The off-diagonal coherence itself directly reduces the power in the finite-time NELC. It imposes a kinetic suppression on the thermalization rate, which rigorously bounds the finite-time power output. 
However, the efficiency and COP are reduced by the eigenbasis mismatch caused by the non-commutativity of the Hamiltonian, which can be termed internal friction. The eigenbasis mismatch also serves as the true microscopic origin of the power-efficiency trade-off in the finite-time NELC. 
By capturing both the mismatch and the kinetic suppression, the internally coupled NELC provides a comprehensive and physically consistent framework for realistic quantum thermodynamic cycles. 

\section{Conclusion}

In this work, we found the detailed roles of the internal coupling in quantum Otto thermal machines. First, the presence of internal coupling updates the Hamiltonian spectrum, leading to a broader operating regime and improved thermal performance compared to the standard Otto cycle. Second, the internal coupling introduces the eigenbasis transformation. On the one hand, the eigenbasis transformation requires the global approach of the GKSL master equation with effective dissipation rates, leading to lower power. On the other hand, when the eigenbasis mismatches across different strokes, the efficiency and the COP are reduced and become time-dependent, leading to the power-efficiency trade-off law. 

We analyzed three types of Otto cycles. The standard Otto cycle, given by the bare energy levels $\omega_\alpha$ without internal coupling, is well studied and describes the situation where the internal coupling is entirely neglected. The dressed-spectrum Otto cycle runs similarly to a standard Otto cycle without internal coupling based on the effective energy gaps $\tilde{\omega}_\alpha$, which only consider the internal coupling as an update of the Hamiltonian spectrum. 
This work focuses on the internally coupled Otto cycle, a realistic cycle that accounts for the influence of internal coupling.

By comparing these three systems from thermalized conditions in the Gibbs-state limit cycle (GSLC) to non-equilibrium conditions in the non-equilibrating limit cycle (NELC), we identified the detailed roles of the internal coupling. 

Under thermalized conditions in GSLCs, the systems are assumed to thermalize rapidly. The dual roles of the internal coupling are revealed. On the one hand, the internal coupling updates the Hamiltonian spectrum, enabling the internally coupled GSLC to surpass the standard GSLC, as given by relations~\eqref{higherefficiency} and \eqref{higherCOP}. On the other hand, non-commutativity results in the eigenbasis mismatch and passively influences the thermal machines. 
We found that the non-commutativity manifests as quantum internal friction and implies that the realistic internally coupled system lies below the dressed-spectrum GSLC. 
When we adjust the coupling strength to vanish the internal friction and make the Hamiltonians commutative, the three systems give equal efficiency and COP, which means the internal friction is a necessary cost when the internal coupling improves the thermal performance by updating the system Hamiltonian. 

In the NELC with finite-time interaction in the isochoric processes, we first validated the global approach of the GKSL master equation by demonstrating that the ELC with infinite interaction time perfectly reproduces the thermal performance of the GSLC. Moving to the finite-time regime, we derived the non-equilibrating limit cycle (NELC) for the internally coupled system and the dressed-spectrum systems, and compared the two. 
Our comparison overturns the conventional analogy that finite-time operation universally necessitates a power-efficiency trade-off in quantum thermal machines~\cite{Koyuk_2019, PhysRevLett.120.190602, PhysRevLett.117.190601}. 
Instead, we found that the eigenbasis mismatch caused by the non-commutativity of the Hamiltonians across different strokes is essential to the power-efficiency trade-off law, as it introduces dynamic quantum friction. 

Unlike the power, which is reduced by the eigenbasis transformation and the off-diagonal coherence itself, the efficiency and COP are reduced by the eigenbasis mismatch. In other words, even with the internal coupling, if the eigenbasis matches and the Hamiltonians are commutable, the time-independent optimal efficiency is restored, and the power-efficiency trade-off law is suppressed. 
However, even without the eigenbasis mismatch, the power output of the internally coupled system remains strictly lower than the dressed-spectrum bound, due to the extra coefficient $\cos^2(2\theta_\alpha)$ of the dissipative rate as in Eq.~\eqref{tildegamma} in the global GKSL master equation as in Eq.~\eqref{globalLiouvillian}. This exposes a purely kinetic suppression effect: the rotation of the internal energy eigenbasis misaligns with the local system-bath interaction, thereby fundamentally restricting the thermalization rate and the finite-time work accumulation.

Consequently, the internally coupled Otto cycle is not deemed more realistic simply because it phenomenologically mimics classical macroscopic friction. Rather, it is a rigorously valid framework and plays dual roles in the Otto thermal machines. On one hand, it broadens the operation regimes and improves the thermal performance of the thermal machines by updating the Hamiltonian. On the other hand, it captures both the dynamic coherence-induced dissipation and the suppression of kinetic thermalization. These results clearly clarify the multifaceted roles of internal coupling and establish physically plausible upper bounds on the performance of internally coupled Otto cycles across non-equilibrating to thermalized conditions.

\appendix

\section{Operation Difference Between GSLC and dressed-spectrum Boundaries}
\label{A4}

In Sec.~\ref{Sec2-1} of the main text, we present that the boundaries of different thermal modes in the internally coupled Gibbs-state limit cycle (GSLC) are different from those in the dressed-spectrum GSLC. In this section, we utilize the factor $\Lambda$ in Eq.~\ref{Lambda_definition}, which describes the mismatch of the eigenbasis across different isochoric processes, to analytically confirm it.

Firstly, we compare the engine boundary $W=0$ of the GSLC to the two boundaries as in Eqs.~\eqref{dressedbound1} and \eqref{dressedbound2} of the dressed-spectrum Otto cycle. 
Introducing the factor $\Lambda$ in Eq.~\ref{Lambda_definition}, we rewrite the work in Eq.~\eqref{WGibbs} of the internally coupled GSLC using
\begin{align}
g_\bathc(g_\bathc-g_\bathh)+\frac{\omega_\bathc(\omega_\bathc-\omega_\bathh)}{4}
=\Delta_\bathc^2-\Lambda\Delta_\bathc\Delta_\bathh
\end{align}
and 
\begin{align}
g_\bathh(g_\bathh-g_\bathc)+\frac{\omega_\bathh(\omega_\bathh-\omega_\bathc)}{4}
=\Delta_\bathh^2-\Lambda\Delta_\bathc\Delta_\bathh, 
\end{align}
and obtain
\begin{align}
W=&\frac{\left[4g_\bathc(g_\bathc-g_\bathh)+\omega_\bathc(\omega_\bathc-\omega_\bathh)\right]}{4\Delta_\bathc}\tanh(\beta_\bathc\Delta_\bathc)\nonumber  \\
&+\frac{\left[4g_\bathh(g_\bathh-g_\bathc)+\omega_\bathh(\omega_\bathh-\omega_\bathc)\right]}{4\Delta_\bathh}\tanh(\beta_\bathh\Delta_\bathh) \\
\label{eq:W_final}
=&\left(\Delta_\bathc-\Lambda\Delta_\bathh\right)\tanh\left(\beta_\bathc\Delta_\bathc\right)
+\left(\Delta_\bathh-\Lambda\Delta_\bathc\right)\tanh\left(\beta_\bathh\Delta_\bathh\right).
\end{align}

On one hand, one of the dressed-spectrum Otto boundaries, as in Eq.~\eqref{dressedbound1}, gives
\begin{align}
\label{dressedbound1_Theta}
\tanh\left(\beta_\bathc\Delta_\bathc\right)=\tanh\left(\beta_\bathh\Delta_\bathh\right)=\tanh\Theta.
\end{align}
Note that the other dressed-spectrum Otto boundary, as in Eq.~\eqref{dressedbound2}, should not be satisfied here, which means $\Delta_\bathh \neq \Delta_\bathc$.
Substituting this condition into Eq.~\eqref{eq:W_final}, the work of the internally coupled GSLC becomes
\begin{align}
W=&\left[\left(\Delta_\bathc-\Lambda\Delta_\bathh\right)
+\left(\Delta_\bathh-\Lambda\Delta_\bathc\right)\right]\tanh\Theta\\
=&(1-\Lambda)(\Delta_\bathc+\Delta_\bathh)\tanh\Theta.
\end{align}
The angle $\theta_\alpha$ in Eq.~\eqref{theta} and the twice energy gap $\Delta_\alpha$ in Eq.~\eqref{Delta} are positive.
\begin{align}
\tanh\Theta \neq 0,
\qquad
\Delta_\bathc+\Delta_\bathh \neq 0.
\end{align}
The engine boundary $W=0$ of the GSLC coincides with this dressed-spectrum Otto boundary, as in Eq.~\eqref{dressedbound1}, only when
\begin{align}
\Lambda=1.
\end{align}

On the other hand, the other boundary of the dressed-spectrum Otto cycle, as in Eq.~\eqref{dressedbound2}, gives
\begin{align}
\label{dressedbound2_Delta}
\Delta_\bathc = \Delta_\bathh = \Delta.
\end{align}
Substituting this condition into Eq.~\eqref{eq:W_final}, the work of the internally coupled GSLC becomes
\begin{align}
W=&(1-\Lambda)\Delta\left [ \tanh(\beta_\bathc\Delta_\bathc) + \tanh(\beta_\bathh \Delta_\bathh) \right ].
\end{align}
Note that the previous dressed-spectrum Otto boundary in Eq.~\eqref{dressedbound1} should not be satisfied here, which means $\Delta_\bathh \beta_\bathc \neq \Delta_\bathc \beta_\bathh$. 
Similarly, in this case, we have 
\begin{align}
\Delta \neq 0,
\qquad
\tanh(\beta_\bathc\Delta_\bathc) + \tanh(\beta_\bathh \Delta_\bathh) \neq 0,
\end{align}
and hence, the engine boundary $W=0$ of the GSLC also coincides with this dressed-spectrum Otto boundary in Eq.~\eqref{dressedbound2} only when
\begin{align}
\Lambda=1.
\end{align}

Similarly, we can find the boundary of the refrigerator by imposing the condition ${Q_\bathc^\mathrm{cp}}=0$.
Using Eq.~\eqref{QcGibbs} together with the definition of $\Lambda$ in Eq.~\eqref{Lambdacos}, the heat exchanged with the cold bath can be rewritten as
\begin{align}
{Q_\bathc^\mathrm{cp}}=&\frac{4g_\bathc g_\bathh+\omega_\bathc\omega_\bathh}{4\Delta_\bathh}
\tanh(\beta_\bathh\Delta_\bathh)
-\Delta_\bathc \tanh(\beta_\bathc\Delta_\bathc)\\
=&\Lambda\Delta_\bathc\tanh\left(\beta_\bathh\Delta_\bathh\right)
-\Delta_\bathc\tanh\left(\beta_\bathc\Delta_\bathc\right)
\\
\label{eq:Qc_final}
=&\Delta_\bathc\left[\Lambda\tanh\left(\beta_\bathh\Delta_\bathh\right)
-\tanh\left(\beta_\bathc\Delta_\bathc\right)\right].
\end{align}
Hence, the refrigerator boundary is determined by
\begin{equation}
\Lambda\tanh\left(\beta_\bathh\Delta_\bathh\right)
=
\tanh\left(\beta_\bathc\Delta_\bathc\right).
\label{fridge_general_boundary}
\end{equation}

The condition for the boundary of the dressed-spectrum Otto cycle, as in Eq.~\eqref{dressedbound1}, can be rewritten as in Eq.~\eqref{dressedbound1_Theta}. Hence, the refrigerator boundary ${Q_\bathc^\mathrm{cp}} = 0$ as in Eq.~\eqref{fridge_general_boundary} is equivalent to $\Lambda=1$ when it coincides with the dressed-spectrum bound in Eq.~\eqref{dressedbound1}.

When the refrigerator boundary ${Q_\bathc^\mathrm{cp}} = 0$ coincides with the other dressed-spectrum bound in Eq.~\eqref{dressedbound2}, we can rewritten the condition in Eq.~\ref{fridge_general_boundary} as
\begin{align}
\label{Lambdarefrigerator2}
\Lambda = \frac{\tanh\left(\beta_\bathc\Delta\right)}{\tanh\left(\beta_\bathh\Delta\right)},
\end{align}
where the $\Delta$ is given in Eq.~\eqref{dressedbound2_Delta}. The bath temperatures should satisfy $\beta_\bathc>\beta_\bathh$, and Eq.~\eqref{Lambdarefrigerator2} should be given by $\Lambda>1$. However, as in Eq.~\eqref{Lambdacos}, the mismatch factor satisfies $0 < \Lambda \leq 1$. Therefore, the refrigerator boundary ${Q_\bathc^\mathrm{cp}} = 0$ cannot coincide with the dressed-spectrum bound in Eq.~\eqref{dressedbound2}.

Similarly, we find the boundary of the accelerator by imposing the condition ${Q_\bathh^\mathrm{cp}}=0$.
\begin{align}
{Q_\bathh^\mathrm{cp}}
=&\frac{4 g_\bathc g_\bathh+\omega_\bathc \omega_\bathh}{4\Delta_\bathc}\tanh(\beta_\bathc \Delta_\bathc)-\Delta_\bathh \tanh(\beta_{\bathh} \Delta_\bathh)\\ \nonumber
\label{QhGibbsLambda}
 =& \Lambda \Delta_\bathh \tanh\left(\beta_\bathc\Delta_\bathc\right)-\Delta_\bathh\tanh\left(\beta_\bathh\Delta_\bathh\right)
\end{align}
Hence, the accelerator boundary is determined by
\begin{equation}
\Lambda\tanh\left(\beta_\bathc\Delta_\bathc\right)
=
\tanh\left(\beta_\bathh\Delta_\bathh\right).
\label{accelerator_general_boundary}
\end{equation}
When $\Lambda=1$, it coincides with one bound of the dressed-spectrum Otto cycle, as in Eq.~\eqref{dressedbound1}. When $\Lambda = \tanh\left(\beta_\bathh\Delta\right)/\tanh\left(\beta_\bathc\Delta\right)$, it coincides with the other bound of the dressed-spectrum Otto cycle, as in Eq.~\eqref{dressedbound2}.

Unlike the dressed-spectrum Otto cycle, in which the operation regime is only given by the relation between the bath temperatures $\beta_\alpha$ and the energy gaps $\tilde{\omega}_\alpha$, the refrigerator boundary of the GSLC depends not only on the dressed energy gaps $\tilde{\omega}_\alpha$ and the bath temperatures $\beta_\alpha$, but also on the mismatch between the dressed eigenbases characterized by $\Lambda$. 

Mostly, the operational boundaries of the GSLC converge to those of the dressed-spectrum cycle when there is no mismatch between the dressed eigenbases, corresponding to $\Lambda=1$. Uniquely, when the coherence-induced friction factor reaches $\Lambda = \tanh\left(\beta_\bathh\Delta\right)/\tanh\left(\beta_\bathc\Delta\right)$, the accelerator boundary (${Q_\bathh^\mathrm{cp}} = 0$) of the GSLC intersects with the second boundary, as in Eq.~\eqref{dressedbound2}, of the dressed-spectrum Otto cycle. This decoupling of characteristic boundaries fundamentally gives rise to a new thermal phase: the heater mode, as illustrated in Table~\ref{table:GSLCregimes}.

\section{Existence and convergence to a non-equilibrating limit cycle}
\label{A2}

In Sec.~\ref{Sec3} of the main text, we analyze the dynamics of the non-equilibrium limit cycle (NELC) with a finite interaction time. In this appendix, we prove that the Otto cycle with a finite interaction time converges to the NELC that alternates between two non-equilibrium states, which differ from the steady states of the equilibrium limit cycle (DLC) with a long-time interaction.

Rather than focusing on the continuous-time dynamics within each stroke, it is convenient to describe the evolution stroboscopically. Let $\rho_n$ denote the system state at the beginning of the $n$th cycle. The evolution over one complete Otto cycle can be written as a discrete map
\begin{equation}
\rho_{n+1} = \mathcal{M}(\rho_n),
\end{equation}
where the map $\mathcal{M}$ for one cycle is given by
\begin{equation}
\mathcal{M} = \mathcal{E}_\bathc \circ \mathcal{U}_\bathc \circ  \mathcal{U}_\bathh^{-1} \circ \mathcal{E}_\bathh \circ \mathcal{U}_\bathh \circ \mathcal{U}_\bathc^{-1},
\end{equation}
where $\mathcal{U}_{\alpha}(\rho) = U_{\alpha}^\dag \rho U_{\alpha}$ denotes the unitary eigenbasis transformation from the original representation basis of $H_s^\alpha$ to its eigenbasis, while $\mathcal{E}_{\alpha}=\exp[\tilde{\mathcal{L}}^\alpha t_\alpha]$ represents the quantum time evotution generated by the Liouvillian superoperator $\tilde{\mathcal{L}}^\alpha$ of the global GKSL master equation during the hot or cold isochoric stroke over the finite interaction time $t_\alpha$.
Note that $\tilde{\mathcal{L}}^\alpha$ is represented in the eigenbasis of $H_s^\alpha$.
This corresponds to the one-cycle operator in Eq.~\eqref{onecycleoperator}.


Each component of the cycle map $\mathcal{M}$ is completely positive and trace preserving (CPTP). Consequently, their concatenation $\mathcal{M}$ is also a CPTP map acting on a finite-dimensional Hilbert space. Moreover, due to the presence of finite system-bath coupling and internal coupling during the isochoric strokes, the population transfer between any relevant energy eigenstates is allowed, and no nontrivial invariant subspace is preserved. As a result, the one-cycle map $\mathcal{M}$ is primitive.

According to the quantum Perron--Frobenius theorem, a primitive CPTP map admits a unique fixed point $\rho^\star$, corresponding to the unity eigenvalue, while moduli of all other eigenvalues are less than unity, and hence it is a strict contraction mapping on the quantum state space. 
Therefore, for any initial state $\rho_0$, we obtain
\begin{align}
\label{steadycycle}
\lim_{n \to \infty} \mathcal{M}^n(\rho_0) = \rho^\star .
\end{align}
This fixed point $\rho^\star$ comprises a non-equilibrium limit cycle (NELC) of the Otto engine.
The states associated with different strokes in the NELC are given by
\begin{align}
\label{CtoH}
\tilde{\rho}_{\mathrm{NELC}}^\bathh &=\mathcal{E}_\bathh \circ \mathcal{U}_\bathh \circ \mathcal{U}_\bathc^{-1}(\rho^\star),\\
\label{HtoC}
\tilde{\rho}_{\mathrm{NELC}}^\bathc &= \rho^\star
\end{align}
In Sec.~\ref{Sec3-1}, these equations are rewritten as in Eqs.~(\ref{globalsteadycycleH}) and (\ref{globalsteadycycleC}). 

%

\section{Comparison Between Local and Global Master Equations}
\label{A1}
In Sec.~\ref{Sec3} of the main text, we utilize the Gorini--Kossakowski--Sudarshan--Lindblad (GKSL) master equation in the isochoric processes to describe the Markovian quantum open system. In this section, we derive the system states in the isochoric processes using the local and global master equations. By comparing these approaches, we confirm the validity of the global master equation and the invalidity of the local one.

The standard approach of the GKSL master equation neglects the influence of the internal coupling, which is called the local master equation. Considering the influence of the internal coupling, the GKSL master equation is updated using a global approach, yielding the global master equation~\cite{Hofer_2017, Scali2021localmaster, Cattaneo_2019, PhysRevResearch.6.023172}.

\subsection{Local Master Equation}
\label{A1-1}
For comparison between the local and global cases, we briefly summarize the formulation of the former, in which the internal coupling is not included in the dissipative part. The system dynamics is governed by the standard GKSL form
\begin{align}
\label{Liouvillianlocal}
\dot{\rho}_\alpha (t) = -i[H_S^\alpha,\rho_\alpha]
+ \gamma_\alpha^+ \mathcal{D}(\sigma_S^-)
+ \gamma_\alpha^- \mathcal{D}(\sigma_S^+),
\end{align}
with the dissipator as in Eq.~\eqref{dissipator}
\begin{align}
\mathcal{D}(\hat{O}) = \hat{O}\rho \hat{O}^\dagger -\frac{1}{2}\{\hat{O}^\dagger\hat{O},\rho\}, \nonumber
\end{align}
the system Hamiltonian $H_S^\alpha$ as in Eq.~\eqref{HS}, the Pauli matrices $\sigma_S^\pm$, and the dissipation rate $\gamma_\alpha^\pm$.

We set the dissipation rates to
\begin{align}
\gamma_\alpha^\pm &= \gamma(\pm\omega_\alpha,\beta_\alpha),
\end{align}
with the bath correlation function as in Eq.~\eqref{gamma}
\begin{align}
\gamma(\omega,\beta) =
\begin{cases}
J(\omega)\, n(\beta,\omega), & \omega < 0,\\
J(\omega)\,[n(\beta,\omega)+1], & \omega > 0,
\end{cases}\nonumber
\end{align}
with $J(\omega)$ being the spectral density and $n(\beta,\omega)$ the Bose-Einstein distribution:
\begin{align}
&J(\omega) = \Gamma \omega e^{-\omega^s/\omega_{ct}^{1-s}},\nonumber\\
&n(\beta,\omega) = \frac{1}{e^{\beta \omega}-1},\nonumber
\end{align}
where $\omega_\mathrm{ct}$ is the cut-off frequency and $\Gamma$ indicates the coupling strength of the spectral density. The exponent $s$ characterizes the low-frequency behavior of the bath spectral density and determines whether the environment is sub-Ohmic ($s<1$), Ohmic ($s=1$), or super-Ohmic ($s>1$). In the analysis of the Markovian dynamical system in Sec.~\ref{Sec3}, we assume an Ohmic environment and set $s=1$.

Flattening the density matrix $\rho_\alpha$ into a vector $\ket{\psi_\alpha}=(\rhogg^\alpha,\rhoge^\alpha,\rhoeg^\alpha,\rhoee^\alpha)^{T}$, we can write the master equation in terms of a $4\times4$ Liouvillian superoperator:
\begin{align}
\frac{d}{dt}\ket{\psi_\alpha(t)}=\mathcal{L}_\alpha\ket{\psi_\alpha(t)},
\end{align}
with
\begin{align}
\mathcal{L}_\alpha=
\begin{pmatrix}
-\gamma_\alpha^- & i g_\alpha & -i g_\alpha & \gamma_\alpha^+ \\
i g_\alpha & -\frac{1}{2}(\gamma_\alpha^++\gamma_\alpha^- -2 i\omega_\alpha) & 0 & -i g_\alpha \\
- i g_\alpha & 0 & -\frac{1}{2}(\gamma_\alpha^++\gamma_\alpha^- +2 i\omega_\alpha) & i g_\alpha \\
\gamma_\alpha^- & -i g_\alpha & i g_\alpha & -\gamma_\alpha^+
\end{pmatrix}.
\end{align}

The spectral properties of $\mathcal{L}_\alpha$ can be obtained analytically. In particular, the Liouvillian admits a unique zero eigenvalue $\lambda_0^\alpha$ corresponding to a stationary state. The remaining eigenvalues $\lambda_{i=1,2,3}^\alpha$ have negative real parts, ensuring relaxation toward this steady state, given by the three roots of the cubic equation:
\begin{align}
F_3(\lambda) = 8 g^2 (\gamma_\alpha^- + \gamma_\alpha^+ + 2 \lambda) + (\gamma_\alpha^- + \gamma_\alpha^+ + \lambda) ({\gamma_\alpha^-}^2 + \
{\gamma_\alpha^+}^2 + 4 \gamma_\alpha^+ \lambda + 2 \gamma_\alpha^- (\gamma_\alpha^+ + 2 \lambda) + 4 (\lambda^2 + \omega_\alpha^2))
\end{align}
Corresponding to the $i$th eigenvalue $\lambda_i^\alpha$, the left eigenvectors 
\begin{align}
\begin{pmatrix} \rhogg^\alpha & \rhoge^\alpha & \rhoeg^\alpha & \rhoee^\alpha \end{pmatrix}
\end{align}
and the right eigenvectors
\begin{align}
\begin{pmatrix} \rhogg^\alpha & \rhoge^\alpha & \rhoeg^\alpha & \rhoee^\alpha \end{pmatrix}^T
\end{align}
are bi-orthogonal to each other.

The left eigenvectors corresponding to $i$th eigenvalue $\lambda_i^\alpha$ are given by
\begin{align}
\bra{L_0^\alpha} &= \begin{pmatrix} 1 &0 &0 &1 \end{pmatrix},\\
\bra{L_j^\alpha} &= \begin{pmatrix} 
\frac{ \gamma_\alpha^- - \gamma_\alpha^+ +\lambda_j^\alpha}{\gamma_\alpha^- - \gamma_\alpha^+ -\lambda_j^\alpha} \\
-\frac{ i (2 \lambda_j^\alpha)  (8  g_\alpha^2 + 2  (\gamma_\alpha^- + \gamma_\alpha^+)  (\gamma_\alpha^- + \gamma_\alpha^+ - 2  i  \omega_\alpha) + (3  \gamma_\alpha^- + 3  \gamma_\alpha^+ -  2  i  \omega_\alpha)  (2*\lambda_j^\alpha) + (2 \lambda_j^\alpha)^2)}{2  g_\alpha  (-2  \gamma_\alpha^- + 2  \gamma_\alpha^+ + (2 \lambda_j^\alpha))  (\gamma_\alpha^- + \gamma_\alpha^+ - 2  i \omega_\alpha + (2 \lambda_j^\alpha))}\\
-\frac{4  i g_\alpha  (2 \lambda_j^\alpha)}{(-2  \gamma_\alpha^- + 2  \gamma_\alpha^+ + (2 \lambda_j^\alpha))  (\gamma_\alpha^- +\gamma_\alpha^+ - 2  i  \omega + (2*\lambda_j^\alpha))}\\
1 \end{pmatrix}^T.
\end{align}
The right eigenvectors are
\begin{small}
\begin{align}
\label{localR0}
\ket{R_0^\alpha} &= \frac{1}{N_0^\alpha}\begin{pmatrix} 4 g_\alpha^2 (\gamma_\alpha^- + \gamma_\alpha^+) + \gamma_\alpha^- ({\gamma_\alpha^-}^2 + 
    2 \gamma_\alpha^- \gamma_\alpha^+ + {\gamma_\alpha^+}^2 + 4 \omega_\alpha^2) \\
     2 i g (\gamma_\alpha^- - \gamma_\alpha^+) (\gamma_\alpha^- + \gamma_\alpha^+ + 2 i \omega_\alpha) \\
      -2 i g (\gamma_\alpha^- - \gamma_\alpha^+) (\gamma_\alpha^- + \gamma_\alpha^+ - 2 i \omega_\alpha) \\
       4 g_\alpha^2 (\gamma_\alpha^- + \gamma_\alpha^+) + \gamma_\alpha^+ ({\gamma_\alpha^-}^2 + 
    2 \gamma_\alpha^- \gamma_\alpha^+ + {\gamma_\alpha^+}^2 + 4 \omega_\alpha^2) \end{pmatrix}, \\
&\resizebox{0.95\hsize}{!}{$ 
\ket{R_j^\alpha} = \frac{1}{N_j^\alpha}
\begin{pmatrix} 
g_\alpha^2 (\gamma_\alpha^- - \gamma_\alpha^+ - \lambda_j^\alpha) (\gamma_\alpha^- + \gamma_\alpha^+ + 2 \lambda_j^\alpha - 2 i \omega_\alpha) (\gamma_\alpha^- + \gamma_\alpha^+ + 2 \lambda_j^\alpha + 2 i \omega_\alpha) \\
-(i g (\gamma_\alpha^- - \gamma_\alpha^+ - \lambda_j^\alpha) (4 g_\alpha^2 + (\gamma_\alpha^- + \gamma_\alpha^+ + \lambda_j^\alpha) (\gamma_\alpha^- + \gamma_\alpha^+ + 2 \lambda_j^\alpha + 2 i \omega_\alpha)) (\gamma_\alpha^- + \gamma_\alpha^+ + 2 \lambda_j^\alpha -  2 i \omega_\alpha)) \\
-(4 i g^3 (\gamma_\alpha^- - \gamma_\alpha^+ - \lambda_j^\alpha) (\gamma_\alpha^- + \gamma_\alpha^+ + 2 \lambda_j^\alpha - 2  i  \omega_\alpha)) \\
-(g_\alpha^2 (\gamma_\alpha^- - \gamma_\alpha^+ - \lambda_j^\alpha)(\gamma_\alpha^- + \gamma_\alpha^+ + 2 \lambda_j^\alpha - 2 i \omega_\alpha) (\gamma_\alpha^- + \gamma_\alpha^+ + 2 \lambda_j^\alpha + 2 i \omega_\alpha)) 
\end{pmatrix}
$},
\end{align}
\end{small}
where
\begin{align}
N_0^\alpha =& (\gamma_\alpha^- + \gamma_\alpha^+) (8 g_\alpha^2 + {\gamma_\alpha^-}^2 + 2 \gamma_\alpha^- \gamma_\alpha^+ + {\gamma_\alpha^+}^2 + 4 \omega^2),\\
\begin{split}
N_j^\alpha =& \lambda_j^\alpha (32 g_\alpha^4 + (\gamma_\alpha^- + \gamma_\alpha^+ +\lambda_j^\alpha)^2 ({\gamma_\alpha^-}^2 + {\gamma_\alpha^+}^2 + 4 \gamma_\alpha^+\lambda_j^\alpha + 2 \gamma_\alpha^- (\gamma_\alpha^+ + 2\lambda_j^\alpha) + 4 ({\lambda_j^\alpha}^2+ \omega^2))\\ & + 2 g_\alpha^2 (5 {\gamma_\alpha^-}^2 + 5 {\gamma_\alpha^+}^2 + 16 \gamma_\alpha^+\lambda_j^\alpha + 2 \gamma_\alpha^- (5 \gamma_\alpha^+ + 8\lambda_j^\alpha) + 4 (3{\lambda_j^\alpha}^2 + \omega^2)))
\end{split}.
\end{align}

The right eigenvector $\ket{R_0^\alpha}$ is the equilibrium state, corresponding to the zero eigenvalue $\lambda_0^\alpha$ of the local Liouvillian superoperator $\mathcal{L}_\alpha$. Due to the neglect of the internal coupling in the construction of the dissipator, the zero-eigenvalue eigenvector $\ket{R_0^\alpha}$ does not give the Gibbs state of the interacting system Hamiltonian. The local master equation yields a thermodynamically inconsistent limit cycle in the presence of internal coupling and is therefore invalid in our model. 

\subsection{Global Master Equation}
\label{A1-2}
Similar to the derivation of the global master equation in the multi-qubit system~\cite{Hofer_2017, Scali2021localmaster, Cattaneo_2019, PhysRevResearch.6.023172}, to generalize the local master equation to the global one in our single-qubit system, we rewrite the interaction Hamiltonian in the diagonalized basis:
\begin{align}
\tilde{H}_I^\alpha(t) =&e^{i \tilde{H}_S^\alpha t } U_\alpha^\dagger H_I U_\alpha e^{-i \tilde{H}_S^\alpha t }\\
=&e^{i \tilde{H}_S^\alpha t } (U_\alpha^\dagger \sigma_S^x U_\alpha) e^{-i \tilde{H}_S^\alpha t }\sum_{k,\alpha} (V_{k,\alpha} \hat{b}_\alpha + V_{k,\alpha}^* \hat{b}_\alpha^\dagger)\\
=& (-\sin(2\theta)\tilde{\sigma}_S^z +\cos(2\theta) e^{i \tilde{\omega}_\alpha t } \tilde{\sigma}_S^+ +\cos(2\theta) e^{-i \tilde{\omega}_\alpha t } \tilde{\sigma}_S^-) \sum_{k,\alpha} (V_{k,\alpha} \hat{b}_\alpha + V_{k,\alpha}^* \hat{b}_\alpha^\dagger),
\end{align}
where $\tilde{\omega}_\alpha = \epsilon_\alpha^+ - \epsilon_\alpha^- = \sqrt{4g_\alpha^2+\omega_\alpha^2}$. The angle $\theta$, the diagonalized system Hamiltonian, and the diagonalizing transformation $U_\alpha$ are defined in Eqs.~(\ref{theta}), ($\ref{tildeHS}$) and (\ref{Ualpha}), respectively. The Pauli matrices in the diagonalized map are $\tilde{\sigma}_S^z$ and $\tilde{\sigma}_S^\pm$.

In deriving the GKSL master equation, the system operators are decomposed into their Fourier components, and the bath spectral density assigns a weight to each transition frequency. Hence, we obtain the global master equation of the internally coupled system, as in Eq.~\eqref{globalLiouvillian} in the main text:
\[
\frac{d\tilde{\rho}_\alpha}{dt} = \tilde{\mathcal{L}}_\alpha \tilde{\rho}_\alpha = -i [\tilde{H}_S^\alpha, \tilde{\rho}_\alpha] +\tilde{\gamma}_\alpha^- \mathcal{D}(\cos(2\theta)\tilde{\sigma}_S^+) + \tilde{\gamma}_\alpha^+ \mathcal{D}(\cos(2\theta)\tilde{\sigma}_S^-) + \tilde{\gamma}_\alpha(0) \mathcal{D}(-\sin(2\theta)\tilde{\sigma}_S^z),
\]
Under the diagonalized system basis \[\tilde{\rho}_\alpha = \begin{pmatrix} \tilderhogg^\alpha &\tilderhoge^\alpha \\\tilderhoeg^\alpha &\tilderhoee^\alpha \end{pmatrix},\] as in Eq.~(\ref{tildeHS}), there is no energy gap for the transition \[\tilde{\sigma}_S^z = \begin{pmatrix}1 &0\\ 0 & -1 \end{pmatrix},\] while the excitation frequency $\Delta \omega = \tilde{\omega}_\alpha$ is associated with \[\tilde{\sigma}_S^+ = \begin{pmatrix}0 & 0 \\ 1 & 0\end{pmatrix},\] and the de-excitation frequency $\Delta \omega = -\tilde{\omega}_\alpha$ is associated with \[\tilde{\sigma}_S^- = \begin{pmatrix}0 & 1 \\ 0 & 0\end{pmatrix}.\]
 
Assuming the Ohmic spectral density $J(\omega) \propto \omega$ again, we have $\tilde{\gamma}_\alpha(0) = 0$, and the other dissipative rates are as in Eq.~\eqref{tildegamma}
\begin{align}
&\tilde{\gamma}_\alpha^\pm = \gamma(\pm\tilde{\omega}_\alpha,\beta_\alpha),\nonumber
\end{align}
where the dissipator $\mathcal{D}(\hat{O})$ and the dissipation function $\gamma(\omega,\beta)$ are defined in Eqs.~(\ref{dissipator}) and (\ref{gamma}). Hence, the global Liouvillian can be expressed as Eq.~\eqref{globalmatrix} in the main text:
\[
\tilde{\mathcal{L}}_\alpha = \begin{pmatrix}
-\tilde{\Gamma}_\alpha^- &0 &0 &\tilde{\Gamma}_\alpha^+\\
0 &-\frac{1}{2}(\tilde{\Gamma}_\alpha^+ + \tilde{\Gamma}_\alpha^-)+i \tilde{\omega}_\alpha &0 &0 \\
0 &0 &-\frac{1}{2}(\tilde{\Gamma}_\alpha^+ + \tilde{\Gamma}_\alpha^-)-i \tilde{\omega}_\alpha &0\\
\tilde{\Gamma}_\alpha^- &0 &0 &-\tilde{\Gamma}_\alpha^+,
\end{pmatrix}
\]
with
\[
\tilde{\Gamma}_\alpha^\pm = \cos^2(2 \theta_\alpha) \tilde{\gamma}_\alpha^\pm
\]
as in Eq.~\eqref{globalGamma} in the main text. 

The basis is $ \begin{pmatrix} \tilderhogg & \tilderhoge &\tilderhoeg & \tilderhoee \end{pmatrix}^T$. 
For the subsequent analysis, we rewrite the left and right eigenvectors $\bra{\tilde{L}_i^\alpha}$ and $\ket{\tilde{R}_i^\alpha}$ of the $i$th eigenvalue $\tilde{\lambda}_i^\alpha$ as the left and right eigenmatrices 
\begin{align}
\tilde{\sigma}_i^\alpha = \begin{pmatrix} \tilderhogg & \tilderhoeg \\ \tilderhoge & \tilderhoee \end{pmatrix}\nonumber
\end{align} 
and 
\begin{align}
\tilde{\rho}_i^\alpha = \begin{pmatrix} \tilderhogg & \tilderhoge \\ \tilderhoeg & \tilderhoee \end{pmatrix},\nonumber
\end{align} respectively.
The left and right eigenmatrices are bi-orthogonal $\Tr[\tilde{\sigma}_i^\alpha \tilde{\rho}_j^\alpha]=\delta_{i,j}$.

The spectral properties of the global Liouvillian superoperator $\tilde{\mathcal{L}}_\alpha$ can be obtained analytically. The eigenvalues $\tilde{\lambda}_i^\alpha$ are 
\begin{align}
&\tilde{\lambda}_0^\alpha = 0,\nonumber \\
&\tilde{\lambda}_1^\alpha = -\frac{1}{2} (\tilde{\Gamma}_\alpha^- + \tilde{\Gamma}_\alpha^+) + i \tilde{\omega}_\alpha,\nonumber \\
&\tilde{\lambda}_2^\alpha = -\frac{1}{2} (\tilde{\Gamma}_\alpha^- + \tilde{\Gamma}_\alpha^+) - i \tilde{\omega}_\alpha,\nonumber \\
&\tilde{\lambda}_3^\alpha = -(\tilde{\Gamma}_\alpha^- + \tilde{\Gamma}_\alpha^+),
\end{align}
as in Eqs.~\eqref{globallambda0}--\eqref{globallambda3} in the main text. 
In particular, the zero eigenvalue $\tilde{\lambda}_0^\alpha$ corresponds to a stationary state.

Corresponding to the $i$th eigenvalue $\tilde{\lambda}_i^\alpha$, the left eigenmatrices $\tilde{\sigma}_i^\alpha$ are
\begin{align}
\label{globalL0}
& \tilde{\sigma}_0^\alpha = \begin{pmatrix}
1 &0 \\ 0 &1
\end{pmatrix},\\
& \tilde{\sigma}_1^\alpha = \begin{pmatrix}
0 &0\\ 1 &0
\end{pmatrix},\\
& \tilde{\sigma}_2^\alpha = \begin{pmatrix}
0 &1\\ 0 &0
\end{pmatrix},\\
\label{globalL3}
& \tilde{\sigma}_3^\alpha = \begin{pmatrix}
\tilde{\Gamma}_\alpha^- &0 \\ 0 &-\tilde{\Gamma}_\alpha^+
\end{pmatrix}.
\end{align}
The right eigenmatrices $\tilde{\rho}_i^\alpha$ are
\begin{align}
\label{globalR0}
& \tilde{\rho}_0^\alpha = \frac{1}{\tilde{\Gamma}_\alpha^- + \tilde{\Gamma}_\alpha^+}\begin{pmatrix}
\tilde{\Gamma}_\alpha^+ &0 \\ 0 &\tilde{\Gamma}_\alpha^-
\end{pmatrix},\\
& \tilde{\rho}_1^\alpha = \begin{pmatrix}
0 &1 \\ 0 &0
\end{pmatrix},\\
& \tilde{\rho}_2^\alpha = \begin{pmatrix}
0 &0 \\ 1 &0
\end{pmatrix},\\
\label{globalR3}
& \tilde{\rho}_3^\alpha = \frac{1}{\tilde{\Gamma}_\alpha^- + \tilde{\Gamma}_\alpha^+}\begin{pmatrix}
1 &0 \\ 0 &-1
\end{pmatrix}.
\end{align}

The right eigenmatrix $\tilde{\rho}_0^\alpha$ is the equilibrium steady state in each isochoric process $\alpha$ ($\alpha=\bathh,\bathc$), corresponding to the zero eigenvalue $\tilde{\lambda}_0^\alpha=0$ of the global Liouvillian superoperator $\tilde{\mathcal{L}}$. Obviously, it is a thermalized equilibrium state and exhibits properties similar to those of the Gibbs state.
Hence, in the main text, we focus on the global master equation, which properly incorporates the system eigenstructure and yields physically consistent steady-cycle behavior.

\section{Dressed-Spectrum Non-Equilibrating Limit Cycle}
\label{A3}

For an internally coupled non-equilibrating limit cycle (NELC), the dynamics of the local and global GKSL master equations differ due to the internal coupling, as discussed in Appendix~\ref{A1}. However, the global and local approaches are equivalent in the dressed-spectrum NELC, since it can be regarded as a system without internal coupling governed by the effective Hamiltonian. 
In this section, we derive the non-equilibrating limit cycle (NELC) of the dressed-spectrum system.

The master equation of the dressed-spectrum system in the isochoric process is given by
\begin{equation}
\label{Liouvilliandressed}
\dot{\rho}_{\rm ds}^\alpha = \mathcal{L}^\alpha_{\rm ds}(\rho_{\rm ds}) = -i[\tilde{H}_S^\alpha, \rho_{\rm ds}] + \tilde{\gamma}_\alpha^+ \mathcal{D}(\sigma_S^-) +\tilde{\gamma}_\alpha^- \mathcal{D}(\sigma_S^+).
\end{equation}
The dissipator $\mathcal{D}(\hat{O})^\alpha$ is defined in Eq.~\eqref{dissipator}. The effective Hamiltonian is defined in Eq.~\eqref{tildeHS}. The excitation frequency $\Delta \omega = \tilde{\omega}_\alpha$ is associated with \[\sigma_S^+ = \begin{pmatrix}0 & 0 \\ 1 & 0\end{pmatrix},\] and the de-excitation frequency $\Delta \omega = -\tilde{\omega}_\alpha$ is associated with \[\sigma_S^- = \begin{pmatrix}0 & 1 \\ 0 & 0\end{pmatrix}.\]

Different from the global Liouvillian superoperator $\tilde{\mathcal{L}}^\alpha$ in Eq.~\eqref{globalmatrix} of the internally coupled NELC, in which the dissipation rates $\tilde{\Gamma}_\alpha^\pm$ in Eq.~\eqref{globalGamma} are influenced by an extra term $\cos^2(2\theta_\alpha)$, the dissipation rates $\tilde{\gamma}_\alpha^\pm$ here is defined in Eq.~\eqref{tildegamma}, which only depend on the effective energy gap $\tilde{\omega}_\alpha$ and the bath temperatures $\beta_\alpha$.

By vectorizing the density matrix as $\vec{\rho}_{\rm ds} = (\rhogg, \rhoge, \rhoeg, \rhoee)^T$ in the Liouville space, the superoperator $\mathcal{L}^\alpha$ can be represented as a $4 \times 4$ matrix
\begin{align}
\label{dressedmatrix}
\mathcal{L}^\alpha_{\rm ds} = \begin{pmatrix}
-\tilde{\gamma}_\alpha^- &0 &0 &\tilde{\gamma}_\alpha^+\\
0 &-\frac{1}{2}(\tilde{\gamma}_\alpha^+ + \tilde{\gamma}_\alpha^-)+i \tilde{\omega}_\alpha &0 &0 \\
0 &0 &-\frac{1}{2}(\tilde{\gamma}_\alpha^+ + \tilde{\gamma}_\alpha^-)-i \tilde{\omega}_\alpha &0\\
\tilde{\gamma}_\alpha^- &0 &0 &-\tilde{\gamma}_\alpha^+,
\end{pmatrix}.
\end{align}
By diagonalizing this matrix, we obtain the four eigenvalues $\lambda_i^\alpha$:
\begin{align}
\label{dressedeigenvalues}
\lambda_0^\alpha &= 0,  \\
\lambda_1^\alpha &= -\frac{\tilde{\gamma}_\alpha^+ + \tilde{\gamma}_\alpha^-}{2} + i\tilde{\omega}_\alpha, \\
\lambda_2^\alpha &=  -\frac{\tilde{\gamma}_\alpha^+ + \tilde{\gamma}_\alpha^-}{2} - i\tilde{\omega}_\alpha, \\
\lambda_3^\alpha &= -(\tilde{\gamma}_\alpha^+ + \tilde{\gamma}_\alpha^-).
\end{align}

Corresponding to the $i$th eigenvalue $\lambda_i^\alpha$, the left eigenmatrices $\sigma_i^\alpha$ are
\begin{align}
\label{dressedsigma0}
& \sigma_0^\alpha = \begin{pmatrix}
1 &0 \\ 0 &1
\end{pmatrix},\\
\label{dressedsigma1}
& \sigma_1^\alpha = \begin{pmatrix}
0 &0\\ 1 &0
\end{pmatrix},\\
\label{dressedsigma2}
& \sigma_2^\alpha = \begin{pmatrix}
0 &1\\ 0 &0
\end{pmatrix},\\
\label{dressedsigma3}
& \sigma_3^\alpha = \begin{pmatrix}
\tilde{\gamma}_\alpha^- &0 \\ 0 &-\tilde{\gamma}_\alpha^+
\end{pmatrix}.
\end{align}
The right eigenmatrices $\rho_i^\alpha$ are
\begin{align}
\label{dressedrho0}
& \rho_0^\alpha = \frac{1}{\tilde{\gamma}_\alpha^- + \tilde{\gamma}_\alpha^+}\begin{pmatrix}
\tilde{\gamma}_\alpha^+ &0 \\ 0 &\tilde{\gamma}_\alpha^-
\end{pmatrix},\\
\label{dressedrho1}
& \rho_1^\alpha = \begin{pmatrix}
0 &1 \\ 0 &0
\end{pmatrix},\\
\label{dressedrho2}
& \rho_2^\alpha = \begin{pmatrix}
0 &0 \\ 1 &0
\end{pmatrix},\\
\label{dressedrho3}
& \rho_3^\alpha = \frac{1}{\tilde{\gamma}_\alpha^- + \tilde{\gamma}_\alpha^+}\begin{pmatrix}
1 &0 \\ 0 &-1
\end{pmatrix}.
\end{align}
The right eigenmatrix $\rho_0^\alpha$ is the equilibrium steady state in each isochoric process $\alpha$ ($\alpha=\bathh,\bathc$), corresponding to the zero eigenvalue $\lambda_0^\alpha=0$ of the dressed Liouvillian superoperator $\mathcal{L}^\alpha_{\rm ds}$ in Eq.~\eqref{dressedmatrix}. 

We can expand the state using the complete eigenbasis as
\begin{equation}
\rho_{\rm ds}^\alpha = \sum_{i} c_i^\alpha \rho_i^\alpha.
\end{equation}
When the interaction time is infinite, the dressed-spectrum NELC approaches the ELC, in which $\rho_{\rm ds}^\alpha = \rho_0^\alpha$. If the interaction time is finite, the non-equilibrium terms $\rho_i^\alpha$ ($i\neq 0)$ are not negligible. 

Similarly to the internally coupled NELC, we can impose the dressed-spectrum NELC conditions for the coefficients $c_k^\alpha$. Without the internal coupling, the eigenbasis remains the same among all the processes in the limit cycle. The periodically steady relation can be simplified as
\begin{align}
\label{dressedsteadycycleH}
\rho_{\rm ds}^\bathh = \sum_{i=0}^3 \mathrm{Tr}[\sigma_i^\bathh  \rho_{\rm ds}^\bathc ] \rho_i^\bathh,\\
\label{dressedsteadycycleC}
\rho_{\rm ds}^\bathc = \sum_{i=0}^3 \mathrm{Tr}[\sigma_i^\bathc \rho_{\rm ds}^\bathh ] \rho_i^\bathc.
\end{align}
Based on these simplified periodically steady conditions, the relations between the expansion coefficients $c_k^\alpha$ are given directly by 
\begin{align}
\label{dressedcH1}
c_k^\bathh =& \sum_{j=0}^3 c_j^\bathc \mathrm{Tr} [\sigma_k^\bathh \rho_j^\bathc] e^{\lambda_k^\bathh t_\bathh} \\
\label{dressedcH2}
=& \sum_{i=0}^3 c_i^\bathh \sum_{j=0}^3 \mathrm{Tr}[\sigma_j^\bathc \rho_i^\bathh] \mathrm{Tr}[\sigma_k^\bathh \rho_j^\bathc] e^{\lambda_j^\bathc t_\bathc + \lambda_k^\bathh t_\bathh},\\
\label{dressedcC1}
c_k^\bathc =& \sum_{j=0}^3 c_j^\bathh \mathrm{Tr} [\sigma_k^\bathc \rho_j^\bathh] e^{\lambda_k^\bathc t_\bathc} \\
\label{dressedcC2}
=&\sum_{i=0}^3  c_i^\bathc \sum_{j=0}^3 \mathrm{Tr}[\sigma_j^\bathh \rho_i^\bathc] \mathrm{Tr}[\sigma_k^\bathc \rho_j^\bathh] e^{\lambda_j^\bathh t_\bathh + \lambda_k^\bathc t_\bathc}.
\end{align}

As in Eqs.~\eqref{dressedsigma0}--\eqref{dressedrho3}, we have $\sigma_i^\bathh  = \sigma_i^\bathc$ and $\rho_i^\bathh  = \rho_i^\bathc$ for $i=1,2$. 
Consequently, the steady cycle condition for the coherence coefficient $c_1^\alpha$ drastically simplifies to
\begin{equation}
c_1^\alpha \left( 1 - e^{\lambda_1^\bathh t_\bathh + \lambda_1^\bathc t_\bathc} \right) = 0.
\end{equation}
Since the real parts of the eigenvalues $\lambda_1^\alpha$ are strictly negative, in which $-(\tilde{\gamma}_\alpha^+ + \tilde{\gamma}_\alpha^-) / 2 < 0$, the exponential term strictly decays $|e^{\lambda_1^\bathh t_\bathh + \lambda_1^\bathc t_\bathc}| < 1$ for finite interaction time. Therefore, the only physically permissible solution is 
\begin{equation}
\label{dressedc1final}
c_1^\alpha  = 0.
\end{equation}
Similarly, we also confirm that
\begin{align}
\label{dressedc2final}
c_2^\alpha = 0.
\end{align}
Hence, the dynamics are entirely constrained within the population subspace spanned by $\rho_0^\alpha$, which has a zero eigenvalue, and $\rho_3^\alpha$, which has eigenvalue $\lambda_3^\alpha$: 
\begin{align}
\rho_{\rm ds}^\alpha = c_0^\alpha \rho_0^\alpha +c_3^\alpha \rho_3^\alpha.
\end{align}
Both of these eigenvalues are real, leading to classical monotonic charging and discharging of populations without transient power peaks or efficiency oscillations.

In Eqs.~\eqref{dressedsigma0}, we also have $\sigma_0^\bathh  = \sigma_0^\bathc$. The trace of the system should be unity. In Eqs.~\eqref{dressedrho0} and \eqref{dressedrho3}, we have $\mathrm{Tr}[\rho_0^\alpha]=1$ and $\mathrm{Tr}[\rho_3^\alpha]=0$.
Consequently, the steady cycle condition for the equilibrium-term coefficient $c_0^\alpha$ simplifies to
\begin{equation}
\label{dressedc0final}
c_0^\alpha = 1.
\end{equation}
Hence, the state of the dressed-spectrum NELC is given by
\begin{align}
\label{dressedNELCstatefinal}
\rho_{\rm ds}^\alpha = \rho_0^\alpha + c_3^\alpha \rho_3^\alpha,
\end{align}
where
\begin{align}
\label{dressedc3hfinal}
c_3^\bathh =&\frac{\mathrm{Tr}[\sigma_3^\bathh \rho_0^\bathc] e^{\lambda_3^\bathh t_\bathh} + \mathrm{Tr}[\sigma_3^\bathh \rho_3^\bathc]  \mathrm{Tr}[\sigma_3^\bathc \rho_0^\bathh] e^{\lambda_3^\bathh t_\bathh+\lambda_3^\bathc t_\bathc}}{1- \mathrm{Tr}[\sigma_3^\bathh \rho_3^\bathc]  \mathrm{Tr}[\sigma_3^\bathc \rho_3^\bathh] e^{\lambda_3^\bathh t_\bathh+\lambda_3^\bathc t_\bathc}},\\
\label{dressedc3cfinal}
c_3^\bathc =&\frac{\mathrm{Tr}[\sigma_3^\bathc \rho_0^\bathh] e^{\lambda_3^\bathc t_\bathc} + \mathrm{Tr}[\sigma_3^\bathc \rho_3^\bathh]  \mathrm{Tr}[\sigma_3^\bathh \rho_0^\bathc] e^{\lambda_3^\bathc t_\bathc+\lambda_3^\bathh t_\bathh}}{1- \mathrm{Tr}[\sigma_3^\bathc \rho_3^\bathh]  \mathrm{Tr}[\sigma_3^\bathh \rho_3^\bathc] e^{\lambda_3^\bathc t_\bathc+\lambda_3^\bathh t_\bathh}}.
\end{align}

Similarly to Eqs.~(\ref{QhNELC})--(\ref{WNELC}), the total energy flows during the isochoric processes are given by the system Hamiltonian and the states in the original basis. For the dressed-spectrum NELC, we can rewrite them as the functions of the interaction times $t_\bathh$ and $t_\bathc$, given by
\begin{align}
\label{QhdressedNELC}
{Q_\bathh^\mathrm{cp}}(t_\bathh,t_\bathc) &= \Tr[\tilde{H}_S^\bathh (\rho_{\rm ds}^\bathh - \rho_{\rm ds}^\bathc)],\\
\label{QcdressedNELC}	
{Q_\bathc^\mathrm{cp}}(t_\bathh,t_\bathc) &= \Tr[\tilde{H}_S^\bathc (\rho_{\rm ds}^\bathc - \rho_{\rm ds}^\bathh)].
\end{align}
In each work-production process, the total work extracted from the environment is given by
\begin{align}
\label{W1dressedNELC}
{W_1^\mathrm{cp}}(t_\bathh,t_\bathc)&= \Tr[(\tilde{H}_S^\bathc-\tilde{H}_S^\bathh)\rho_{\rm ds}^\bathh], \\ 
\label{W2dressedNELC}
{W_2^\mathrm{cp}}(t_\bathh,t_\bathc)&= \Tr[(\tilde{H}_S^\bathh-\tilde{H}_S^\bathc)\rho_{\rm ds}^\bathc],
\end{align}
and the total work is given by
\begin{align}
\label{WdressedNELC}
W(t_\bathh,t_\bathc) ={W_1^\mathrm{cp}}(t_\bathh,t_\bathc) + {W_2^\mathrm{cp}}(t_\bathh,t_\bathc) =  -\left [ {Q_\bathh^\mathrm{cp}}(t_\bathh,t_\bathc)+{Q_\bathc^\mathrm{cp}}(t_\bathh,t_\bathc)\right ],
\end{align}
which still obeys the first law of thermodynamics. 
Hence, the efficiency and the power of the dressed-spectrum NELC are given by
\begin{align}
\label{efficiencydressedNELC}
\eta(t_\bathh,t_\bathc) = -\frac{W(t_\bathh,t_\bathc)}{{Q_\bathh^\mathrm{cp}}(t_\bathh,t_\bathc)},\\
\label{powerdressedNELC}
P(t_\bathh,t_\bathc) = -\frac{W(t_\bathh,t_\bathc)}{t_\bathh+t_\bathc}.
\end{align}
and the coefficient of performance (COP) is given by
\begin{align}
\label{COPdressedNELC}
\xi(t_\bathh,t_\bathc) = \frac{{Q_\bathc^\mathrm{cp}}(t_\bathh,t_\bathc)}{W(t_\bathh,t_\bathc)}.
\end{align}
With infinite-time interaction, the dressed-spectrum NELC approaches the dressed-spectrum ELC, as discussed in Sec.~\ref{Sec3-2}, which shows the same thermal performance as the dressed-spectrum GSLC in Sec.~\ref{Sec2} with the thermal bounds defined in Sec.~\ref{Sec1-2}.

\bibliography{reference}
\end{document}